\documentclass{article}
\pdfoutput=1
\usepackage[utf8]{inputenc}
\usepackage{jcappub}
\usepackage[dvipsnames]{xcolor}
\usepackage{amssymb,amsmath}
\usepackage{subcaption}
\usepackage{graphicx}
\usepackage{multirow}
 \usepackage[stable]{footmisc}

\newcommand{\Rsp}[1]{#1}

\author[a,b]{Chirag Modi,}
\author[a,b]{Emanuele Castorina,}
\author[a,b]{Yu Feng,}
\author[a,b,c]{Martin White}

\affiliation[a]{Department of Physics, University of California, Berkeley, CA 94720}
\affiliation[b]{Berkeley Center for Cosmological Physics, Berkeley, CA 94720}
\affiliation[c]{Department of Astronomy, University of California, Berkeley, CA 94720}

\emailAdd{modichirag@berkeley.edu}
\emailAdd{ecastorina@berkeley.edu}
\emailAdd{yfeng1@berkeley.edu}
\emailAdd{mwhite@berkeley.edu}

\title{Intensity mapping with neutral hydrogen and the Hidden Valley simulations}
\date{December 2018}

\abstract{This paper introduces the HiddenValley simulations, a set of trillion-particle N-body simulations in gigaparsec volumes aimed at intensity mapping science.  We present details of the simulations and their convergence, then specialize to the study of 21-cm fluctuations between redshifts 2 and 6.  Neutral hydrogen is assigned to halos using three prescriptions, and we investigate the clustering in real and redshift-space at the 2-point level.  In common with earlier work we find the bias of HI increases from near 2 at $z=2$ to 4 at $z=6$, becoming more scale dependent at high $z$.  The level of scale-dependence and decorrelation with the matter field are as predicted by perturbation theory.  Due to the low mass of the hosting halos, the impact of fingers of god is small on the range relevant for proposed 21-cm instruments.  We show that baryon acoustic oscillations and redshift-space distortions could be well measured by such instruments.  Taking advantage of the large simulation volume, we assess the impact of fluctuations in the ultraviolet background, which change HI clustering primarily at large scales.}

\begin{document}
\maketitle
\flushbottom

\section{Introduction}

The study of the large-scale structure of the Universe promises to shed light on a wide variety of topics from theories of the early Universe to the formation and evolution of galaxies within the evolving cosmic web.  The high redshift Universe, in particular, combines a large cosmological volume for precise statistical inference with a more linear density field which can be more reliably modeled and which is better correlated with the initial conditions.  Accessing the high redshift Universe in 3D with a high enough density of tracers over a large fraction of the sky is, however, observationally challenging since the galaxies become fainter and their spectroscopic detection harder. In this regard intensity mapping has emerged as a promising means of efficiently mapping the high $z$ Universe.

Intensity mapping methods carry out a low resolution, spectroscopic survey to measure integrated flux from unresolved sources on large areas of sky at different frequencies. They capture the largest elements of the cosmic web and map out the distribution of matter in very large cosmological volumes in a fast and efficient manner, with good radial resolution \cite{Kovetz17}. Emission from cosmic neutral hydrogen (HI) offers one tracer to map out the Universe in such a way. The ground state of neutral hydrogen is split into two energy levels because of the spin-spin interaction between the electron and the proton and when the electron makes a transition from the higher energy level to the lower one, it emits a photon with the rest frame wavelength of $21\,$cm. This 21-cm signal provides an efficient way to probe the spatial distribution of neutral hydrogen, and hence the underlying dark matter from the local Universe to the dark ages \cite{Kovetz17,CVDE-21cm}.

 A number of upcoming (CHIME \cite{CHIME}, HIRAX \cite{HIRAX}, BINGO \cite{Bingo}, Tianlai \cite{Tianlai}, SKA \cite{SKACosmo}) or planned (CVDE \cite{CVDE-21cm}) HI intensity mapping instruments will use the 21-cm signal to probe the redshift window $2<z<6$, which remains largely unexplored on cosmological scales. There is strong theoretical motivation to survey this redshift window since it increases our lever arm for constraining any deviation from the standard paradigm of $\Lambda$CDM model, such as alternative models of dark energy and modified gravity. For example, one of the primary goal of the aforementioned experiments is to constrain the expansion history of the Universe in the pre-acceleration era by measuring the baryon-acoustic-oscillations (BAO) in the clustering of the 21-cm signal.

State-of-the art cosmological analyses of large-scale structure rely on high fidelity numerical simulations and on accurate analytical modeling of the expected signal, the former frequently being used to test the latter.  The goal of sampling the largest possible volume while simultaneously maintaining high enough mass resolution to resolve the low-mass dark matter halos which host the majority of the neutral hydrogen makes simulating 21-cm signal quite challenging. At the same time the very large survey volumes and low noise levels expected in 21-cm measurements will require characterization of the theory to unprecedented levels.

In this paper we present a series of large, N-body simulations designed to model the signals relevant to intensity mapping surveys.  Each simulation evolves $10240^3$ particles in a $1\,h^{-1}$Gpc volume, simultaneously providing large volumes for precise measurements while resolving the halos most relevant for intensity mapping surveys.  This makes it possible to address several modeling issues with higher precision than has been possible to date and at the same time analyze the response of the observed signal to these modeling choices.

Our initial focus will be on modeling the 21-cm signal over the range $2<z<6$, using a halo model formalism.  The halo model combines a large degree of astrophysical realism while maintaining a flexibility which is highly desirable in our current state of ignorance about the high-$z$, 21-cm signal.  We will present several different models motivated by existing theoretical ideas and simulations and assess their commonalities and differences, with the expectation that these models will need to be revised as further observational information becomes available.

This paper is organized as follows.  We begin by outlining the requirements imposed upon the simulations by the challenge of modeling 21-cm emission in Section \ref{sec:requirements}. 
In Section \ref{sec:sims} we present the `Hidden Valley' simulations, describing briefly the N-body code used, validation of the results and basic performance metrics with the details of code development in the Appendix \ref{app:scaling}. 
Section \ref{sec:model} describes the different semi-analytic and halo based models we use to populate the simulated halos with HI as well as their calibration. The details of the satellite model used for this purpose are discussed in Appendix \ref{app:satellites}. 
We then use these models to look at the clustering of HI in real- and redshift space in Section \ref{sec:clustering}, with separate focus on redshift space distortions to the clustering and baryon acoustic oscillations.  This section also presents a preliminary comparison with perturbative, analytic models.
In Section \ref{sec:response}, we study the response of HI clustering to different parameters and show how going beyond linear theory breaks denegeneracies amongst them, as anticipated by ref.~\cite{CasWhi19}.
Then we use our simulations in Section \ref{sec:intensitybias} to model spatially fluctuating ultraviolet ionizing background and study the response of the 21-cm signal to these background fluctuations.  Finally we summarize our key results and some further avenues for investigation in Section \ref{sec:conclusions}.

%%%%%%%%%%%%%%%%%%%%%%%%%%%%%%%%%%%%%%%%%
\section{Setting the requirements}
\label{sec:requirements}

We begin by discussing the physical and observational constraints that drive the requirements on the numerical simulation suite.  We start with a discussion of the required mass resolution (set by ideas about how HI inhabits dark matter halos) and then turn to the epochs and length scales that will be probed by proposed experiments.

\subsection{Mass resolution: the HI-halo connection}

In the post-reionization era most of the hydrogen in the Universe is ionized, and the 21-cm signal comes from self-shielded regions such as galaxies (between the outskirts of disks until where the gas becomes molecular within star-forming regions).
Unfortunately, there are not many observational constraints on the manner in which HI traces galaxies and halos in the high-$z$ Universe, so we are forced to rely on numerical simulations \cite{Dave13,Eagle,VN18} and inferences from other observations \cite{Padmanabhan17,Castorina17}.

Both physical intuition and numerical simulations suggest that there is a minimum halo mass ($M_{\rm cut}$) below which neutral hydrogen will not be self-shielded from UV photons.  Above this mass the amount of HI should increase as the halo mass increases, though not necessarily linearly.   A reasonable model of the distribution of HI in halos is thus \cite{Padmanabhan15,Castorina17}
\begin{equation}
    M_{HI} (M_h;z)  = A(z) (M_h/M_{\rm cut})^{\alpha(z)} e^{-M_{\rm cut}(z)/M_h}
\label{eqn:HI_HOD}
\end{equation}
where we have explicitly written the redshift dependence of the model parameters.
The normalization constant can be fixed by matching the global HI abundance (see below).  Simulations suggest $\alpha\approx 1$ at $z>2$.
The characteristic halo mass scale ($M_{\rm cut}$) is determined by the clustering of HI.  \Rsp{At $z\simeq0$ the HOD parameters are known from a joint analysis of HI-selected galaxies and groups in the Sloan Digital Sky Survey \cite{Obuljen2019}.} At $z\simeq 1$ this number is known approximately from cross-correlation with optical galaxies \cite{Masui13,Switzer13,Anderson18}, however at higher $z$ there are no direct measurements.  The clustering of Damped Lyman$\alpha$ systems (DLAs) has been measured at $z\simeq 2-3$ by ref.~\cite{Rafols}, and since the DLAs contain the majority of the HI at those redshifts this can be used as a proxy for the HI clustering amplitude.
For the model in Eq.~(\ref{eqn:HI_HOD}) the DLA measurement implies $M_{\rm cut} = \mathrm{few}\times 10^9\,h^{-1}M_\odot$ if $\alpha\simeq 1$ \cite{Castorina17}.
These numbers agree with an analysis of the most recent hydrodynamical simulations (see Table 6 of ref.~\cite{VN18}).
This requirement sets a particle mass resolution of about $10^8\,h^{-1}M_\odot$ or better in order to be able to resolve the majority of the HI.

%%%%%%%%%%%%%%%%%%%%%
\subsection{Volume and force resolution: relevant scales}

The scope for the simulations and model has been set by considering the Stage {\sc ii} 21-cm experiment suggested by ref.~\cite{CVDE-21cm}, i.e.~a compact, square array of $256\times 256$, fully illuminated $6\,$m dishes observing half the sky with frequencies corresponding to $2<z<6$.  We shall use this strawman proposal in order to motivate the range of scales our simulations should aim to resolve and the level of accuracy desired. At $z<2.5$ the simulations could also be useful for CHIME and HIRAX, and at $z>2$ for experiments targeting CO and/or C-II lines.

The Stage {\sc ii} experiment is an interferometer and as such it makes measurements directly in the Fourier domain.  The correlation between every pair of feeds, $i$ and $j$, measures the Fourier transform of the sky emission times the primary beam at a wavenumber, $k_\perp = 2\pi \vec{u}_{ij}/ \chi(z)$, set by the spacing of the two feeds in units of the observing wavelength ($\vec{u}_{ij}$) \cite{TMS17}.
The visibility noise is inversely proportional to the number (density) of such baselines \cite{ZalFurHer04,McQ06,Seo2010,Bull2015,SeoHir16,Wol17,Alonso17,White17,Obuljen18}.
Where necessary, we take the noise parameters from Refs.~\cite{CVDE-21cm,Chen19}.
A defining feature of the Stage {\sc ii} experiment is that the total noise is dominated by shot noise at low $k$, in contrast with previous surveys that were thermal-noise dominated \cite{Cohn16,Obuljen18,CHIME,HIRAX,Tianlai}.

\begin{figure}
    \centering
    \resizebox{\columnwidth}{!}{\includegraphics{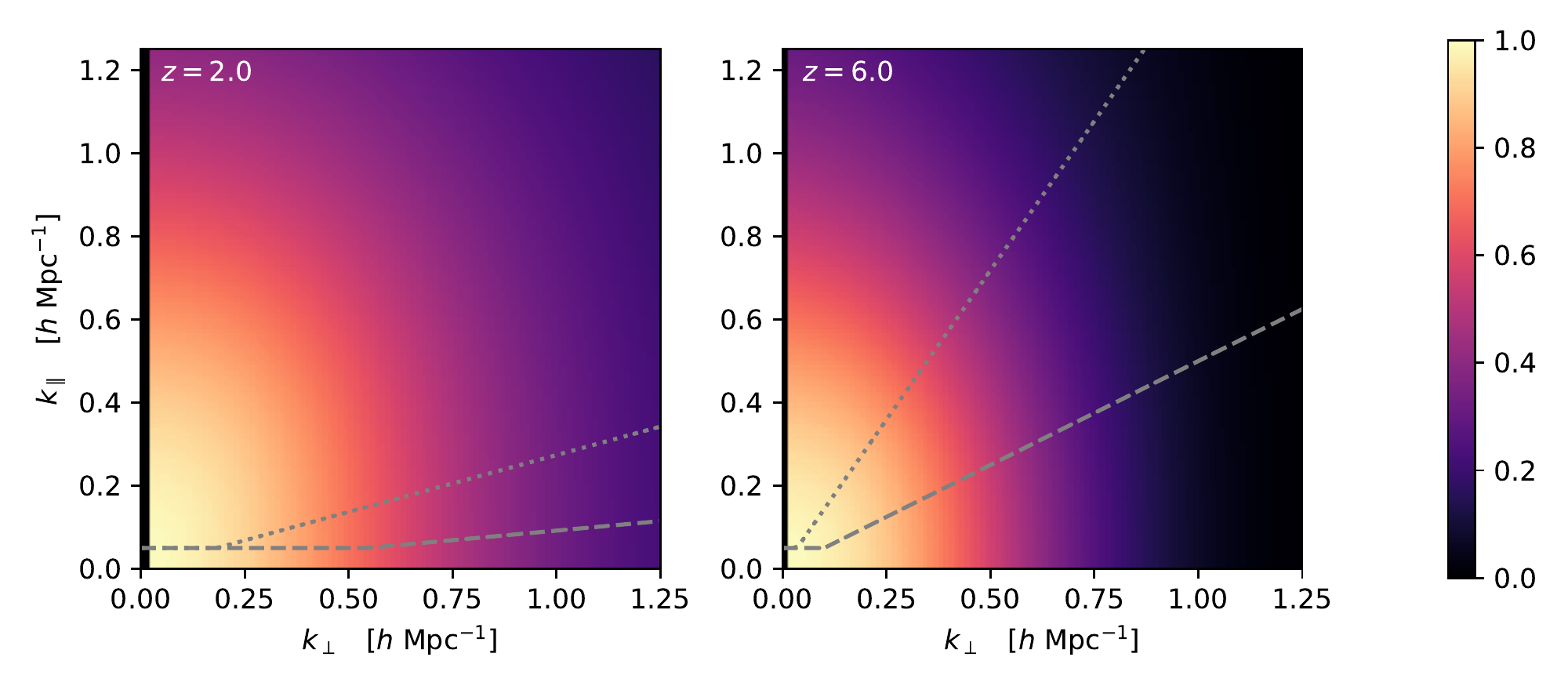}}
    \caption{The region of $k_\perp-k_\parallel$ space well probed by the Stage {\sc ii} 21-cm experiment described in ref.~\cite{CVDE-21cm}.  The color scale shows $P_{\rm sig}/P_{\rm tot}$ at $z=2$ (left) and $z=6$ (right), including thermal noise and shot-noise appropriate to our fiducial model.  Modes at very low $k_\perp$ are lost because dishes cannot be closer together than their diameters, imposing a minimum baseline length.  Dashed lines show the foreground wedge assuming $k_\parallel^{\rm min}=0.05\,h\,{\rm Mpc}^{-1}$ and the `primary beam wedge' while the dotted line shows the equivalent cut for $3\times$ the primary beam (see ref.~\cite{Chen19} for more information).  Modes below and to the right of those lines become increasingly contaminated by foregrounds (see text).}
    \label{fig:CVDE}
\end{figure}

While the Stage {\sc ii} experiment has very low thermal noise, and as we shall see below the HI signal has intrinsically low shot-noise, it must still contend with bright foregrounds: primarily free-free and synchrotron emission from the Galaxy and unresolved point sources \cite{Furlanetto06,Shaw14,Pober15,Seo16,Cohn16}.
Foregrounds make it very difficult to recover low $k_\parallel$ modes, i.e.~modes close to transverse to the line-of-sight.  The precise value below which recovery becomes impossible is currently unknown (see Refs.~\cite{Shaw14,Shaw15,Pober15,Byrne18} for a range of opinions).
In addition to low $k_\parallel$, non-idealities in the instrument lead to leakage of foreground information into higher $k_\parallel$ modes.  This is usually phrased in terms of a foreground ``wedge'' which renders modes with $k_\parallel/k_\perp<\sin\theta\,\mathcal{R}$ unusable, with \cite{Datta10,Morales12,Parsons12,Shaw14,Shaw15,Liu14,Pober15,Seo16,Cohn16,Byrne18}
\begin{equation}
    \mathcal{R} \equiv \frac{\chi(z)\,H(z)}{c(1+z)}
    = \frac{E(z)}{1+z}\int_0^z\frac{dz'}{E(z')} \qquad \textrm{(spatially flat)}
\end{equation}
where $\theta\approx\theta_{FOV}$.  Information in this wedge is not irretrievably lost, but it becomes more and more difficult to extract the deeper into the wedge one pushes.  The better the instrument can be calibrated and characterized the smaller the impact of the wedge.

Fig.~\ref{fig:CVDE} shows this information in graphical form.  The color scale shows the fraction of the total power which is signal, as a function of $k_\perp$ and $k_\parallel$.  The modes lost to foregrounds are illustrated by the gray dashed and dotted lines assuming a fiducial $k_\parallel^{\rm min}=0.05\,h\,{\rm Mpc}^{-1}$ and $\theta$ equal to the primary beam (i.e.~field of view) or $3\times$ the primary beam.  At $z=2$ and low $k$ the signal dominates, at intermediate $k$ the shot-noise starts to become important and at high $k_\parallel$ the thermal noise from the instrument dominates.  At $z=6$ the thermal noise is a larger fraction of the total for all $k$, but the fraction of the total which is signal is also increased by the higher bias of the HI at high $z$.

To benefit from the wide range of linear and quasi-linear scales that will be probed with high fidelity, we need to simulate a large cosmological volume.  Resolving $k\simeq 0.01-0.05\,h\,{\rm Mpc}^{-1}$ well requires a box size of $1\,h^{-1}$Gpc or larger.  However, in the spirit of intensity mapping, the inteferometer will not resolve individual dark matter halos or galaxies.  This suggests that the detailed profiles and subhalo properties of dark matter halos are less likely to be important than their relation to the evolving cosmic web.  For this reason we choose to set our force resolution by the requirement that halos be properly formed, not that their internal properties be resolved or that we can track substructures within them.

%%%%%%%%%%%%%%%%%%%%%%%%%%%%%%%%%%%%%%%%%
\section{Hidden Valley simulation suite}
\label{sec:sims}

\begin{table}[t]
    \centering
    \begin{tabular}{l|c|c|c|c|c|c}
    \hline\hline
         Run     & $\Omega_M$ & $h$ & $\sigma_8$ & $\Omega_B$ & $n_s$  & Amplitude \\
    \hline
      HV10240/F  &  0.309167 & 0.677  &  0.8222    &  0.04903 &  0.96824 & Fixed \\
      HV10240/F+  &  -- & --  &  0.8633    &  -- & -- & Fixed \\
      HV10240/F-  &  -- & --  &  0.7811    &  -- & -- & Fixed \\
      HV10240/R  &  -- & --  &  0.8222    &  -- & -- & Rayleigh \\
       HV2560/F  &  0.309167 & 0.677  &  0.8222    &  0.04903 &  0.96824 & Fixed \\
       HV2560/F+  &  -- & --  &  0.8633    &  -- & -- & Fixed \\
       HV2560/F-  &  -- & --  &  0.7811    &  -- & -- & Fixed \\
       HV2560/R  &  -- & --  &  0.8222    &  -- & -- & Rayleigh \\
    \hline\hline
    \end{tabular}
    \caption{The cosmological parameters used in the simulations.  The number in the name of each simulation represents $N_{\rm part}^{1/3}$ and the box size, in $0.1\,h^{-1}$Mpc (comoving).}
    \label{tab:params}
\end{table}

As outlined above, the simulation requirements for this project are quite demanding, since we need to simultaneously resolve the small-mass halos in which the majority of the neutral hydrogen signal lives while covering a large cosmological volume to make precise predictions for the clustering statistics that will be exquisitely measured by future surveys.  To meet this challenge we employed the FastPM code \cite{FengChuEtAl16} to evolve $10240^3$ particles in a periodic, cubic volume of $1024\,h^{-1}$Mpc resulting in a mean inter-particle separation of $100\,h^{-1}$kpc (comoving) and a mass resolution of $8.58\times 10^7\,h^{-1}M_\odot$.  The code used a $20480^3$ mesh  (i.e.~ force resolution factor $B=2$) for the gravity calculation and took 35 time steps from $z=99$ to $z=2$.  We simulated $\Lambda$CDM cosmologies with parameters close to the latest Planck results, as given in Table \ref{tab:params}.  The linear power spectrum\footnote{We provide the CAMB parameter file and matter power spectrum as part of the public data release.} was computed using CAMB \cite{CAMB} at $z=2$ and initial conditions were generated from this using second-order Lagrangian perturbation theory at $z=99$.  We saved halo catalogs and a 4\% subset of the dark matter particles at $z=6$ to $2$ every $0.5$ in $z$. The subsampling ensures the shotnoise level in the particles is less than 1\% of the non-linear power spectrum at $k=1\,h\,\mathrm{Mpc}^{-1}$ at redshift $z\le 6$. In addition to the large volume we also ran a smaller, $256 h^{-1} \mathrm{Mpc}$ simulation with the same spatial resolution as the larger box. The smaller volume is used for development of the numerical model and software tools.

We ran several realizations which differed only in the construction of the initial conditions. The realizations have identical phases, but the amplitudes are drawn differently. Our primary focus in this work is the `F' simulations, which we ran with initial conditions chosen such that $|\delta_k|=\sqrt{P(k)}$ with no scatter in order to reduce sampling variance (i.e.~Fixed amplitude). This technique has been shown to produce unbiased estimates of a large variety of 2 point and 3 point statistics
\cite{Angulo16,Chuang18}. The Hidden Valley simulation suite also contains three other amplitude settings: `R', where the amplitude is sampled from a Rayleigh distribution, corresponding to Gaussian initial conditions, and two additional fixed amplitude simulations, `F+' and `F-' where the power spectrum amplitude is changed by $\pm 5\%$, corresponding to a change in the $\sigma_8$ parameter which we can use to assess sensitivity of our results to $\sigma_8$.

\begin{figure}
    \centering
    \resizebox{\columnwidth}{!}{\includegraphics{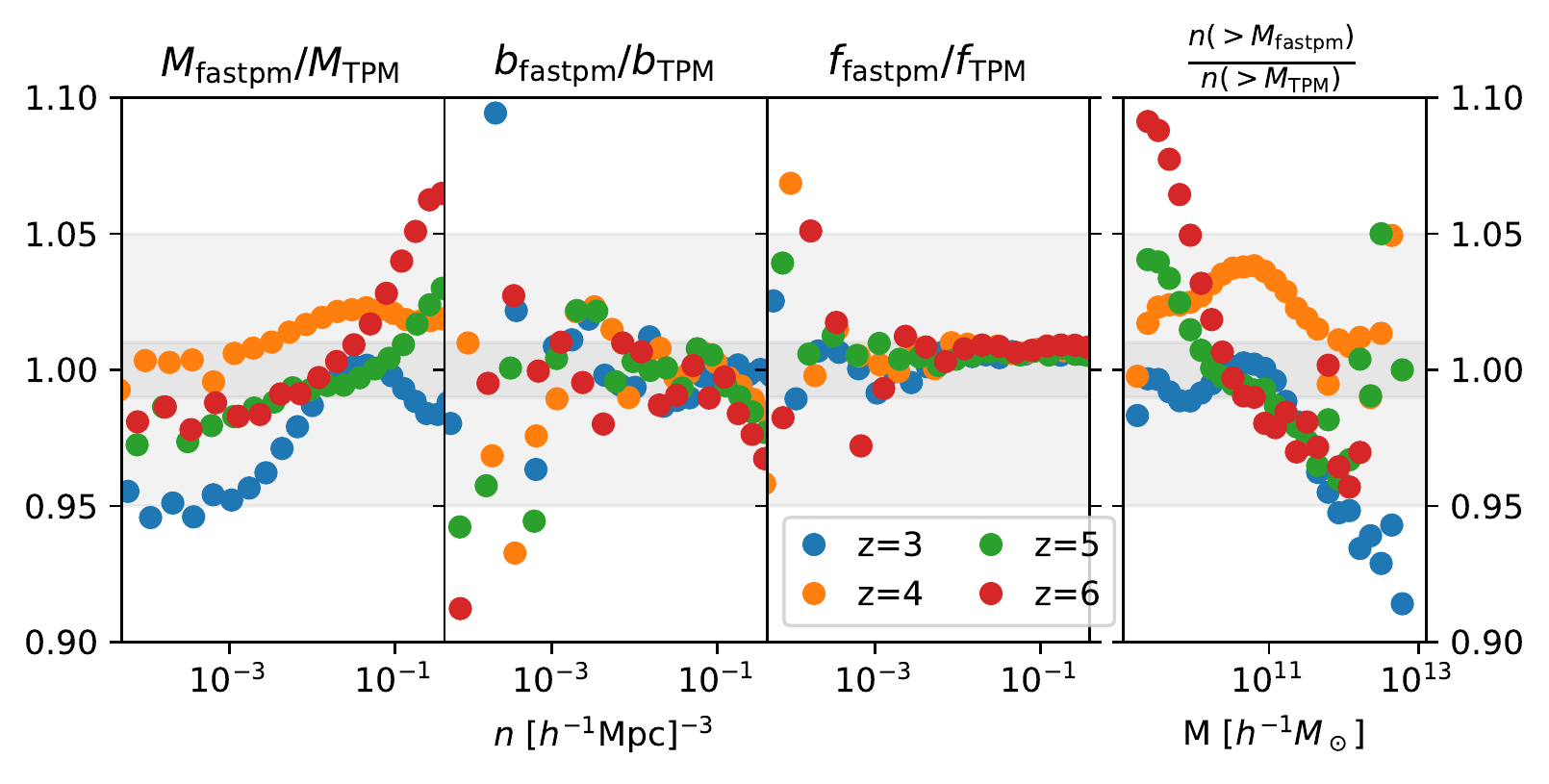}}
    \caption{Linear bias and power spectrum anisotropy (expressed as an effective growth rate, $f$) as a function of abundance, \Rsp{and cumulative mass functions}; for halos from this FastPM scheme compared against those of TreePM in the $256\,h^{-1}$Mpc calibration simulation.  The halos masses are within $5\%$, but the bias and growth rate at fixed abundance agree at the percent level except at all the number densities of interest. \Rsp{The mass function agreement is within 5\% on most masses, with larger discrepancy for large halo masses which anyway contribute very little to the total amount of HI.} This level of agreement is perfectly satisfactory for HOD modeling.}
    \label{fig:compare-tpm-linear}
\end{figure}

\Rsp{At low redshift FastPM predicts summary statistics of the halo and HOD modeled galaxy fields at the percent level \cite{FengChuEtAl16,Ding}.}
To demonstrate the accuracy of the scheme at these high redshifts, we compare the FastPM simulation to a TreePM simulation run with the same initial conditions and mass resolution \cite{TreePM,Stark15a,Stark15b} in a smaller box.  The comparison is performed on a box with side length of $256\,h^{-1}$Mpc utilizing $2560^3$ particles. In Fig.~\ref{fig:compare-tpm-linear} we compare the halo mass, \Rsp{halo mass function}, large-scale bias ($b$) and power spectrum anisotropy from the two simulations.  We see that at fixed abundance, the halo mass reported by FastPM is within $5\%$ of that reported by TreePM though with a mass-dependent trend.  \Rsp{The halo mass function is also predicted within $10\%$ across all masses and redshifts, and within $5\%$ for the mass ranges that contribute majority of the HI signal}. The bias and power spectrum anisotropy (expressed as an equivalent growth rate, $f$) are consistent at the $1\%$ level, except at very low abundance ($\bar{n}\simeq 10^{-3}\,h^3{\rm Mpc}^{-3}$) where the fitting is strongly affected by shot-noise.  \Rsp{This level of agreement is perfectly adequate for modeling of the HI field, given the scatter in halo-HI relation (\cite{VN18}) and the uncertainties in the HOD parameters are certainly larger than the discrepancy between FastPM and TPM. Thus for this work, we use the output FastPM masses of halos directly as HOD inputs to estimate HI signal, but in principle, its possible to preprocess these masses to abundance match against TreePM masses (or theoretical mass functions of choice (such as Sheth-Tormen mass function \cite{ST99} compared with in Figure \ref{fig:mf})).}

\begin{figure}
    \centering
    \includegraphics[width=0.95\textwidth]{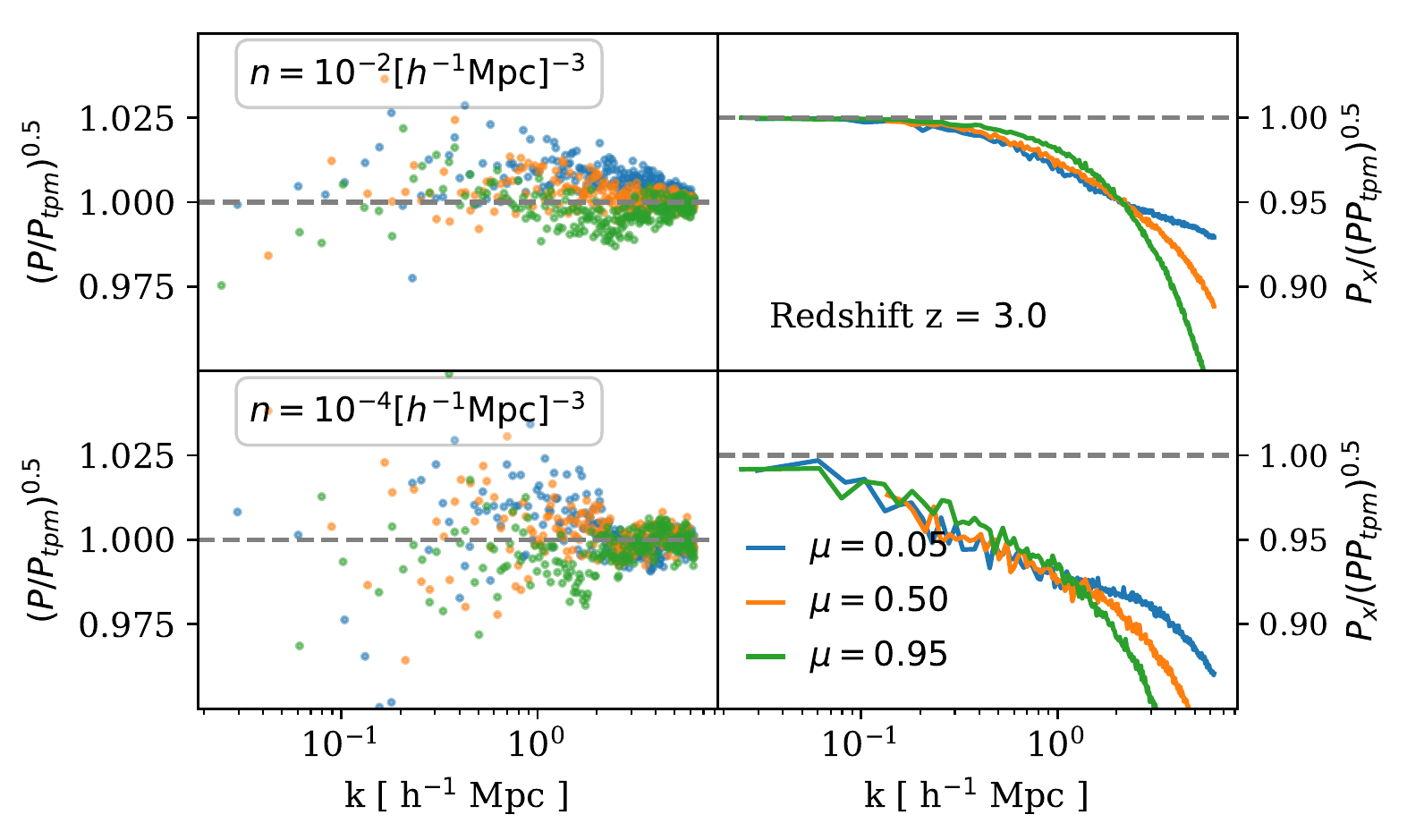}
    \caption{The transfer function, $T(k,\mu)=\sqrt{P/P_{tpm}}$, and cross-correlation coefficient, $r(k,\mu)$, between TreePM and FastPM at $z=3$ for two halo abundances ($\bar{n}=10^{-2}$ and $10^{-4}\,h^3\,{\rm Mpc}^{-3}$). The level of agreement is similar at other redshifts and even better for halos of lower abundance. In the transfer function plots (left), we show the power ratio for individual $(k, \mu)$ bins to give an idea of the level of agreement.}
    \label{fig:compare-tpm-halo-pkmu1}
\end{figure}

In terms of scale and anisotropy, we find that the halo positions and velocities produced by FastPM and TreePM are highly correlated and the redshift-space power spectra agree at the few percent level up to $k=1\,h\,\mathrm{Mpc}^{-1}$. In Fig.~\ref{fig:compare-tpm-halo-pkmu1} we show the transfer function, $T(k, \mu)=\sqrt{P/P_{tpm}}$, and the cross-correlation coefficient, $r_\times(k,\mu)$, at $z=3$ for two halo abundances ($\bar{n}=10^{-2}$ and $10^{-4}\,h^3\,{\rm Mpc}^{-3}$).  The agreement is very similar at other redshifts. % and even better at lower abundances than shown. 

The majority of the HI in a halo lies in the central galaxy, however a fraction can be distributed to larger radii.  If a non-negligible fraction of the HI moves with close to virial velocity within the halo, this can impact the observed redshift-space clustering through a finger-of-god effect.  Unfortunately, the observational constraints on this distribution are almost non-existent and so we are forced to model the satellites with simple prescriptions (described below and in Appendix \ref{app:satellites}).

Fig.~\ref{fig:slices} shows the real-space distribution of dark matter and HI in the simulation at our central redshift, $z=4$.  Each panel in the first column shows a $10\,h^{-1}$Mpc thick slice of the matter or HI field.  Moving from left to right the panels show successive zooms by a factor of 10, which highlights the high dynamic range of the simulation in mass and force resolution.  The cosmic web is clearly visible in the left and middle panels, while the complex environment around massive halos at high $z$ is obvious in the right-most panels.  Due to the paucity of HI in low mass halos, and the sub-linear scaling at higher mass, the low density regions are more empty in the HI than in the matter and the highest density regions have lower contrast.  Overall the structure is more pronounced in the HI than in the matter, as the former has a high bias.

\begin{figure}[t]
    \centering
    \resizebox{\columnwidth}{!}{\includegraphics{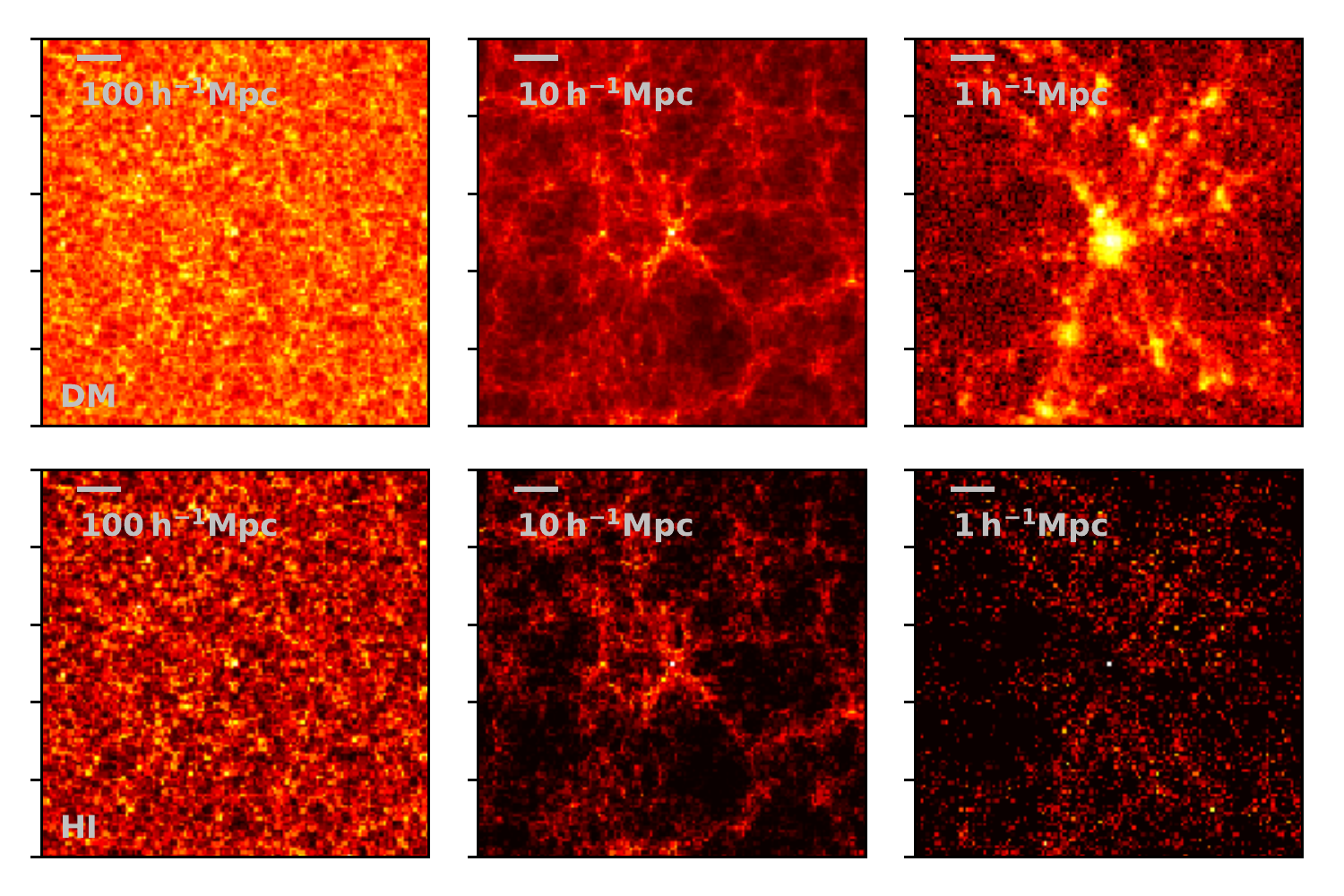}}
    \caption{Slices through the simulation showing the projected dark matter (top) and HI (bottom) densities at our middle output, $z=4$.  The left panels show a $100\,h^{-1}$Mpc thick slice through the full simulation box, while the middle and right panels show successive zooms by a factor of $10$, centered around the $5^{\rm th}$ most massive halo in the simulation.  The solid line in each panel indicates the linear scale while the color scale is log of the density.  The HI clearly traces the large-scale structure seen in the matter.
    }
    \label{fig:slices}
\end{figure}

%%%%%%%%%%%%%%%%%%%%%%%%%%%%%%%%%%%%%%%%%
\section{The HI model}
\label{sec:model}

There\footnote{This section differs from the published version in that the HI content in Model A and Model C is scaled up with $(1+z)^3$ and Model B with $(1+3)^3$. This does not change the clustering of HI or S/N of any signal evaluated in the later sections due to consistent change in definition on signal and noise. It only changes the amount of HI in a given halo and hence $\Omega_{HI}$. The corresponding figures (Fig. \ref{fig:calibration} and Fig. \ref{fig:distribution}) and tables (Table \ref{tab:basic_prop}) have been updated.}
 is lot of uncertainty in how hot and cold gas inhabit halos at high $z$. This is further complicated by the lack of any direct observational evidence on the mass scales of relevance. Thus the flexibility of being able to alter the prescription for assigning HI to the halos as we build our mock catalog upon a dark matter simulation is highly desirable. In what follows we shall consider three methods for assigning HI to our halos and subhalos (generated by Monte Carlo as described in Appendix \ref{app:satellites}). Given the dearth of high redshift clustering observations, we will use the following two pieces of data to calibrate these models. 

Our first calibration will be to the amount of neutral hydrogen at high $z$.  We follow standard convention\footnote{Some previous work (e.g.~ref.~\cite{Crighton15,Castorina17,VN18}) introduced a related quantity where the density at $z$ is instead normalized by the present day critical density, $\Omega_{HI,0}\equiv\rho_{HI}(z)/\rho_{c,0}$.  This differs by a factor of $\rho_c(z)/\rho_c(0)=E^2(z)/(1+z)^3$ from our definition.} and write the neutral hydrogen density as a fraction of critical at redshift $z$ as $\Omega_{HI}(z) = \rho_{HI}(z)/\rho_{c}(z)$.  Existing constraints on $\Omega_{HI}$ are obtained by integrating the HI column density distribution function inferred from QSO spectra.  We take the compilation of $\Omega_{HI}$ from Table 5 of ref.~\cite{Crighton15}, as shown in Fig.~\ref{fig:calibration}.

In order to fit for the characteristic mass scale and relative occupancy (e.g.~$M_{\rm cut}$ and $\alpha$ in Eq.~\ref{eqn:HI_HOD}) we need some information about the clustering of the HI. Lacking such measurements in the redshift window of interest ($2<z<6$), we will use the clustering of Damped Lyman-$\alpha$ systems, which contain more than $90\%$ of the neutral hydrogen in the Universe, as a proxy for the HI distribution.  If the HI column density distribution is only weakly dependent on $M_h$, then $b_{DLA}\approx b_{HI}$ and we shall assume this is true for $z\simeq 2-3$ (there is support for this from the hydrodynamics simulation of ref.~\cite{VN18} and data \cite{Rafols}).
The BOSS collaboration has measured $b_{\rm DLA}= 2.0 \pm 0.1$ in a broad bin centered at $z=2.3$ \cite{Rafols}.  As the clustering of DLAs is measured only at one redshift with sufficient accuracy, extrapolating the model in Eq.~(\ref{eqn:HI_HOD}) requires some assumptions.  We detail our fiducial model, and possible variants, below. Future data from high redshift QSOs in DESI \cite{DESI}, or direct measurement of the 21-cm signal, will be required to more accurately model $M_{HI}(M_h)$.

\subsection{Model A}

Our first model, which we shall also use as a fiducial model for most of our plots, assigns HI to the halos using the $M_{HI}-M_h$ relation of Eq.~(\ref{eqn:HI_HOD}).
This HI is then distributed between the central and the satellite galaxies such that the total HI in galaxies adds up to that in the halo.
Our primary motivation for this split is to include Finger of God (FOG) effects in the redshift-space power spectrum.
To that end, assuming that all the HI in halos is associated with either centrals or satellites and ignoring the free HI content inside the halo not associated with any galaxy is a fair assumption \cite{VN18}.

\begin{table}[t]
    \centering
    \begin{tabular}{c|c|cccc}
        Model &  Redshift    & $10^3\Omega_{HI}$  &  $b_1$       & $P_{sn}$  & $f_{\rm sat}$ \\ \hline
        \multirow{5}{*}{A}
        & $z=2$       & 2.01               &  1.91        & 53.29     & 0.07  \\
        & $z=3$       & 3.07               &  2.15        & 15.81     & 0.07  \\
        & $z=4$       & 3.31               &  2.58        & 10.1      & 0.06  \\
        & $z=5$       & 3.28              &  3.12        & 9.00      & 0.04  \\
        & $z=6$       & 3.12               &  3.72        & 9.24      & 0.03  \\ \hline
        \multirow{5}{*}{B}
        & $z=2$       & 5.17               &  1.79        & 40.42     & 0.03  \\
        & $z=3$       & 3.20               &  2.45        & 42.44     & 0.02  \\
        & $z=4$       & 1.83               &  3.10        & 42.93     & 0.02  \\
        & $z=5$       & 0.96               &  3.72        & 43.62     & 0.01  \\
        & $z=6$       & 0.51               &  4.33        & 43.6      & 0.01  \\ \hline
        \multirow{5}{*}{C}
        & $z=2$       & 2.02               &  1.81        & 73.70     & --  \\
        & $z=3$       & 2.98               &  2.30        & 39.74     & --  \\
        & $z=4$       & 3.42               &  2.89        & 29.35     & --  \\
        & $z=5$       & 3.17               &  3.54        & 27.44     & --  \\
        & $z=6$       & 2.67               &  4.27        & 29.63     & --  \\ \hline
    \end{tabular}
    \caption{Properties of the HI distribution in different models.  The bias ($b_1$) is estimated from the (real-space) HI-matter cross-spectrum at low $k$.  The shot noise $P_{\rm SN}$ is in $(h^{-1}{\rm Mpc})^{3}$.
    % {\color{red} This represents an upper limit, since we have not included scatter in the HI-$M_h$ relation.}. \MW{Lower limit?}
    $f_{\rm sat}$ is the fraction of total HI that is in the satellites.
    }
\label{tab:basic_prop}
\end{table}

\begin{figure}
    \centering
    % \resizebox{\columnwidth}{!}{\includegraphics{./figs/calib}}
    % \resizebox{\columnwidth}{!}{\includegraphics{./figs/calib_full}}
    \resizebox{\columnwidth}{!}{\includegraphics{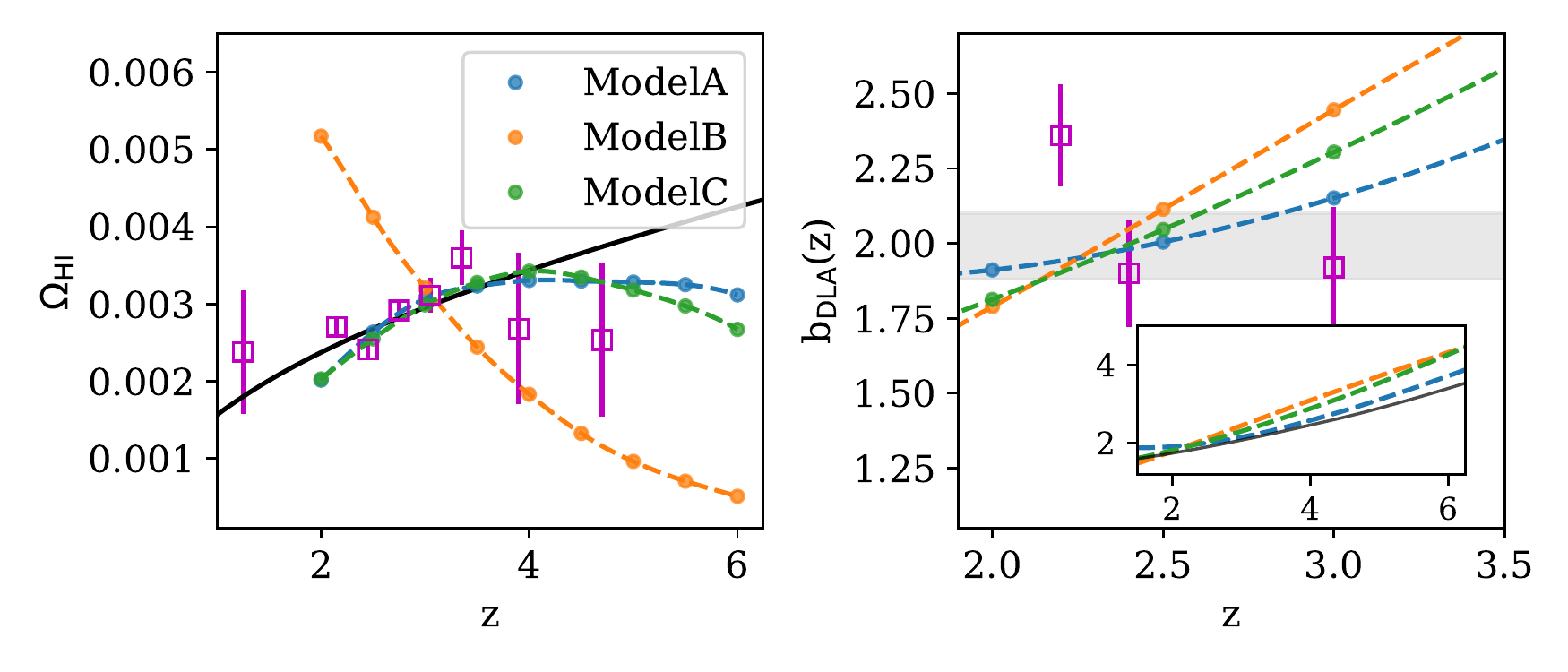}}
    \caption{(Left) The neutral hydrogen density, in units of the critical density, as a function of $z$.  The squares with error bars are from the compilation of ref.~\cite{Crighton15} (Table 5) and the line is their best fit. The circles show the results from our models. (Right) The bias of DLAs, which contain most of the HI in the high-$z$ Universe, at $z\simeq 2-3$.  Squares with error bars are the measurements from ref.~\cite{Rafols} while the circles and line show the predictions from our models. The horizontal grey band is the redshift-average value from ref.~\cite{Rafols}.  The inset shows the same bias prediction from our models up to $z=6$. For comparison, we also show the bias for mass-weighted halo field with gray line in the inset.
    }
    \label{fig:calibration}
\end{figure}

The $M_{HI}-M_h$ relation we choose is motivated by the desire that the total HI content of a halo be given (at least approximately) by Eq.~(\ref{eqn:HI_HOD}), as this has some support from recent numerical simulations.
To calibrate this model, we are guided by Table 6 of ref.~\cite{VN18} and the data shown in Fig.~\ref{fig:calibration}. We take the parameters for this model to be
\begin{equation}
    \alpha(z) = \frac{1+2z}{2+2z} \quad {\rm and}\quad
    M_{\rm cut}(z) =  3\times10^9\,h^{-1}M_\odot
    \ \left[1+10\left(\frac{3}{1+z}\right)^8\right]
    \label{eqn:modelAcutoff}
\end{equation}
valid \Rsp{only} over the range $2\le z\le 6$.  The steep drop of $M_{\rm cut}$ with $z$ is required in order to flatten the increasing bias with $z$ at lower redshifts to better match the DLA data.  To fit the abundance of HI we set
\begin{equation}
    A_h(z) =
    8\times 10^5\,h^{-1}M_\odot
    \left[1+\left(\frac{3.5}{z}\right)^6 \right] \times (1+z)^3
\end{equation}

This leads to the predictions shown in Fig.~\ref{fig:calibration} and Table \ref{tab:basic_prop}, where we see relatively good agreement with the available data at high $z$.
The parameters in Eq.~\ref{eqn:modelAcutoff} are slightly different than those in ref.~\cite{VN18}. The reason is that the hydrodynamical simulations in ref.~\cite{VN18}, despite being in better agreement with the HI data than their predecessors, still underestimate both the measured $b_{DLA}$ (by $2\,\sigma$) and $\Omega_{\rm HI}$. In a halo based approach we have enough freedom to tune our parameters to better match the data.

Next we want to split this HI between centrals and satellites. We use the same relation for the satellites as for halos, albeit with a different normalization where we put more HI in satellites at the same (sub)halo mass and increasingly so with increasing redshift.
\begin{equation}
    A_s(z) = A_h(z)\times(1.75 + 0.25\,z) 
\end{equation}
The centrals then take up the residual HI from the halos that is not assigned to satellites. With our simple prescription, we are able to match the HI fraction in satellites observed by \cite{VN18} in Illustris simulations.

We show the distribution of the HI for our model in halos of different mass and at different redshifts in Fig.~\ref{fig:distribution}.  The top row shows the $M_{HI}-M_h$ relation as given by Eq.~\ref{eqn:HI_HOD} for this model. The middle row shows the fraction of total HI that resides in the halos of a given mass. Towards highest halo masses the contribution decreases because of the exponential drop in the halo mass function, while at the lower end the HI hosted in the individual halos drops due to the cutoff (Eq.~\ref{eqn:modelAcutoff}). Hence most of the signal contribution comes from halos of intermediate masses: $\sim 10^{11} \,h^{-1}M_\odot$ at $z=2$ and decreasing to $\sim 10^{10} \,h^{-1}M_\odot$ at $z=6$. This also illustrates the convergence of our simulations with respect to the HI signal i.e.\ for this model our simulations have enough resolution to capture essentially all the expected HI signal across all the redshifts of interest. 

The last row of Fig.~\ref{fig:distribution} show the fraction of total HI inside a halo that is embedded in the satellites as a function of halo mass. This fraction increases with the halo mass, with as much as $\sim 50 \%$ HI residing in satellites for $\sim 10^{14} \,h^{-1}M_\odot$ halos. As described above, the remaining HI of the halos is assigned to centrals. This is in qualitative agreement with satellite HI fraction seen in the Illustris simulations in ref.~\cite{VN18}.

While it would be straightforward to include a random scatter in HI at fixed $M_h$, we have no observational constraints on the form or evolution of the scatter.  Thus we choose to neglect it, noting where this impacts our results.

\subsection{Model B}

\begin{figure}
    \centering
    \resizebox{\columnwidth}{!}{\includegraphics{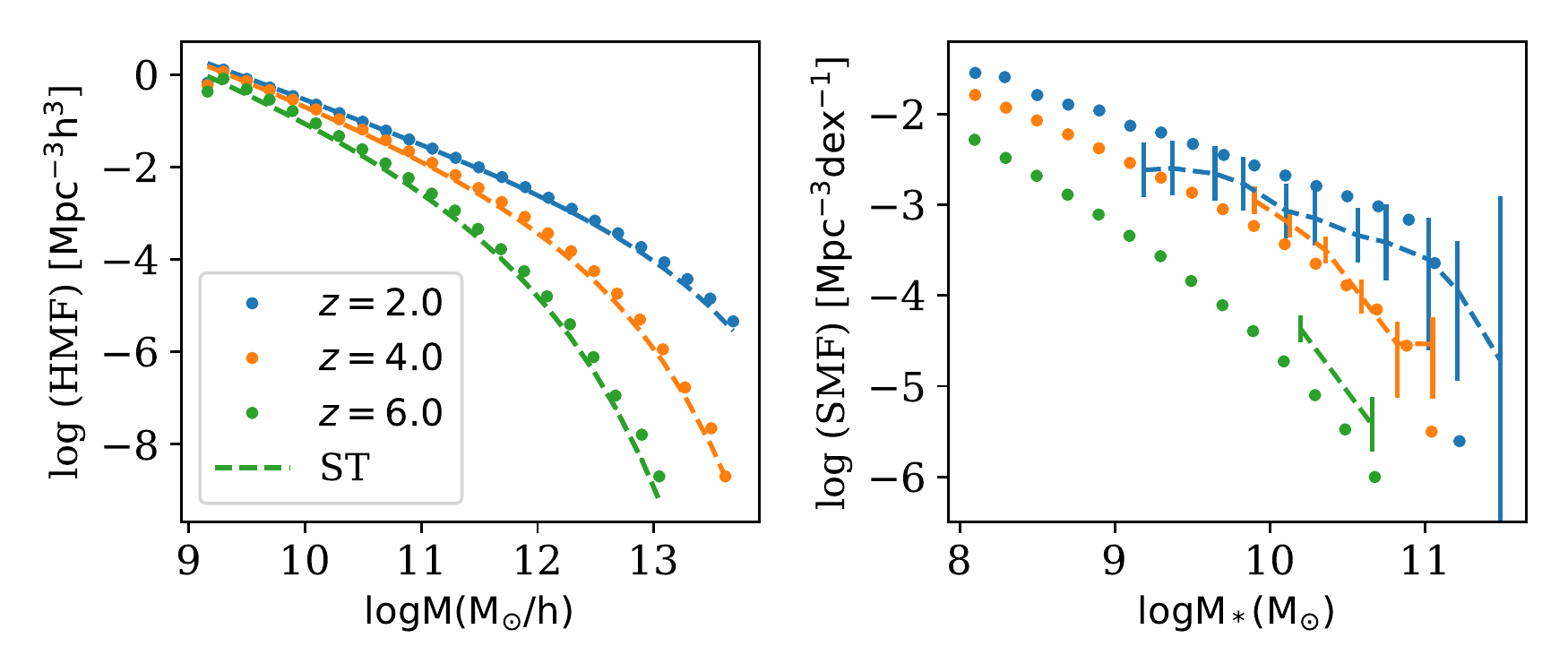}}
    \caption{(left) The halo mass function from our simulations at $z=2$, 4 and 6. The dashed lines are the Sheth-Tormen mass function prediction \cite{ST99}.  (right) Stellar mass function from our simulations (points) at the same redshifts. The data (dashed line with error bars) are taken from the compilation of ref.~\cite{Behroozi18}\protect\footnotemark.
    }
    \label{fig:mf}
\end{figure}
\footnotetext{https://www.peterbehroozi.com/data.html}

Inspired by ref.~\cite{Dave13}, our second model assumes that the HI content of galaxies depends upon stellar mass. We use the fit of ref.~\cite{Moster13} to assign stellar masses ($M_\star$) to our central and satellite halos. The corresponding stellar mass function is shown in Figure \ref{fig:mf} which agrees fairly well with the data \cite{Behroozi18}.  We could likely improve the fit by adjusting the parameters in the stellar-mass halo-mass relation, but this is sufficient for our purposes so we keep the fiducial parameters of ref.~\cite{Moster13}. Then we assign an HI mass to each of these galaxies based on the same `HI richness' ($M_{HI}/M_\star$) relation. Ref.~\cite{Dave13} find that $M_{HI}/M_\star$ evolves relatively little with redshift (e.g.\ see their Fig.~4) implying that the redshift evolution of $M_h-M_{HI}$ is due to the evolution of $M_\star$. We take
\begin{equation}
    \frac{M_{HI}}{M_\star} = f\left( \frac{M_1}{M_1 + M_\star}\right)^\alpha %\times (1+3)^3
    \label{eqn:MHI_Mstar}
\end{equation}
where by calibrating the against the observations we get $f=11.52$, $\alpha=0.4$, and $M_1=3\times 10^8 M_\odot$. Similar to the $M_\star-M_h$ relation, it would be straightforward to include a random scatter in HI at fixed $M_h$, however we have no observational constraints on the form or evolution of the scatter. Thus we choose to neglect it. 

\begin{figure}
    \centering
    \resizebox{\columnwidth}{!}{\includegraphics{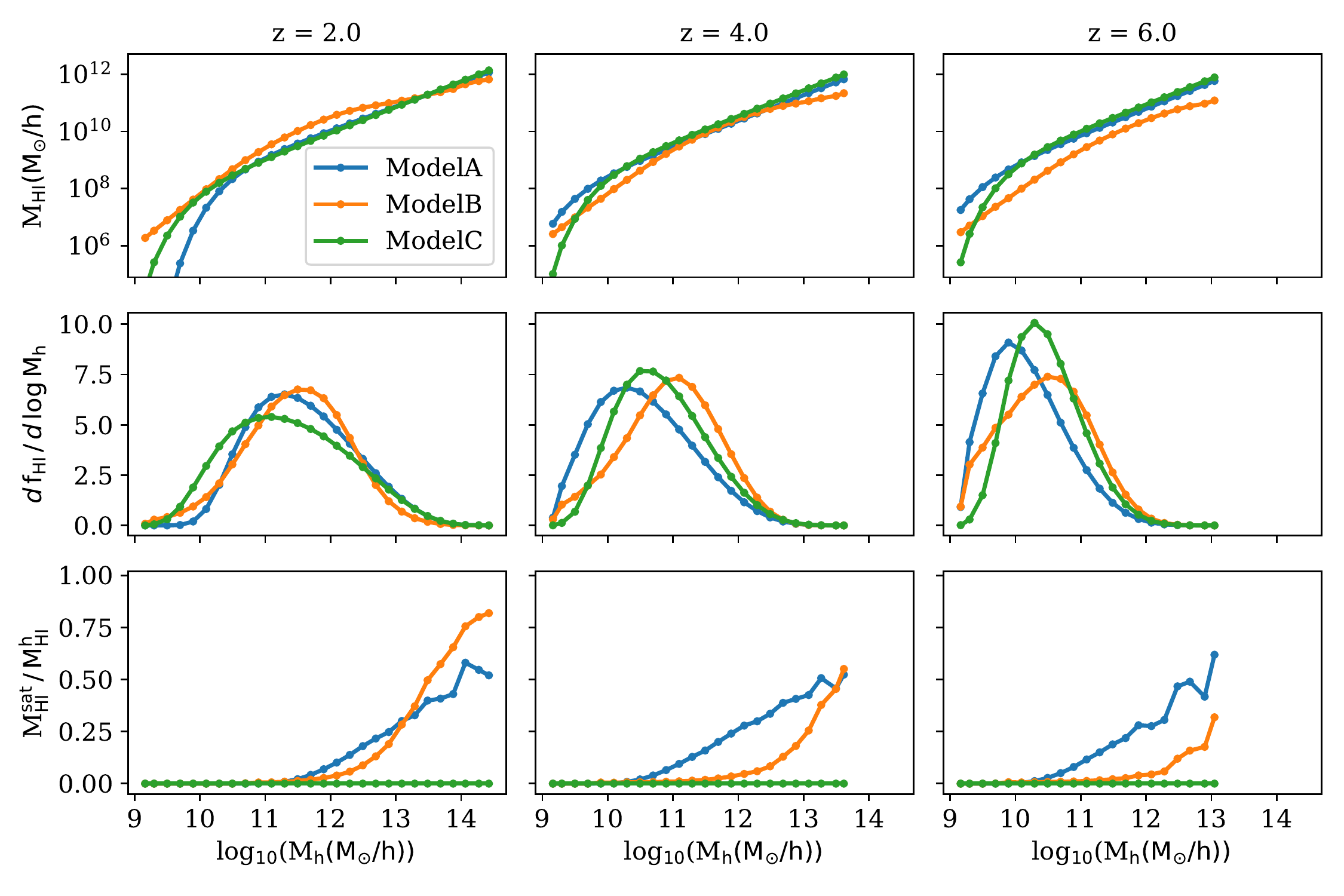}}
    \caption{Distribution of HI for different models (colors) at different redshifts (columns):
    i) The top row shows the halo HI mass function i.e.\ the average HI mass inside halos as a function of halo mass.
    ii) The middle row shows the fraction of the total HI that resides in halos of a given mass. This illustrates the convergence of our simulations with respect to HI signal on the low halo mass end.
    iii) The bottom row shows the fraction of HI mass inside the halos that resides in the satellites (this is zero for Model C by assumption). }
    \label{fig:distribution}
\end{figure}

Unlike our fiducial model, this model does not have an explicit lower mass cut-off. Instead the HI distribution in small halos is implicitly suppressed (but not exponentially) by the stellar-mass halo-mass relation.
Due to this lack of flexibility, it performs slightly worse in matching the DLA bias at low redshifts (Fig.~\ref{fig:calibration}).
Similarly, due to its lack of explicit redshift dependence beyond the evolution of stellar mass, it also has very different evolution of $\Omega_{HI}$ across redshifts.
This model also puts relatively more HI in the intermediate mass halos and less in the highest mass halos than our fiducial Model A (Fig.~\ref{fig:distribution}, top row) due to the flattening of the $M_\star-M_h$ relation at high mass. Hence the peak of the signal contribution as a function of halo masses (middle row) is shifted to slightly higher mass, and is lower compared to that of Model A.  Since we use the same relation (Eq.~\ref{eqn:MHI_Mstar}) for both centrals and satellites, we lack the flexibility of Model A in distributing HI inside halos. As a result, we find that the satellite HI fraction for this model declines sharply with redshift and is in qualitative disagreement with ref.~\cite{VN18} at higher redshifts.

One advantage of this model is that the redshift dependence of the $M_h-M_{HI}$ distribution is entirely determined by the stellar mass evolution, which is fairly well constrained. Thus we have effectively only two free parameters, $M_1$ and $\alpha$, that impact the impact the shape of $b(z)$ and $\Omega_{\rm HI}(z)$ with $f$ impacting the overall amplitude. 
However in its current implementation, this lack of explicit redshift dependence leads to a different evolution in total HI content despite having similar clustering to Model A.
Thus precise observations at high redshift in the future will serve to test the validity of this model.

\subsection{Model C}

As our last model, we use a simplistic alternative which assumes a constant $M_{\rm cut}$ and $\alpha$ at all redshifts to assign HI to halos. This has been used in the literature in refs.~\cite{Castorina17,Chen19}. We set $\alpha=0.9$ (which is roughly the average of the values at the extreme redshifts used in Model A) and $M_{\rm cut} = 10^{10}\, h^{-1}M_\odot$. Given the current state of the high $z$ clustering observations this model is difficult to completely rule out. However, like Model B, due to the lack of the flexibility in varying the lower mass cut-off this model to performs poorly in matching the DLA bias (Fig.~\ref{fig:calibration}).  With these parameters fixed, one can match the observed HI density up to high redshifts by making the normalization redshift dependent. We find that the following form fits the data well:
\begin{equation}
    A_h(z) = 3.5\times 10^6\,h^{-1}M_\odot
    \ \left(1+\frac{1}{z}\right) \times (1+z)^3
\end{equation}
As can be seen from Fig.~\ref{fig:distribution}, despite having minimal redshift evolution in the parameters, this model qualitatively follows Model  A and Model B in the distribution of HI as a function of halo masses.  However unlike the other two models, we deliberately do not include any satellites to distribute HI inside an individual halo. This is to avoid having any 1-halo finger-of-god effects in the redshift space. Thus when compared with other models, this will serve to illustrate their impact over and above the redshift space distortions that are captured by the halo velocities themselves.

%%%%%%%%%%%%%%%%%%%%%%%%%%%%%%%%%%%%%%%%%
\section{HI clustering}
\label{sec:clustering}

The amplitude of the 21-cm signal depends on the abundance and the clustering of the HI. In the previous section, we considered three different models for distributing HI in halos and measured its abundance. In this section, we look in detail at the clustering of HI. We will mostly show the results from our fiducial Model A and simply comment on similarities or differences with the prediction of other models, unless explicitly showing the differences is more insightful.

We will look only at the 2-point statistics of HI clustering. Even though it's not observable, we will begin with real space clustering of HI as a function of redshift since it helps elucidate its relation to the underlying dark matter clustering. Next we will discuss the clustering in redshift space where the signal will actually be observed. Lastly, we will discuss the baryon acoustic oscillations (BAO) in 21-cm signal since its one of the major goals of upcoming 21-cm surveys to detect BAO and measure distances to high redshifts with them.

In order to optimize what can be learned from the surveys mentioned above, theoretical predictions in the mildly and fully non-linear regimes are also needed. Previous work has suggested that perturbative models can accurately predict the clustering of biased tracers to a significant fraction of the nonlinear scale $k_{\rm nl} = \Sigma^{-1}$ at high $z$ \cite{Matsubara08,Carlson13,Vlah16,Foreman16,Modi17}, where $\Sigma$ is mean square one-dimensional displacement in the Zeldovich approximation given by
\begin{equation}
    k_{\rm nl}=\Sigma^{-1} = \left[ \frac{1}{6\pi^2} \int P_L(k)\ dk \right]^{-1/2} \quad .
\label{eqn:Sigma}
\end{equation}
with $P_L$ the linear theory power spectrum. For high redshift tracers such as HI, perturbative models with their sophisticated biasing schemes become more useful due to increasing complexity of bias and decreasing non-linearity of the matter field as we go further back in time \cite{Modi17}. While we leave a detailed study of modeling 21-cm signal with such models for future work, when more sophisticated models and more simulation volume could become available, here we will compare the clustering signal to the predictions of linear theory and `Zeldovich effective field theory' (ZEFT \cite{Vlah15}).  ZEFT is a combination of lowest order Lagrangian perturbation theory (i.e.~the Zeldovich approximation \cite{Zel70}) plus a counter-term going as $k^2$ (it can also be thought of as a $k^2$ component to the bias).  

\subsection{Real-space clustering}

\begin{figure}
    \centering
    \resizebox{\columnwidth}{!}{\includegraphics{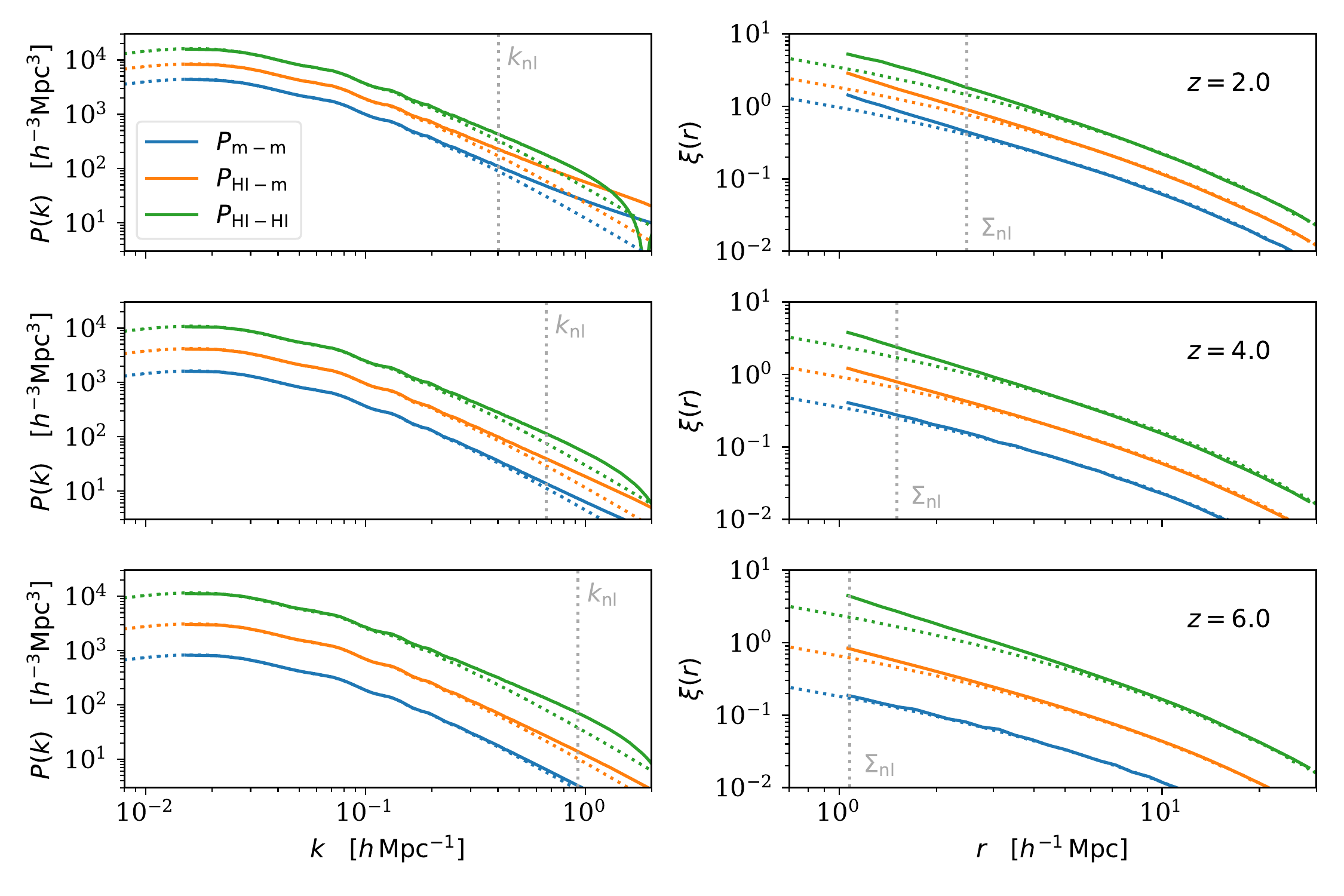}}
    \caption{The real-space power spectrum (left) and cross correlation function (right) of the dark matter and HI in our fiducial model at $z=2$, $4$, and $6$.  Solid lines show the N-body results while dotted lines show linear theory with scale-independent bias (set by the amplitude of the cross-spectra at low $k$ or cross-correlation at large $r$).  Blue, orange and green line show the matter clustering, matter-HI cross-clustering and the HI auto-clustering respectively.  The downturn in the HI auto-spectrum (green) at high $k$ is due to our subtraction of a constant `shot-noise' component (see text).  The vertical grey lines mark $k_{\rm nl}$ and $\Sigma_{\rm nl}$ of Eq.~(\ref{eqn:Sigma}).
    }
    \label{fig:real_pk}
\end{figure}

We begin with a discussion of the real-space clustering of the HI as a function of redshift, and its relation to the underlying dark matter clustering and to linear theory. Fig.~\ref{fig:real_pk} shows the real-space power spectra on the left, and the two-point functions (i.e.~correlation functions) as a function of separation, $r$, on the right at some selected redshifts. The blue, orange and green lines show matter auto-clustering (spectra + correlation), matter-HI clustering and the HI auto-clustering respectively, and we compare them to the predictions of linear theory with constant bias (i.e.~$b^2P_L(k)$ and $bP_L(k)$ respectively for power spectra) as dotted lines with $b$ matched to the amplitude of the matter-HI cross-spectra at low $k$ (we perform a simple unweighted average of points with $k<0.06\,h\,{\rm Mpc}^{-1}$).  The difference between the simulations and linear theory predictions, coming from a combination of non-linear clustering and scale-dependent bias, is clearly visible at all redshifts. 

\begin{figure}
    \centering
    \resizebox{\columnwidth}{!}{\includegraphics{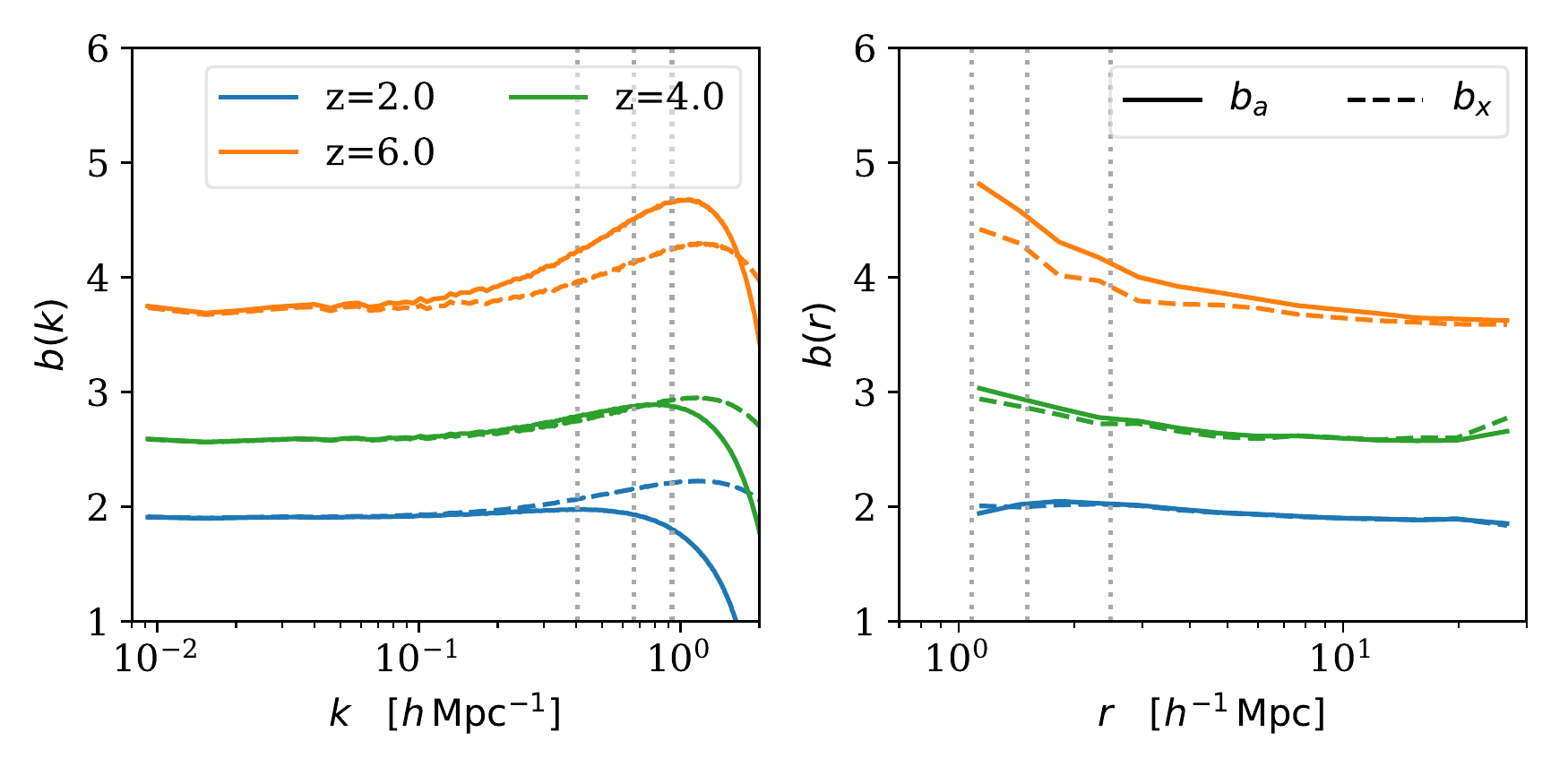}}
    \caption{(Left) The HI bias, in real-space, measured from the cross- and auto-spectra: $b_a(k)\equiv\sqrt{P_{HI,HI}/P_{mm}}$ (solid) and $b_\times(k)\equiv P_{HI,m}/P_{mm}$ (dotted).  As the bias becomes larger, it also becomes more scale-dependent and the difference between $b_a$ and $b_\times$ extends to lower $k$. (Right) The scale-dependent bias defined by the auto- and cross-correlations (solids and dashed, respectively) viz.~$b_a(r)\equiv\sqrt{\xi_{HI,HI}/\xi_{mm}}$ and $b_\times(r)\equiv\xi_{HI,m}/\xi_{mm}$.}
    \label{fig:real_bk}
\end{figure}

Fig.~\ref{fig:real_bk} shows the bias explicitly for both power spectra and correlation functions. The solid lines show the bias estimated from the auto-clustering $b_a(k)\equiv\sqrt{P_{HI,HI}/P_{mm}}$ (and corresponding ratios in configuration space) while the dashed lines are the cross-clustering biases  $b_\times(k)\equiv P_{HI,m}/P_{mm}$. The difference between them indicates the degree to which the HI and matter fields are decorrelated. At larger scales, the two biases converge to same values showing that the HI provides a good tracer of the matter field and the cross-correlation coefficient is close to unity. The decorrelation occurs on the scales of a few Mpc and shifts to larger scales (lower $k$) with increasing bias as we go higher in redshift.

On large scales ($>10\,h^{-1}$Mpc), the bias is scale-independent and our simulations are large enough that we can clearly see the `flat' low $k$ plateau in the Fourier space. However we see the bias becoming scale dependent on small scales in both Fourier and configuration space, with the dependence extending to larger scales and lower $k$ for the more biased objects at higher redshifts. For this figure, we have defined the biases with respect to the non-linear matter spectra (correlation function) -- the scale-dependence of the bias would be larger if we had used the linear theory spectrum in the definition.  
Note also that as the the bias of the HI in our fiducial model becomes larger and more scale dependent at higher redshifts, the non-linear scale shown in dotted gray lines $k_{\rm nl}/\Sigma_{\rm nl}$ simultaneously shifts to smaller scales.  Thus scale-dependence of the bias thus becomes relevant before the non-linear clustering (a similar effect is seen for high-$z$ galaxies in ref.~\cite{Modi17,Wilson19}).

\begin{figure}
    \centering
    \resizebox{\columnwidth}{!}{\includegraphics{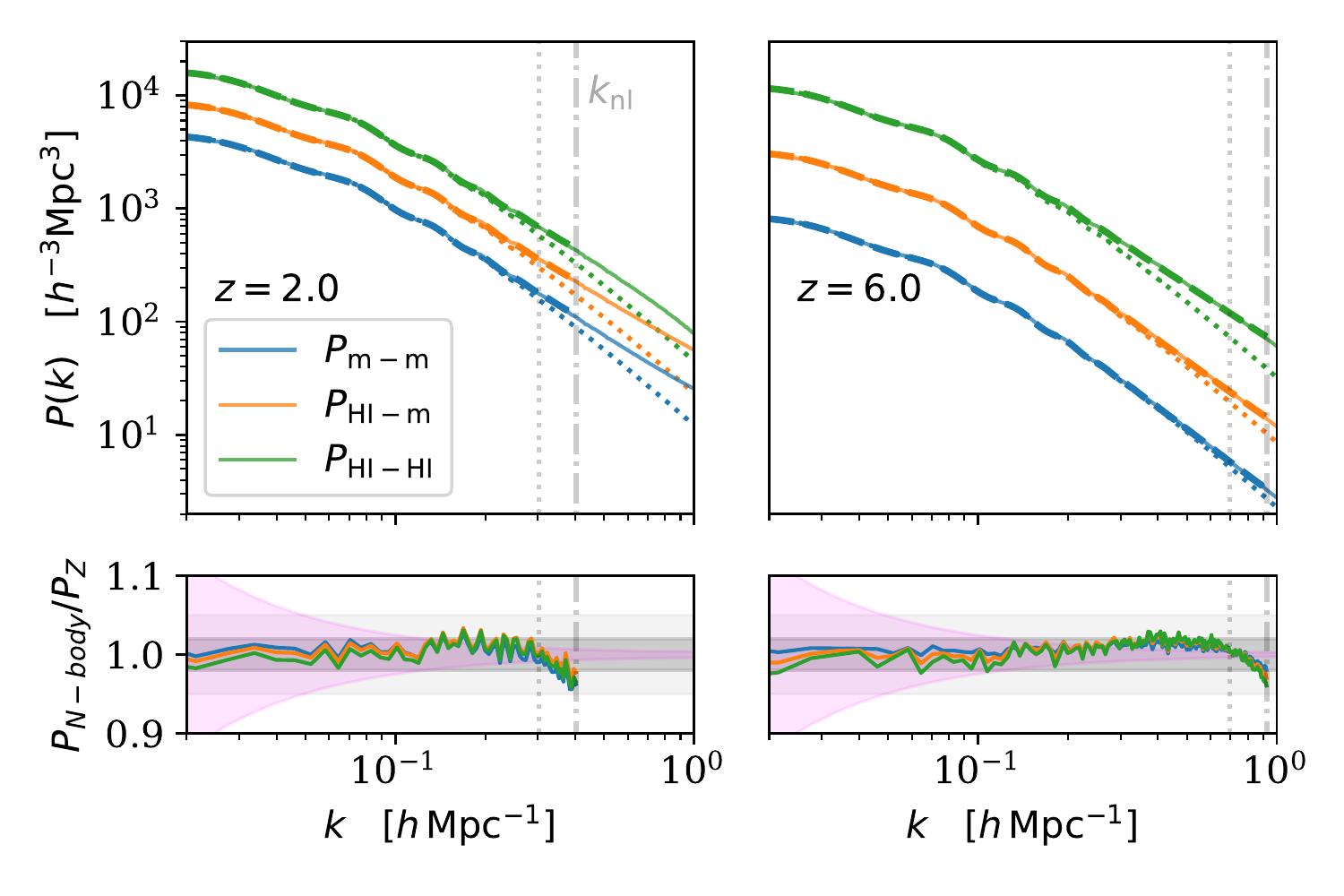}}
    \caption{(Top) The real-space power spectrum of the dark matter and HI in our fiducial model at $z=2$ and 6.  Solid lines show the N-body results, dashed lines show the Zeldovich effective field theory calculation and dotted lines linear theory.  Blue lines show the matter power spectrum, orange lines the matter-HI cross-spectrum and green lines the HI auto-spectrum.  A single set of parameters fits all curves simultaneously.  The dotted and dash-dot vertical grey lines mark $0.75k_{\rm nl}$ and $k_{\rm nl}$ of Eq.~(\ref{eqn:Sigma}).  (Bottom) The ratio of the the N-body results to Zeldovich fit.  The dark and light grey shaded regions show $\pm 2\%$ and $\pm 5\%$.   The shaded magenta region shows the sample variance error in our simulation volume (including shot noise).
    }
    \label{fig:zeld_pkr}
\end{figure}

The trade-off between complex biasing and non-linearity makes modeling high redshift tracers an ideal application for perturbative models with sophisticated biasing models, such as those developed in Refs.~\cite{Vlah15,Vlah16,Modi17} (see also ref.~\cite{VN18}). Fig.~\ref{fig:zeld_pkr} shows real-space power spectrum at $z=2$ and 6 compared to the predictions of ZEFT. We have fit all three spectra simultaneously, e.g.~the same value of $b_2$ that fits the HI auto-spectrum fits the cross-spectrum.  At high redshift the contribution of shear terms becomes suppressed so we have not included shear terms in our fit, though we do include a $b_{\nabla^2}$ contribution, leading to a total of $4$ free parameters in our model.  The agreement is very good up to about $\sim 0.75\, k_{\rm nl}$, as suggested by ref.~\cite{VN18} in a much smaller volume, and within the expected sample variance over the perturbative range.  We expect theoretical models including higher-order effects would do even better.  Over this range the N-body results can depart by order unity from the linear theory predictions.

One complication while fitting HI clustering with any model is to take into account the shot noise. Unlike galaxies, HI is not a discrete tracer of the underlying matter field, hence the Poisson approximation for noise is no longer true. This is also shown in ref.~\cite{VN18} who find that $P(k)$ does not asymptote to a constant at high $k$, likely due to the HI profile inside the halos on small scales. Thus if one knows the shape of $1-$halo term, this can be fit for or subtracted while modeling the signal. In addition the 21-cm signal is dominated by low mass halos, $\sim 10^{10} h^{-1}M_\odot$, whose clustering has a complex shot noise form. While this needs to be investigated in the future, here we approximate the noise as Poisson, i.e. $P_{\rm SN} \sim \sum_h M_{\rm HI}^2/(\sum M_{\rm HI})^2$, where $M_{\rm HI}$ is the HI mass in a halo and the sum is over all the halos. This simplistic form ignores any scale dependence of the noise and ref.~\cite{VN18} show that this over-estimates the noise on small scales. 

\Rsp{Lastly, we have shown the results only for HI distribution with Model A. The results in real space are qualitatively similar for the other models with the only difference being in the amplitude of the clustering i.e.\ Model B and Model C predict slightly higher bias on large scales at higher redshifts. This can be seen from the inset plot in Fig.~\ref{fig:calibration}. As expected, this also leads to greater scale dependence of bias and higher decorrelation with matter clustering on smaller scales. }

\subsection{Redshift-space clustering: supercluster infall}

\begin{figure}
    \centering
    \resizebox{\columnwidth}{!}{\includegraphics{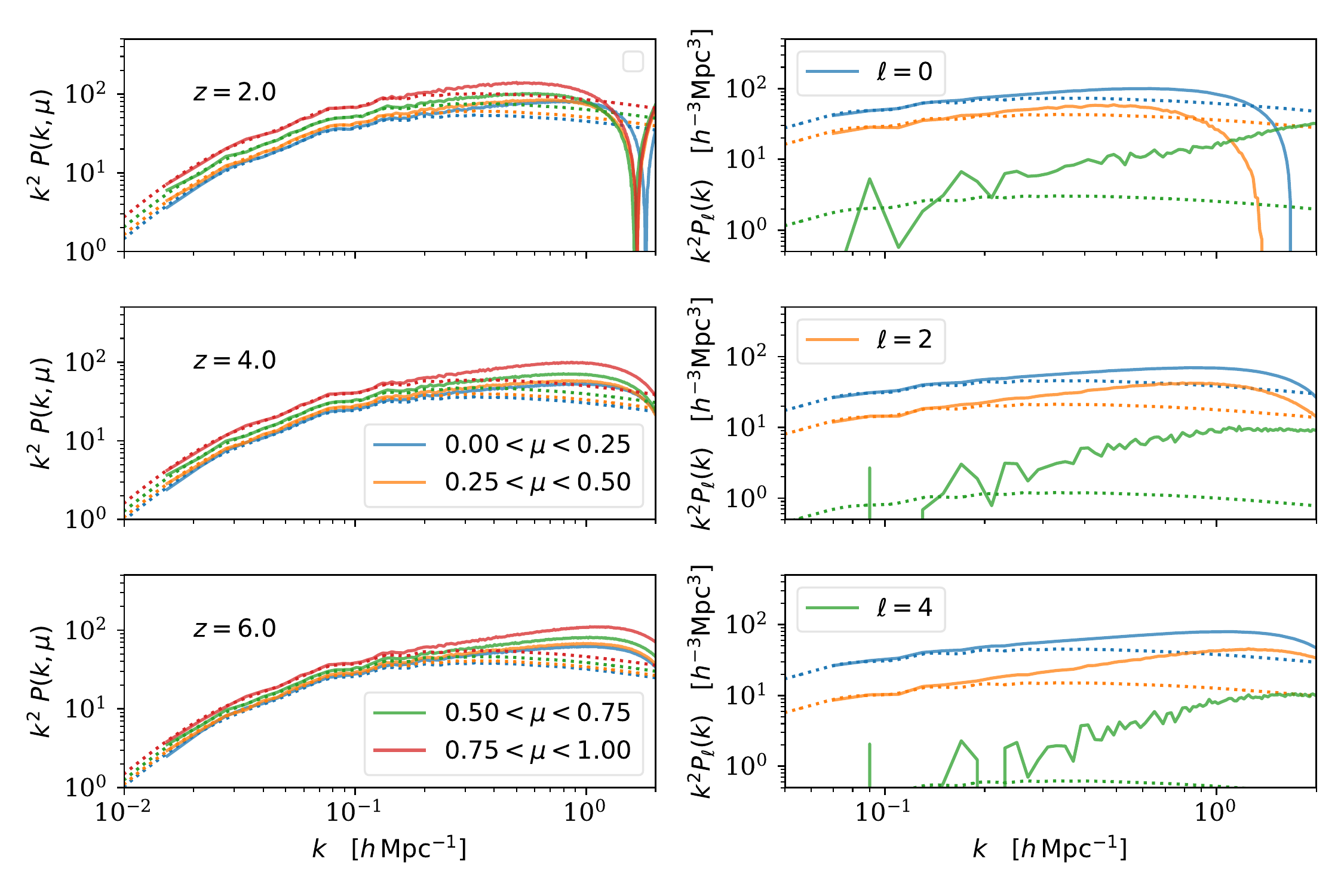}}
    \caption{The redshift-space power spectrum of the HI in our fiducial model at $z=2$, $4$ and $6$.
    (Left) The solid lines show the N-body result for $P(k, \mu)$ wedges while the dotted lines show linear theory prediction with scale-independent bias.
    (Right) The first three, even multipoles of the anisotropic redshift space power spectrum along with the linear theory, constant bias predictions (dotted).}
    \label{fig:red_pk}
\end{figure}

Future experiments will measure the clustering signal in redshift space, not real space,  thus it is useful to look at the impact of peculiar velocities on the signal.  There are several effects which are of interest.  First, the fact that only the line-of-sight peculiar velocity contributes to the measured redshift imprints an anisotropy in the clustering signal \cite{Kai87,H98}.  In the distant observer limit\footnote{Since the observations are at high $z$ we make the distant observer approximation throughout.  From the analysis of Refs.~\cite{CasWhi18a,CasWhi18b} we expect this introduces errors well below our other uncertainties.} our power spectrum becomes a function of $k$ and $|\mu|$, with $\mu\equiv\hat{k}\cdot\hat{z}$ if $\hat{z}$ is the (common) line-of-sight direction.  On large scales, supercluster infall enhances the signal, while on smaller scales inter- and intra-halo virial motions lead to anisotropic power suppression.  The amount of suppression, also referred to as the finger-of-god (FOG) effect, is dependent upon the amount of HI that lives in satellites within massive halos and their virial motions. We focus here on the former and explore the FOG effect in the next section.

Fig.~\ref{fig:red_pk} shows the HI auto-spectrum in redshift space in two different forms and at various redshifts - on the left are the $P(k, \mu)$ wedges in 4 slices of $\mu$ while on the right are the first three multipoles generated by anisotropic clustering along the line of sight. We have multiplied the curves by $k^2$ to reduce the dynamic range in order to better separate the different curves. In addition, we have subtracted the shot noise with the same approximation of weighted means as for the real space clustering. 
In linear theory, the line-of-sight power is enhanced by supercluster infall \cite{Kai87} by a factor $(1+[f/b]\mu^2)^2$, where $f=d\ln D/d\ln a\simeq\Omega_m^{0.55}(z)\approx 1$ is the growth rate. This prediction is shown in dotted lines for different wedges and multipoles and agrees well with the simulations on large scales.

\begin{figure}
    \centering
    \resizebox{\columnwidth}{!}{\includegraphics{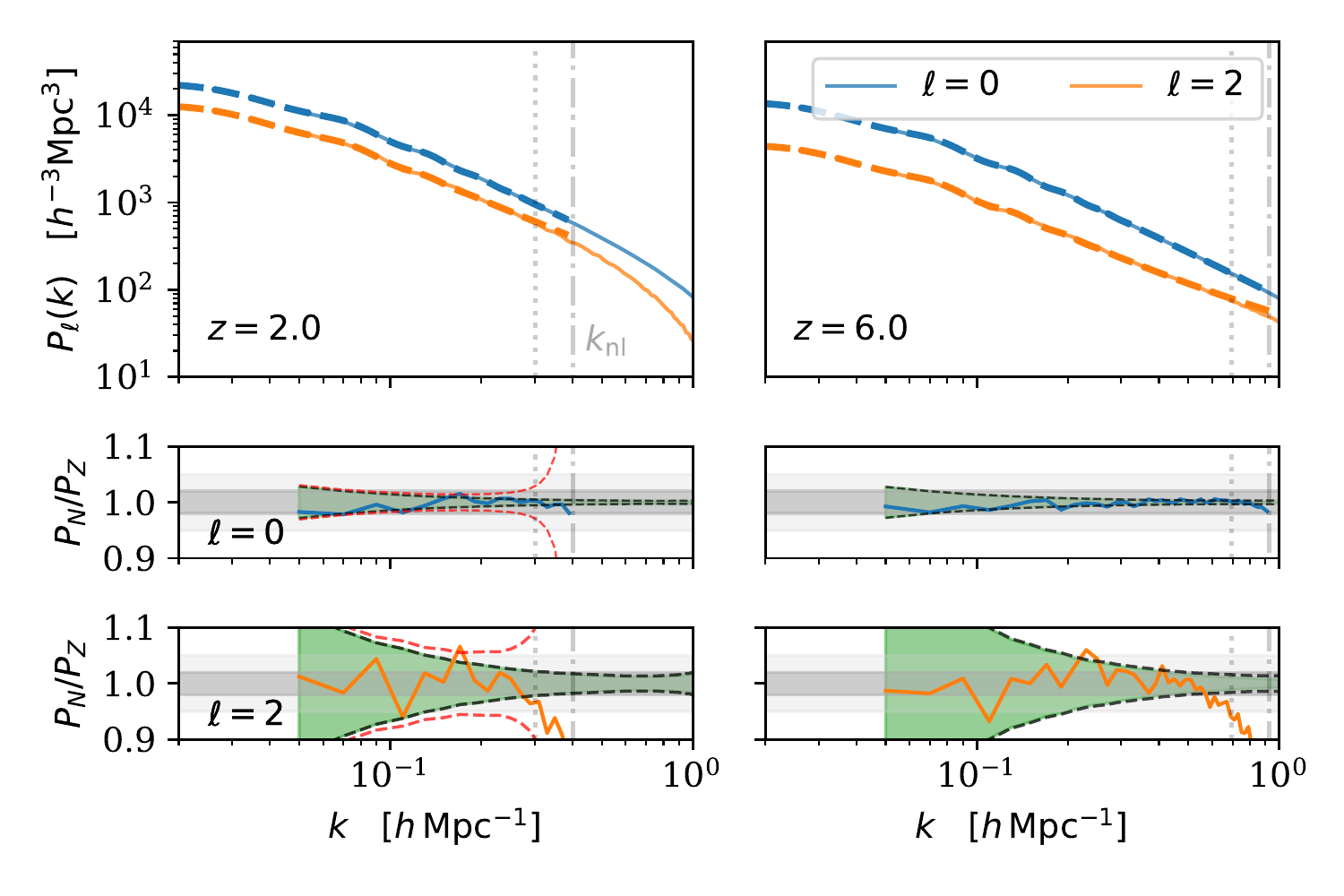}}
    \caption{(Top) The redshift-space, HI power spectrum multipoles in our fiducial model at $z=2$ and 6.  Solid lines show the N-body results while dashed lines show the Zeldovich effective field theory calculation.  Blue lines show the monopole ($\ell=0$) while red lines show the quadrupole ($\ell=2$).  A single set of parameters fits both multipoles simultaneously.  The dotted and dash-dot vertical grey lines mark $0.75k_{\rm nl}$ and $k_{\rm nl}$ of Eq.~(\ref{eqn:Sigma}).  (Bottom rows) The two rows show the ratio of the N-body results to the Zeldovich fit for monopole and quadrupole. The dark and light grey shaded regions show $\pm 2\%$ and $\pm 5\%$. The shaded green region shows the expected sample variance error in our simulation volume, including shot noise, though the actual error in our simulations is expected to be less due to our fixed amplitude construction of initial conditions. The dashed red and black lines show the sample variance error after including the thermal noise for HIRAX-like and Stage {\sc ii} experiment respectively.}
    \label{fig:zeld_pks}
\end{figure}

To model intermediate scales, similar to real space clustering, we turn to perturbative models and compare the power spectrum multipoles with ZEFT prediction. This is shown in Fig.~\ref{fig:zeld_pks}. There are numerous ways of computing the redshift-space power spectrum in the Zeldovich approximation \cite{Vlah18}, here we have performed a Hankel transform of the correlation function multipoles. The N-body multipole power spectra are noisier than their real-space counterparts, but the level of agreement is still quite good for such a simple model. Over the range shown the N-body results depart from linear theory by order unity, while the agreement with our Zeldovich model is always within sample variance.  In addition, we also show the sample variance after including the thermal noise, which dominates on small scales, for a HIRAX-like and Stage {\sc ii} experiment and find that perturbation theory fits the model well within the errors for HIRAX-like experiment on these small scales as well. Higher order perturbation theory and a more sophisticated treatment of redshift-space distortions would almost certainly improve the agreement, however this preliminary investigation suggests that even low order perturbation theory is correctly predicting the signatures of non-linear evolution and complex bias in our simulations. Given the enormous uncertainties associated with the manner in which HI traces the matter field, this suggests the Zeldovich approximation can be used as an efficient and relatively accurate forecasting tool \cite{CasWhi19}.

\begin{figure}
    \centering
    \resizebox{\columnwidth}{!}{\includegraphics{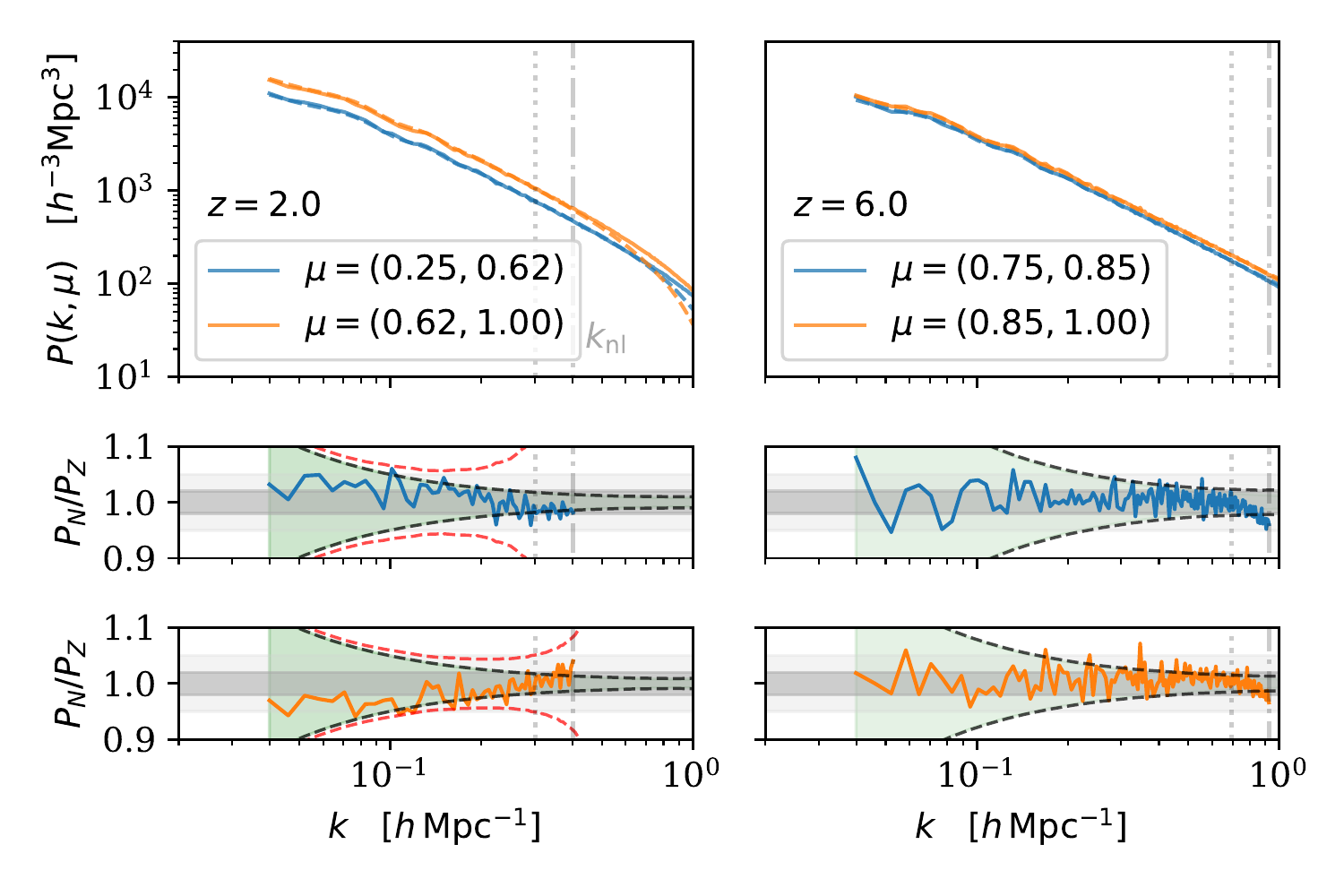}}
    \caption{(Top) The redshift-space, HI power spectrum wedges for our fiducial model at $z=2$ and 6. Solid lines show the N-body results while dashed lines show the Zeldovich effective field theory calculation.  A single set of parameters fits both the wedges simultaneously.  The dotted and dash-dot vertical grey lines mark $0.75k_{\rm nl}$ and $k_{\rm nl}$ of Eq.~(\ref{eqn:Sigma}).  (Bottom rows) The two rows show the ratio of the N-body results to the Zeldovich fit for the two wedges. The dark and light grey shaded regions show $\pm 2\%$ and $\pm 5\%$.  The shaded green region shows the expected sample variance error in our simulation volume, including shot noise, though the actual error in our simulations is expected to be less due to our fixed amplitude construction of initial conditions. The dashed red and black lines show the sample variance error after including the thermal noise for HIRAX-like and Stage {\sc ii} experiment respectively.
    }
    \label{fig:zeld_pkmu}
\end{figure}

In Fig.~\ref{fig:zeld_pkmu}, we also show the comparison for power spectrum wedges in redshift space. To be more realistic here, we have taken into account the ``pessimistic" case of the foreground wedge as outlined in ref.~\cite{Chen19} and split only the observable modes outside this into two equal wedges with respect to the line of sight. This allows us to see the impact of anisotropic thermal noise that is otherwise averaged over in case of multipoles. We clearly see that the thermal noise is larger perpendicular to the line of sight, but is however small in comparison to the sample variance for our $1\,h^{-1}$Gpc volume. To fit Zeldovich perturbation theory to the power spectrum wedges, we have used only first three multipoles and find that it is able to fit the simulations to within $5\%$ up to non-linear scales. Despite being within sample variance, the fit on the large scales for $z=2$ seems biased for the two wedges which is likely due to using only the first three, even multipoles to get the model power spectrum wedges. 

\subsection{Redshift-space clustering: Fingers of God}

\begin{figure}
    \centering
    \resizebox{\columnwidth}{!}{\includegraphics{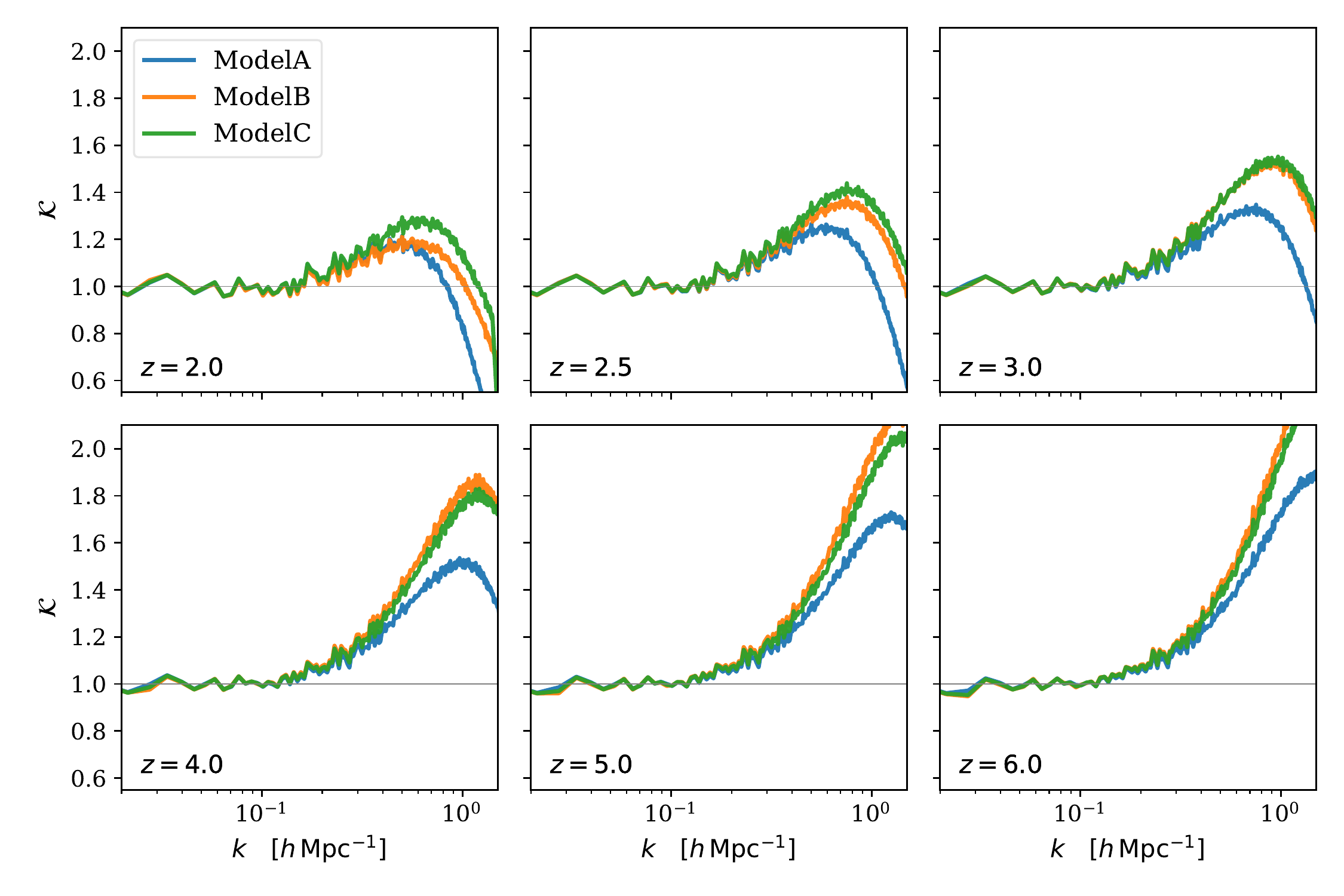}}
    \caption{Fingers of god suppression in the form of the kernel defined in Eq.~\ref{eqn:fogkernel} for the three models of the distribution of HI for various redshifts.
    }
    \label{fig:fog}
\end{figure}

Small-scale clustering in redshift space is suppressed due to FOG effects along the line of sight. This puts an upper limit on the modes which can be used to extract information and hence it is important to look at the amplitude of these effects. The simplest phenomological expression to model the redshift space spectrum using FOG effects is:
\begin{equation}
    P_{\rm HI}^s(k, \mu) = (1+[f/b]\mu^2)^2b^2_LP_{\rm L}(k) \mathcal{K}(k, \mu, \sigma_s) + N(k)
    \label{eqn:foglinear}
\end{equation}
where $b_L$, $P_L$ are the linear bias and linear power spectrum, $N(k)$ is the additive component of shot noise and the kernel $\mathcal{K}(k, \mu, \sigma_s)$ contains both perturbative and non perturbative effects. Usually the FOG effect contribution to this kernel is  modeled using a Gaussian or a Lorentzian. To quantitatively see where FOG lead to suppression of power and loss of information on small scales, we extract the kernel as  
\begin{equation}
    \mathcal{K}(k, \mu, \sigma_s) = \frac{1}{(1+[f/b]\mu^2)^2}\Big[\frac{P_{\rm HI}^s(k, \mu) - P_{\rm HI}^s(k, 0)}{ b^2_LP_{\rm L}(k)} +1\Big]
    \label{eqn:fogkernel}
\end{equation}
where we assumed that the shot noise $N(k)$ is $\mu$ independent and $\mathcal{K}(k, 0, \sigma_s)=1$. This kernel is shown in Fig.~\ref{fig:fog} for different HI models. On large scales, where linear theory is valid, we find the kernel to be close to unity as per our expectations. On intermediate scales, the signal increases due to nonlinear redshift space distortions and biasing. At high-$k$ the FOG suppress the signal,  and this suppression becomes smaller at high redshfit because the low mass halos we probe at high $z$ have low velocity dispersion. As seen in Fig.~\ref{fig:distribution}, since the amount of HI in satellites is higher in Model A, compared to the other two models, the FOG suppression is higher and at slightly larger scales. Interestingly, though Model C does not have satellites and hence does not strictly have 1-halo FOG effects caused due to velocity dispersion, it sees the same suppression as Model B at high redshifts. This suggests that this suppression is due to the combined effect of cluster-infall velocities of halos and the 1-halo contribution to clustering.

Given that FOG effect is subdominant in suppressing clustering at $k\le k_{\rm NL}$, they will not set the limit to the scales along the line of sight from which we can extract cosmological information. This is one advantage of 21-cm surveys over other spectroscopic surveys which can be severely limited by FOG effects (and redshift measurement uncertainties \cite{Wilson19}).  Small FOG also mean that the radial resolution of the analysis will be set by the channel bandwidth or the frequency resolution. The channel bandpass can be approximated with a Gaussian which gives an effective beam in the parallel direction \cite{Bull2015}
\begin{equation}
    B_\parallel(k_\parallel) = {\rm exp}\,\left(-\frac{[r_\nu k_\parallel\ \delta \nu/\nu_{21}]^2}{16{\rm ln}2}\right)
    \label{eqn:bandpass}
\end{equation}
where $r_\nu = c(1+z)^2/H(z)$ and $\delta \nu$ is the channel bandwidth. Based on Fig.~\ref{fig:fog}, if we expect to be able to model the signal up to $k=0.6$, $0.8$ and $1\,h\,{\rm Mpc}^{-1}$ at $z=2$, $3$ and $z\ge 4$, and we require the beam suppression on these scales to not be greater than $e^{-1}$, then this sets the minimum bandwidth to $\delta \nu = 5$, $2$, $1$, $0.8$ and $0.6\,$MHz respectively.

\subsection{Baryon acoustic oscillations}

\begin{figure}
    \centering
    \resizebox{\columnwidth}{!}{\includegraphics{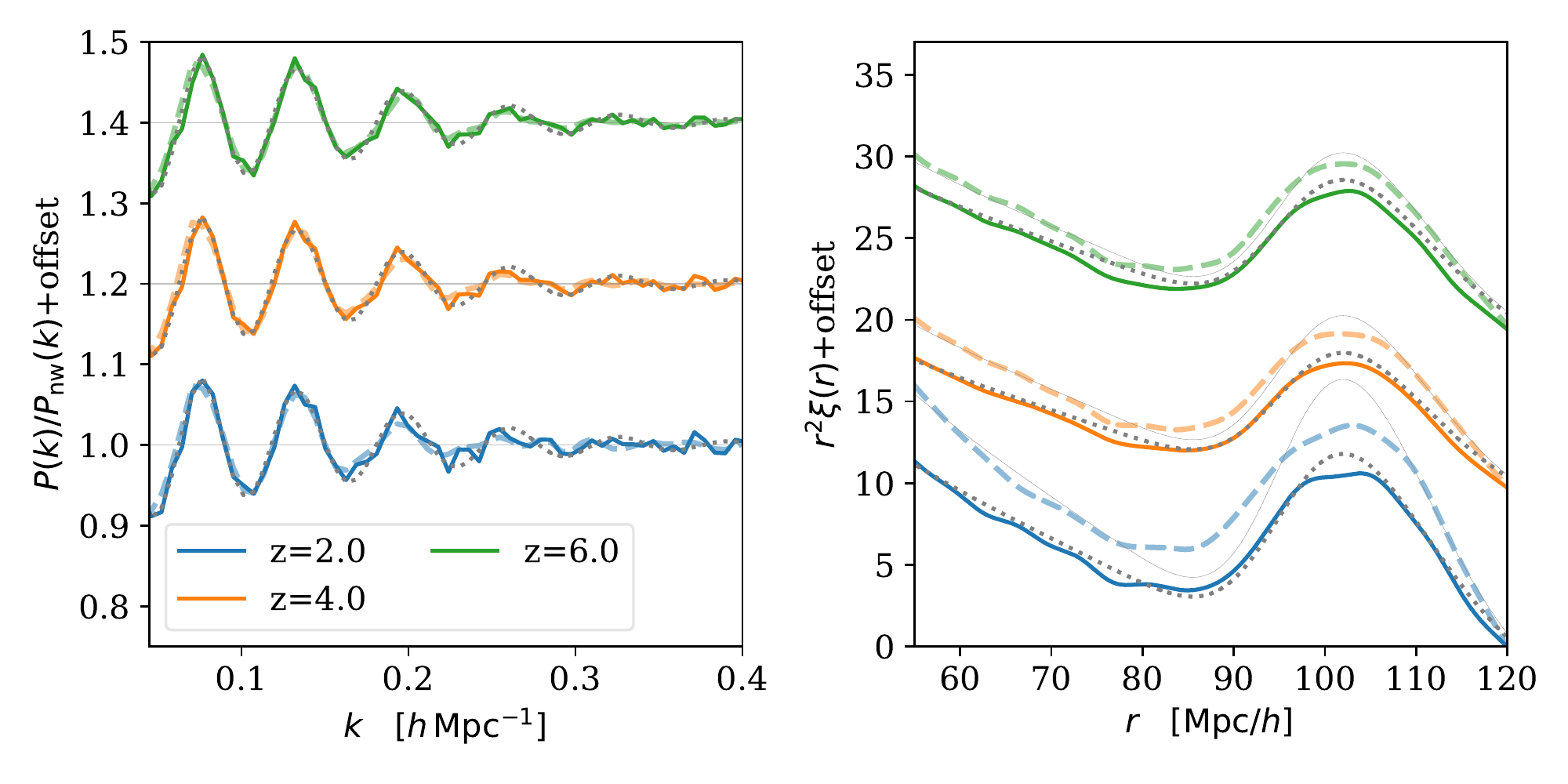}}
    \caption{BAO signal in real (solid lines) and redshift space (monopole, dashed) at various redshift compared with the linear theory in real (gray dots) and redshift space (gray solid line): (Left) Ratio of power spectrum to the ``no-wiggle'' template which is formed by smoothing with a Savitsky-Golay filter. We have moved lines for different redshift vertically by 0.2 to enhance visibility. (Right) Correlation function estimated by Hankel transform of the power spectrum, compared with linear theory i.e.\ linear bias and Kaiser prediction in real and redshift space respectively. We have offset lines for different redshift together vertically by multiples of 10 to enhance visibility.
    }
    \label{fig:bao}
\end{figure}

\begin{figure}
    \centering
    \resizebox{\columnwidth}{!}{\includegraphics{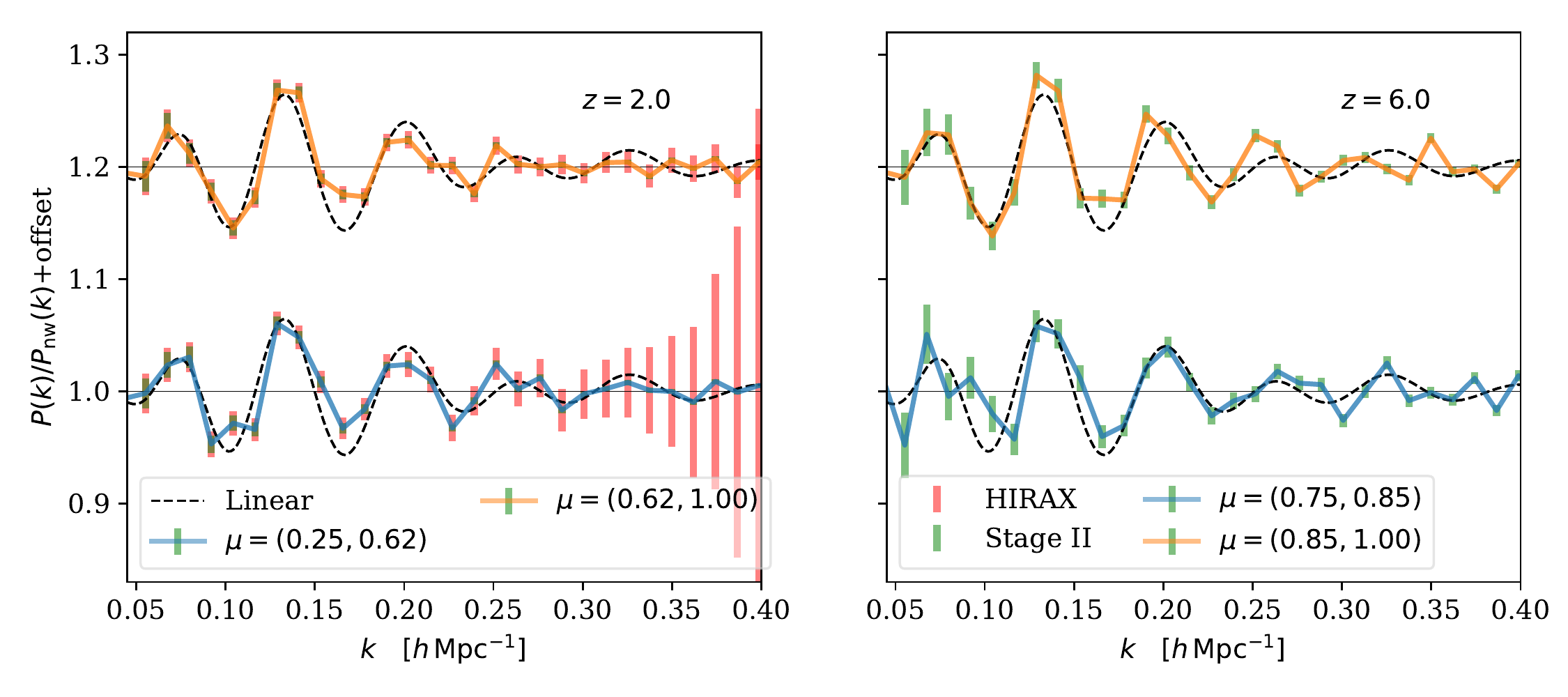}}
    \caption{BAO signal for the power spectrum wedges for the same configuration as Figure \ref{fig:zeld_pkmu} at redshifts $z=2$ and $6$, after taking the ratio of the power spectrum to the ``no-wiggle'' template which is formed by smoothing with a Savitsky-Golay filter. We have moved lines for different $\mu$-bins vertically by 0.2 to enhance visibility. The black dashed line is the linear theory prediction. We show error bars for the two experiments HIRAX-like (red) and Stage {\sc ii} (green). The error bars include the shot noise, thermal noise and sample variance for the volume corresponding to redshift slice of $\Delta z=0.2$ around the the central redshift. We have assumed sky coverage of $15000\, \rm{deg^2}$ and $20000\, \rm{deg^2}$ for the two experiments respectively.}
    \label{fig:bao_mu}
\end{figure}

A major goal of high $z$, 21-cm cosmology is the measurement of distances using the baryon acoustic oscillation method \cite{Weinberg13}.  In Fig.~\ref{fig:bao} we show this signal, in real and redshift space, as power spectrum and correlation function monopole and compare it against linear theory (dotted lines). We find that scale-dependent biasing, redshift-space distortions and non-linear evolution have only a small effect on the BAO signal, and this observation gets stronger as we go higher in redshift. 

In Fourier space the BAO signal appears as a quasi-periodic sequence of damped oscillations superposed upon the smooth shape of the power spectrum.  To enhance the visibility of peaks, we divide $P(k)$ in Fig.~\ref{fig:bao} by a ``no-wiggle'' template which is formed simply by smoothing the power spectrum with a Savitsky-Golay filter. This disentangles effects like scale-dependent bias which leads to only a smooth trend in $P(k)$ and are hence removed by the broad-band division. It is clear from Fig.~\ref{fig:bao} that the BAO peaks would be easily visible in the $k$-band probed by the upcoming 21-cm surveys. Non-linear structure formation damps the BAO peaks and reduces the signal-to-noise ratio for measuring the acoustic scale \cite{Bharadwaj98,Meiksin99,ESW07,Smith08,Crocce08,Matsubara08,Seo08,White14,White15}. Since the non-linear scale shifts to smaller scales at high-redshifts, this damping is expected to be small. In Fig.~\ref{fig:bao} only the fourth BAO peak is slightly damped at $z=2$ as compared to linear power spectrum and no damping is visible at $z=6$, in accordance with the value of $k_{\rm NL}$ at these redshifts.

Since the amount of damping is small at high redshifts and therefore quite difficult to see in the power spectrum, we also show the correlation function where the BAO signal appears as a single, well-localized peak at the BAO scale $\sim 100\,h^{-1}$Mpc. Since our purpose here is to only visually assess the degree of peak broadening, we do not measure the correlation function on these scales via pair-counting in the simulations, but compute it from the power spectrum measurement via Hankel transform. For this purpose, we extrapolate the measurement on large scales with the linear power spectrum and linear bias, and on small scales with a power-law.

The peak broadening over the linear theory prediction is more clearly visible in correlation space, for both real and redshift space clustering, and it decreases significantly between redshift $z=2$ and $4$. While not shown here to maintain clarity of the figure, we find that the Zeldovich approximation provides a good fit to the peak damping in the correlation function at all redshifts. As an aside, we note that the real and redshift space signals at higher redshifts are closer than at low redshifts, illustrating the decreasing effects of redshift space distortions on the broadband due to the increasing bias. Thus though the peak damping is slightly larger in redshift space than in real space at $z=2$, but this is no-longer (noticeably) the case at $z=6$.

The BAO peak can be sharpened, or the high-$k$ oscillations partially restored, by a process known as `reconstruction' \cite{ES3}.  Reconstruction is adversely affected by modes lost to foregrounds and instrument imperfections \cite{Seo16}, but this can be partly overcome by combining 21-cm observations with other surveys \cite{Cohn16}.  Conveniently, the gains from reconstruction are largest at the lowest redshifts, where the overlap with other probes of large-scale structure is also the largest.

To assess the feasibility of detecting BAO signature in presence of foreground wedges, Fig.~\ref{fig:bao_mu} shows BAO wiggles in the power spectrum wedges after removing the broadband. We consider the ``pessimistic" foreground wedge as outlined in the previous section for Fig.~\ref{fig:zeld_pkmu} \cite{Chen19}. We estimate realistic error bars for two experimental setups: HIRAX-like with sky coverage of $15000\, \rm{deg^2}$ and Stage {\sc ii} covering $20000\, \rm{deg^2}$ in the sky. The errors include the shot noise, thermal noise and sample variance for the volume corresponding to redshift slice of $\Delta z=0.2$ around the the central redshift. At redshift $z=2$, both the experiments should be able to detect the BAO signature with equally high significance on all the scales that are not damped due to non-linear evolution. At redshift $z=6$, Stage {\sc ii} is able to detect the signal with high significance despite having access to a very narrow wedge due to foregrounds. 
\Rsp{These results show that 21-cm interferometers are ideal instruments to measure BAO, in principle, and have clear advantages over single dish experiments \cite{FVN2017}, despite the fact that the latter are not affected by the foreground wedge.}

%%%%%%%%%%%%%%%%%%%%%%%%%%%%%%%%%%%%%%%%%
\section{Power spectrum response}
\label{sec:response}

In this section we present the response of the HI clustering to changes in parameters, both cosmological and astrophysical.  We focus on the response ($d\ln P/d\ln\theta$) to $\sigma_8$ and the lowest order Eulerian bias, $b_1$, which are degenerate with each other and with the mean brightness temperature, $\bar{T}$, in linear theory (see discussion in e.g.\ ref.~\cite{Chen19}). For simplicity, we will use Model C to distribute HI in halos since it has the fewest free parameters but still gives similar results to the other models at the 2-point function level. To get the derivatives of the power spectrum with respect to $\sigma_8$ we use our `F+' and `F-' simulations, which have the initial power spectrum amplitude changed by $\pm5\%$. Simply distributing HI with the same HOD parameters as the fiducial cosmology leads to a different linear bias on large scales. Thus we change the value of the parameter $M_{\rm cut}$ (while keeping $\alpha$ fixed to original value of $0.9$) such that the large scale bias, $b_1$, remains the same as in fiducial case. We then use finite central difference to get the response. To estimate the response to $b_1$ at fixed $\sigma_8$, we change $M_{\rm cut}$ by $\pm 5\%$ in dex in the fiducial cosmology and measure the change in clustering as well as change in bias, then evaluate the derivative with respect to bias using the chain-rule. 

\begin{figure}
    \centering
    \resizebox{\columnwidth}{!}{\includegraphics{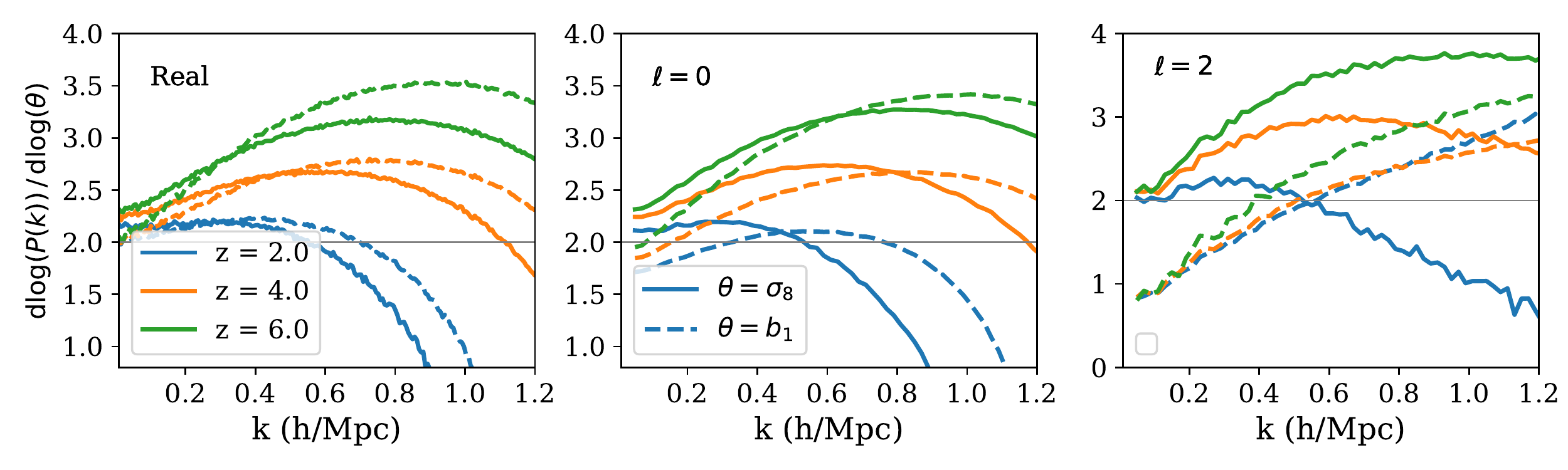}}
    \caption{Response of HI clustering to $\sigma_8$ (solid lines) and (Eulerian) $b_1$ (dashed lines). The gray horizontal line corresponds to derivative with brightness temperature $T_b$. The panels show derivative of real space power spectrum, monopole and quadrapole at three different redshifts
    }
    \label{fig:deriv}
\end{figure}

The power spectrum response, i.e.~the log-derivatives of $P_{HI}$ with respect to the different parameters, is shown in Fig.~\ref{fig:deriv} at three different redshifts.  We present the results for the real space power spectrum, the redshift space monopole and the redshift space quadrupole and for $\sigma_8$, $b_1$ and $\bar{T}$. In linear theory these are all degenerate. This can be seen on large scales in first panel as all the curves converge to $2$. However on mildly non-linear and smaller scales the response of real space spectra to both these parameters is different. As we go higher in redshift, where the bias of HI becomes increasingly non-linear, the shape of the derivatives becomes increasingly different. In redshift space, this degeneracy is also partially broken on linear scales due to redshift space distortions.  While it is not possible to compare our derivatives one-to-one due to the differences in the bias model, our results are very similar to those of ref.~\cite{CasWhi19} who presented a Fisher forecast for 21-cm surveys constraining the growth of large-scale structure.  Since our derivatives are easily distinguishable within the band that could be probed by upcoming 21-cm experiments, we too expect such experiments should be able to make highly precise measurements of the rate-of-growth of large-scale structure.

%%%%%%%%%%%%%%%%%%%%%%%%%%%%%%%%%%%%%%%%%
\section{Ultraviolet background fluctuations}
\label{sec:intensitybias}

Radiation backgrounds in the ultraviolet (UV) regime have the ability to ionize HI and thus modulate the HI content of halos and galaxies.
Spatial fluctuations in the UV background can hence modify the clustering of HI, leading to systematic effects in signals from different large scale structure surveys. This modulation has been shown to be significant for e.g.~3D Ly$\alpha$ forest surveys \cite{McQuinn11a,Pontzen14,Cabass,Gontcho14, Meiksin18}.
Recently ref.~\cite{Sanderbeck18} investigated the importance of this effect for the 21-cm signal and found it to be non-negligible, with an `intensity bias' of HI fluctuations $\sim -0.20$ to $-0.45$ for $1<z<3$ and different photoionization models.

Given the size and resolution of our simulations we have the unprecedented capability to implement these models in the N-body simulations and estimate this response directly from the simulations at high redshifts. To this end we will adopt a simple\footnote{Since our goal is a preliminary study to gauge the impact on the HI power spectrum on the scales relevant for proposed interferometers we also neglect light-cone and radiative transfer effects.  A more accurate calculation would be an interesting avenue for further exploration.} ionization model, motivated by that of ref.~\cite{Miralda00}, henceforth referred to as MHR. In the MHR model one assumes that there is a critical density, $\Delta_S$, above which HI is self-shielded. This critical density is set by the by the photo-ionization rate, with
$\Gamma \propto \Delta_S^{-\kappa}$.  If we further assume a power-law profile for the density of hydrogen inside each absorber ($\rho \propto r^{-\alpha}$), the HI content is modulated by the local ionizing radiation field as:
\begin{equation}
    \frac{M_{\rm HI}}{M_{\rm HI, 0}} = \left(\frac{\Gamma}{\Gamma_0}\right)^{p}
\label{eqn:mh1modulate}
\end{equation}
where $p=(3-\alpha)/(\alpha\kappa)$, ${\Gamma_0}$ is the mean background radiation and $M_{\rm HI, 0}$ is the fiducial HI mass that will be hosted inside the galaxy in case of the mean ionizing background. Here, we will take $M_{\rm HI, 0}$ to be the HI mass assigned to the galaxies previously in our Model A.  In what follows we consider eq.~\ref{eqn:mh1modulate}, with free parameter $p$, as our basic `model' of the impact of UV background fluctuations.  In Appendix \ref{app:uvbg} we give further details of our implementation and motivate reasonable choices for $p$.  Specifically we argue that in the MHR model the ionization index, $\kappa$, can be directly related to the galaxy density profile slope, $\alpha$.  Choosing $\alpha=2$, 2.5, and 2.9 (with $\alpha = 2$ corresponding to an isothermal distribution and $\alpha=2.9$ motivated by the HI profile in halos as seen by ref.~\cite{VN18}) leads to $0.02<p<1/3$ in eq.~\ref{eqn:mh1modulate}.  Regardless of the validity of the model, as we will see this range of $p$ generates an interesting range of signatures to explore.

\begin{figure}
    \centering
    \resizebox{\columnwidth}{!}{\includegraphics{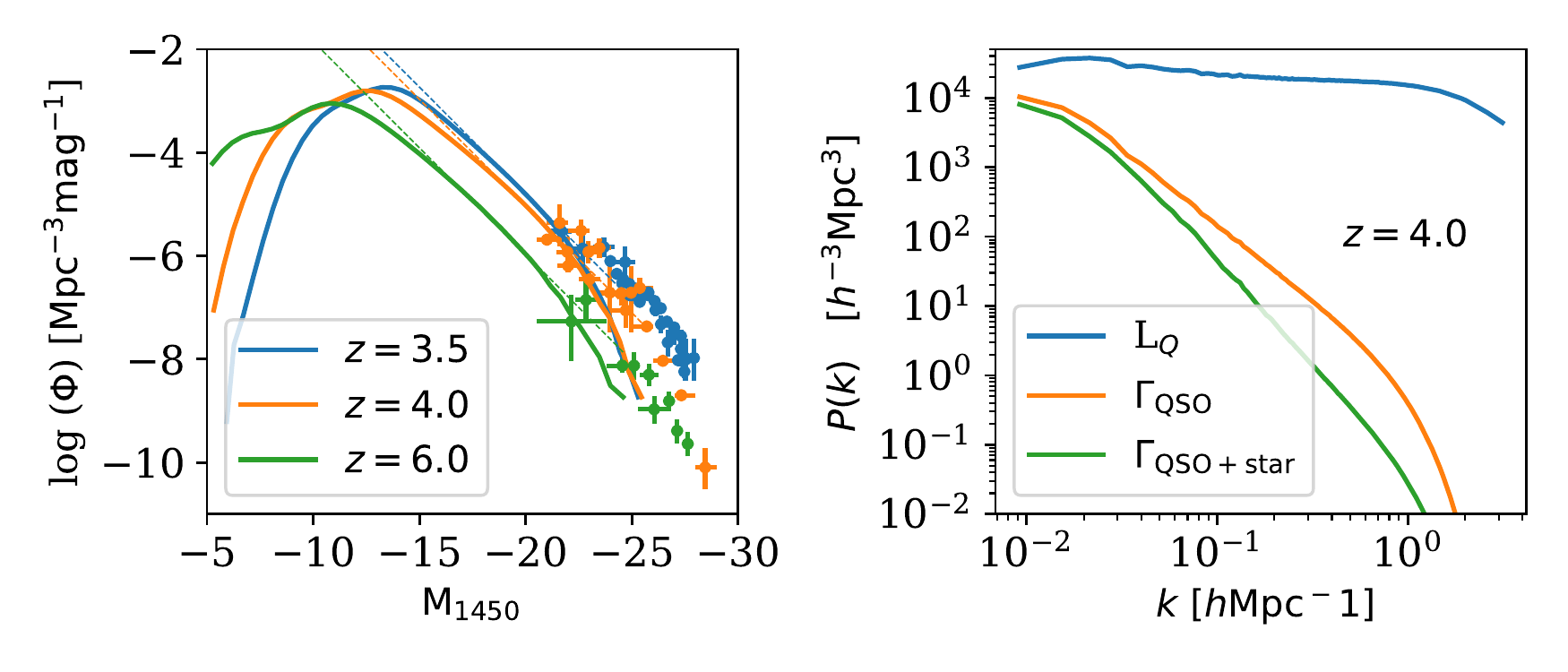}}
    \caption{(left) QSO luminosity function from our simulations (solid lines) at 3 different redshifts. The dashed lines are the double power law fitting function of ref.~\cite{McGreer18}. The points are data compiled by ref.~\cite{Kulkarni18}\protect\footnotemark in bins of $\Delta z=0.2$ around the central redshift. (right) Power spectrum of the QSO luminosity and ionizing UV background, with and without a stellar contribution, at $z=4$.}
\label{fig:qlf}
\end{figure}
\footnotetext{https://github.com/gkulkarni/QLF/blob/master/Data/allqlfs.dat}

We use the model proposed by ref.~\cite{ConWhite13} to make mock QSOs and generate a UV background field from the halos in the simulation (further details are provided in Appendix \ref{app:uvbg}). The mock QSO luminosity function is shown in Fig.~\ref{fig:qlf}, compared to recent observational determinations. Given a mean free path\footnote{We assume the mean free path is constant, neglecting the impact of density-driven fluctuations.  Within linear theory including this effect would serve to increase the intensity bias at fixed $p$ \cite{Sanderbeck18}.} of ionizing photons, $\lambda_{\rm mfp}^{912}$, one can generate a UV background.  The power spectra of the QSO luminosity field, $L_Q$, and of the UV background are shown in Fig.~\ref{fig:qlf}.  For QSOs, we find that the luminosity field $\rho_L$ is dominated by shot noise at all redshifts that we consider.

In addition to QSOs, the ionizing background is also generated by star forming galaxies. In fact, this contribution dominates at high redshifts \cite{Faucher19}. To model this scenario we consider a second case where we add to the QSO background one based on the stellar mass field, under the assumption that the mass-to-light ratio is constant. The amplitude of this redshift-dependent stellar background is estimated from Fig.~1 of ref.~\cite{Faucher19}.  Given our scaling of $L_Q$ with $M_{\rm gal}$, the main difference between the stellar background and the QSO background is stochasticity and scatter.

We are interested in estimating the effect of the fluctuations in the ionizing UV background on HI clustering. This can be measured most simply in the form of an `intensity bias'. To estimate this effect, we make the simple ansatz assuming a direct relation between density and temperature: 
\begin{equation}
    \delta_{\rm HI}^\Gamma = b_1^\Gamma \delta_{\rm m} + b_\Gamma \delta_\Gamma
    \label{eqn:gammabias}
\end{equation}
where $b_1^\Gamma$ is the `correct' HI bias with respect to matter in the presence of a fluctuating UV background, and $b_\Gamma$ is the corresponding intensity bias.
While there is apriori no reason for $b_1^\Gamma$ to be the same as the fiducial HI bias ($b_1$) measured in \S\ref{sec:clustering}, we find that the change in this bias is 
small ($<3\%$). Thus, henceforth, we assume $b_1^\Gamma \equiv b_1$ and so $b_1^\Gamma \delta_m = \delta_{\rm HI}^{\rm fid}$.
% \Rsp{Can we use this fact to argue that fluctuations in the mean free path are higher order and can therefore be neglected in our exploration?}\MW{No, I don't think so.  I believe that $b_\Gamma$ just gets rescaled by $\lambda$-fluctuations within linear theory.}

\begin{figure}
    \centering
    \resizebox{\columnwidth}{!}{\includegraphics{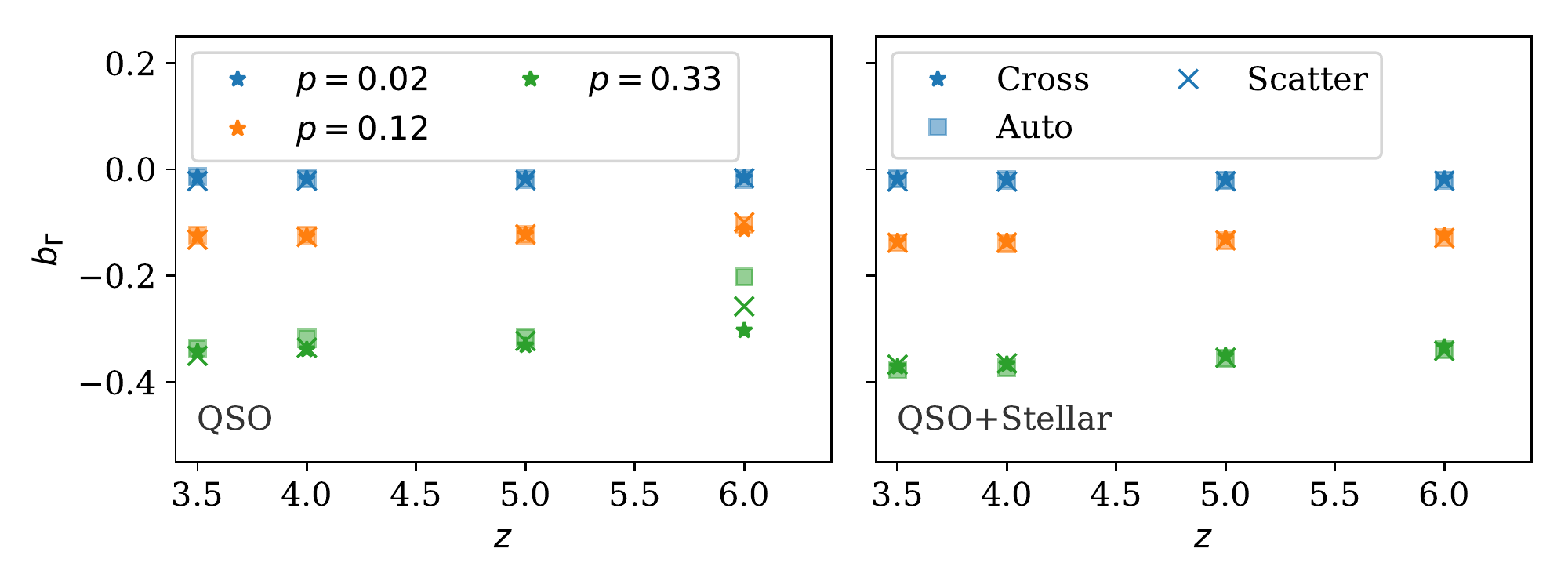}}
    \caption{Intensity bias, $b_\Gamma$, as estimated from the 3 methods outlined in \S\ref{sec:intensitybias} for a QSO-only ionizing background (left) and QSOs with a redshift dependent stellar background (right).}
    \label{fig:bgamma}
\end{figure}

Under this assumption, we estimate $b_\Gamma$ in three different ways
\begin{enumerate}
    \item Regression: we split our $1,024\,h^{-1}$Mpc simulation into voxels of size $128\,h^{-1}$Mpc and measure the change in mean HI overdensity as a function of the mean ionizing background in them. This is akin to the peak-background split method for estimating bias (e.g.\ see ref.~\cite{Modi16}).  We fit for
    \begin{equation}
     b_\Gamma \delta_\Gamma= \delta_{\rm HI}^\Gamma - \delta_{\rm HI}^{\rm fid}
    \label{eqn:scatterbias}
    \end{equation}
    For $z=3.5$ and 4 the default size of the voxels is smaller than the mean free path of the photons, but we have verified that using larger voxels does not change the measurement.
\item Cross correlation: We can also cross-correlate Eq.~\ref{eqn:gammabias} with matter and on large scales fit for the linear bias in the form
    \begin{equation}
         b_\Gamma P_{m \Gamma} + b_1 P_{mm} = P_{m \delta_{\rm HI}^{\Gamma}}
        \label{eqn:fourierbias_1}
    \end{equation}
    where $P_{m \Gamma}$, $P_{mm}$ and $ P_{m \delta_{\rm HI}^{\Gamma}}$ are the cross-spectrum between the UV and matter field, the matter auto-spectrum and the cross-spectrum between the modulated HI field and the matter. Here $b_1$ is the fiducial HI bias measured from \S\ref{sec:clustering} that we have used in place of $b_1^\Gamma$. In principle, one would fit for both $b_1^\Gamma$ and $b_\Gamma$ simultaneously, but we have verified that doing so does not significantly change the value of either bias.
    \item Auto correlation: We can estimate the auto-power spectrum of Eq.~\ref{eqn:gammabias} to estimate bias. This gives us a quadratic equation which can be solved at each scale and then the bias can be estimated from the mean value on large scales:
    \begin{equation}
         b_\Gamma^2(k) P_{\Gamma\Gamma} + 2 b_1 b_\Gamma P_{m\Gamma}  + P_{\delta_{\rm HI}\delta_{\rm HI}} = P_{\delta_{\rm HI}^\Gamma \delta_{\rm HI}^\Gamma}
        \label{eqn:fourierbias_2}
    \end{equation}
    where in addition to previous definitions, we have defined $P_{\Gamma\Gamma}$, $P_{\delta_{\rm HI}\delta_{\rm HI}}$ and $P_{\delta_{\rm HI}^\Gamma \delta_{\rm HI}^\Gamma}$ as the auto-spectra of the UV, fiducial HI and modulated HI fields. 
\end{enumerate}

We show the value of $b_\Gamma$ as estimated from these three methods in Fig.~\ref{fig:bgamma} and find them to be consistent with each other.  Interestingly the size of $b_\Gamma$ does not seem to have much redshift dependence, despite the strongly varying stellar background in the right panels.  This is not too surprising given our $b_\Gamma$ definition and could reflect the simplifying assumptions we have made.  However, $b_\Gamma$ depends strongly on the parameter $p$ with larger values showing larger bias.  This is expected on physical grounds if we associate larger $p$ with shallower HI profiles, since the shallower the profile the more HI mass is affected by a change in the threshold density for ionization.  The modulation of the HI with $\Gamma$ goes away as the power-law slope approaches $3$, since the enclosed mass becomes log-divergent at $r\to 0$ in this case leading to an infinite amount of mass at arbitrarily high density ($>\Delta_S$).

\begin{figure}
    \centering
    \resizebox{\columnwidth}{!}{\includegraphics{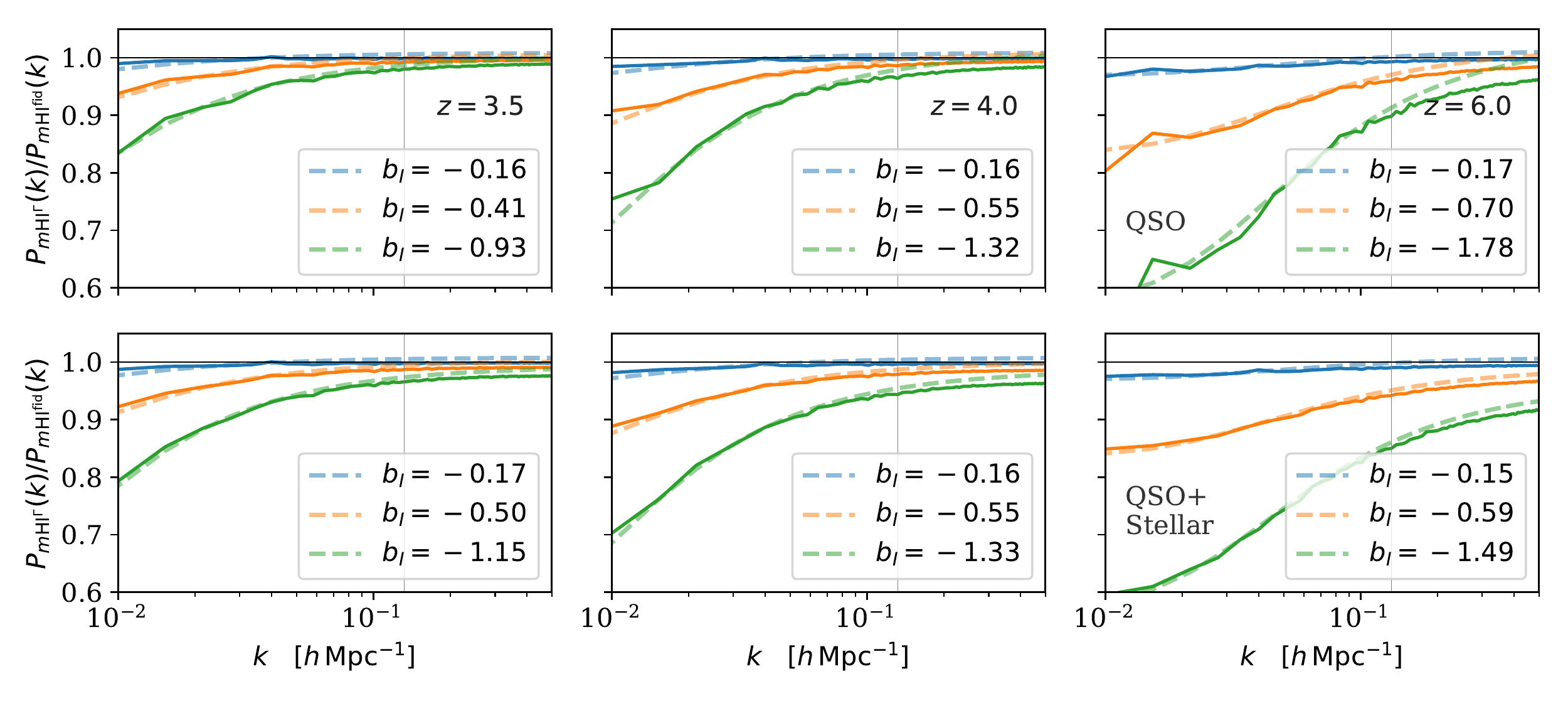}}
    \caption{Ratio of HI-matter cross power spectra in the presence of a fluctuating ionizing background to the fiducial case with uniform ionizing background, for three different redshifts (columns). We use the same color-scheme as Fig.~\ref{fig:bgamma} with blue, orange and green representing profile of $p=0.02$, $1/8$ and $1/3$ respectively. The top row is with a background generated by QSOs only, while the bottom row is includes a fluctuating stellar background as described in the text. Dashed lines are the fit from Eq.~\ref{eq:pmgamma} and legends are best-fit value of corresponding $b_I$ when fit up to $k_{\rm max}$ shown by gray vertical line.}
\label{fig:crossratio}
\end{figure}

To highlight how the HI clustering has changed in the presence of a UV background we measure the ratio of cross-spectra of HI with matter in our fiducial and fluctuating background cases (in real space). This is shown in Fig.~\ref{fig:crossratio} for three different redshifts.

To further quantify this effect, we can model the cross-spectra on large scales as:
\begin{equation}
    P_{m\delta_{\rm HI}^\Gamma} = \Big( b_1^\Gamma + b_I \frac{{\rm arctan} (k \lambda)}{k \lambda} \Big)\,P_{mm}
\label{eq:pmgamma}
\end{equation}
where $b_I$ is the combined bias to take into account $b_\Gamma$ and $b_L$, which is the luminosity bias of the sources of ionizing radiation. We again approximate $b_1^\Gamma$ to the fiducial value of $b_1$ since we find that it does not change the values measurably and improves the fit.  Fig.~\ref{fig:crossratio} shows that this model provides a good fit to our N-body results (solid vs.~dashed lines) and gives the corresponding values of $b_I$.  Since the scale dependent component in Eq.~\ref{eq:pmgamma} becomes unity on large scales, $b_I$ measures the maximum impact that the ionizing background can have on HI clustering over the fiducial case where this clustering is assumed to be only dependent on the underlying matter field.  Although we do not show it here, we verified that the impact on the HI auto-spectrum on large scales is approximately twice the size of the effect shown in Fig.~\ref{fig:crossratio}, as expected since the auto-spectrum scales as $b^2$ rather than $b$.  On small scales however, the auto-spectrum is noisier than the cross-spectrum, possibly due to non-Poisson shot noise and one halo clustering. Further, in our model the signal modulation occurs in real space, so the multipole moments of the redshift-space power spectrum are affected differently.  While the monpole looks similar to the real-space spectrum, the impact on the quadrupole is closer to a scale-independent suppression (at large scales).

Fig.~\ref{fig:crossratio} and eq.~\ref{eq:pmgamma} illustrate that the impact of UV background fluctuations on the HI is predominantly on large scales.  It is larger in the case that QSOs are the only source of UV radiation at $z=6$, since QSOs are rare and highly biased as compared to stellar contribution at that redshift, while at $z=3.5$, the stellar contribution seems to increase the bias slightly.  Based on our simplified model, UV background fluctuations will not be a major concern for RSD measurements, where most of the signal to noise is located at high-$k$, where $k\lambda\gg 1$, and the angular structure gives additional information.  However on large scales a non-zero response to the UV background will likely affect constraints on primordial non-Gaussianities \cite{Sanderbeck18}.  On the other hand the presence of foregrounds will limit 21-cm measurements to $k>0.01\,h\,\text{Mpc}^{-1}$, and most of the constraining power will come from the bispectrum rather than the power spectrum \cite{CVDE-21cm}.  The relative sensitivity of the bispectrum and power spectrum depends, in part, on how UV background fluctuations change the 21-cm bias on non-linear scales.  Based on these preliminary results we concur with the authors of ref.~\cite{Sanderbeck18} that UV background fluctuations need to be investigated more thoroughly, and we anticipate that our simulations may prove valuable in building a more refined model.

%%%%%%%%%%%%%%%%%%%%%%%%%%%%%%%%%%%%%%%%%
\section{Conclusions}
\label{sec:conclusions}

Intensity mapping techniques have emerged as a promising tool to efficiently probe the dark matter distribution at high redshifts, and a number of upcoming surveys plan to use the 21-cm emission signal from neutral hydrogen to probe this largely unexplored territory. The major challenge in accurately simulating 21-cm signal at these redshifts and on cosmological scales is simultaneously achieving high enough mass resolution to properly resolve the 21-cm emission while simultaneously covering a large enough volume to make precise statements about the clustering of the HI.

In this paper, we have introduced the HiddenValley simulations, a set of $10240^3$ particle N-body simulations of a $1\,(h^{-1}\text{Gpc})^3$ volume motivated by intensity mapping science. We have focused on the study of 21-cm fluctuations between redshifts 2 and 6 to demonstrate how the scales and resolution of these simulation allow us to address several modeling issues with higher fidelity than before. 

We used three different prescriptions from the literature to model the cosmological distribution of neutral hydrogen (\S\ref{sec:model}). Our first model is (sub)halo-based, where each halo contains some amount of HI which is then distributed between central and satellites galaxies. The second assumes HI traces the stellar mass of galaxies. The third is a simple halo-only model with constant parameters across the redshift range, primarily as a contrast to the other two models. 
% We find that all the three models are able to match the abundance of neutral hydrogen at all redshifts of interest (Fig.~\ref{fig:calibration}). 
We find that while the Model A and C are able to match the abundance of neutral hydrogen at all redshifts of interest (Fig.~\ref{fig:calibration}), Model B has a different evolution with the redshift due to constrained evolution directed only by stellar mass evolution. 
Despite having very different mechanism for HI distribution, all three models predict that majority of HI occupies halos in the mass range $10^{10}-10^{11.5}\,h^{-1}M_\odot$ (Fig.~\ref{fig:distribution}).

For every model we also investigate the clustering in real and redshift-space at the 2-point level, in both Fourier and configuration space.  While the models broadly predict similar clustering, only the first model (with redshift dependent parameters) has enough flexibility to match the observed DLA bias at $z=2-3$ while the other models predict a sharper evolution of bias with redshift.  In common with earlier work we find the bias of HI increases from near 2 at $z=2$ to 4 at $z=6$, becoming more scale dependent at high $z$ (Fig.~\ref{fig:real_bk}).

Though the contribution of satellites to the 21-cm signal is sub-dominant at all redshifts and in all models, the prediction for its value differs very much between the three models. Despite this, none of the models predict any significant from finger of god effects on the redshift space clustering for the range of scales relevant for proposed 21-cm instruments (Fig.~\ref{fig:fog}). Super-cluster infall on large scales accounts for most of the redshift space distortions observed in 21-cm signal at high redshifts. This is likely due to the low mass of the halos hosting the majority of the neutral hydrogen, which lack satellites and have low velocity dispersion.

The scale of our simulations also allows us to extract the baryon acoustic oscillations in the 21-cm clustering. Non-linear evolution damps the BAO signal and significantly reduces the signal-to-noise ratio at low redshifts. However since the evolution becomes increasingly linear at higher redshifts, we find that the damping of BAO signal decreases and is negligible at $z=6$ (Fig.~\ref{fig:bao}). Moreover, the damping along the line of sight due to redshift space distortions also decreases significantly by $z=6$.  In addition, we estimate uncertainties assuming a ``pessimistic'' wedge and realistic thermal noise to show that a HIRAX-like experiment is able to detect the BAO signal at $z=2$ in power spectrum wedges, and a Stage {\sc ii} experiment can detect the same at $z=6$ with high significance (Fig.~\ref{fig:bao_mu}).

As a preliminary step towards theoretical modeling of observed 21-cm clustering, we fit these signals with `Zeldovich effective field theory' (ZEFT \cite{Vlah15}) which is a simple, tree-level, Lagrangian perturbation theory model. We find that the model is simultaneously able to fit the matter auto-spectra, HI-matter cross spectra and the HI auto-spectra in real space at these redshifts, thus capturing the scale-dependence of the bias and the decorrelation with the matter field (Fig.~\ref{fig:zeld_pkr}). Though the redshift space spectra are noisier in our simulations, the simple model is still able to fit both the multipoles (Fig.~\ref{fig:zeld_pks}) and the wedges (Fig.~\ref{fig:zeld_pkmu}) within sample variance to beyond half $k_{\rm nl}$. We expect the performance can be improved using more sophisticated models, especially an improved treatment of redshift-space distortions, but we leave a more complete exploration for future work. 

Pushing beyond linear scales also allows one to constrain parameters such as the growth of fluctuations, Eulerian bias and the brightness temperature.  These parameters are degenerate in linear theory but non-linear clustering breaks the degeneracy from the auto-spectrum of the 21-cm signal only. Using our simulations, we show explicitly that the clustering in both real and redshift space responds to these parameters differently on mildy non-linear and smaller scales (Fig.~\ref{fig:deriv}), lending support to perturbative treatments which reach a similar conclusion \cite{CasWhi19}.

Lastly, we also demonstrate the reach of our simulations by modeling the intensity bias that can be caused by spatial fluctuations in the ultraviolet background that ionizes HI and hence modulates the HI content of halos. We measure this bias in three different ways: peak-background-split based regression, cross-spectra and auto-spectra and find them to be consistent with each other. In our model the value of intensity bias depends on the profile of HI inside the halos, with shallower profiles leading to increasing bias (Fig.~\ref{fig:bgamma}). While intensity bias measures the response of HI clustering to the UV background, we also explicitly see how the overall clustering of HI changes with respect to the underlying matter field in the presence of such a background (Fig.~\ref{fig:crossratio}). We find that the effect is non-negligible for the largest response we consider. Given our preliminary results and simplifying assumptions (such as constant mean free path for photons and lack of lightcone evolution) we believe that UV background fluctuations need to be investigated more thoroughly in the future, with our simulations providing an opportunity to do so.

The focus of this paper has been upon 21-cm cosmology, but we anticipate that the Hidden Valley simulations will be useful for a wide variety of other science including other intensity mapping probes \cite{Kovetz17}, high-redshift galaxies and QSOs \cite{Wilson19} and modeling the inter-galactic medium \cite{Meiksin09,McQuinn16}.  To facilitate these other investigations we will make the halo catalogs and the subsampled dark matter particles publicly available at {\tt cyril.astro.berkeley.edu} upon acceptance of this paper.

\section*{Acknowledgments}

EC and MW would like to thank the Cosmic Visions 21-cm Collaboration for stimulating discussion on the simulations requirements of 21-cm surveys.
We thank Matt McQuinn for several clarifying discussions about ultraviolet background fluctuations and Avery Meiksin for helpful comments on an earlier draft.
Y.F.~thanks Brandon Cook of NERSC for discussions on performance of MPI reduction operators.
M.W.~is supported by the U.S.~Department of Energy and by NSF grant number 1713791.
This research used resources of the National Energy Research Scientific Computing Center (NERSC), a U.S. Department of Energy Office of Science User Facility operated under Contract No. DE-AC02-05CH11231.
This work made extensive use of the NASA Astrophysics Data System and of the {\tt astro-ph} preprint archive at {\tt arXiv.org}.

%%%%%%%%%%%%%%%%%%%%%%%%%%%%%%%%%%%%%%%%%
\appendix
\section{Code Development and Operational Notes}
\label{app:scaling}

We employ the C/MPI implementation of FastPM \cite{FengChuEtAl16} for the simulations. Several updates to the software were developed and deployed during the prepare of the HiddenValley simulations, and we summarize them in this section. 

First, a Friends-of-Friends (FOF) halo finder \cite{DEFW} is added to FastPM for online data analysis, to avoid the need to write the full particle catalog. The local halo finder uses the fof module from {\tt kdcount}, which is based on a disjoint-set algorithm \cite{FengModi18}. The global merge algorithm is similar to the one in Gadget2 \cite{Gadget2}.
In addition to halo mass, center of mass position and velocity, the halo finder also computes the moments of inertia tensor, the velocity dispersion tensor, and the angular momentum tensor. In addition to the on-the-fly halo finder, a new command-line interface, {\tt fastpm-fof} is added. The command-line interface post-processes an existing particle catalog to produce a FOF catalog of a new linking length. 

Secondly, we identified a slow down in FastPM initial condition code that become severe in the HV10240 runs. FastPM uses a resolution invariant Gaussian random generator scheme, that is identical to {\tt Gadget} / {\tt N-GenIC}. The scheme always samples in the $z$-direction contiguously from a random number sequence. In previous FastPM, the $z$ direction of the initial Gaussian field is distributed to multiple MPI ranks, causing ranks to perform redundant sample draws (in order to preserve the exact sequence).  In this update we switched to a decomposition where the $z$-direction of the initial Gaussian field is contiguously stored with in a single MPI rank. We expect a reduction of factor 256 of wallclock time on relevant sections of the HV10240 run due to this update. The exact improvement was not measured because the runs with the previous initial conditions module were killed after they were stuck in the module for more than 10 minutes. The updated version spent only 107 seconds in the initial condition module.

Thirdly, the sorting module, {\tt mpsort}, and IO module, {\tt bigfile} \cite{mpsort, bigfile}, were modified to simplify the communication pattern in large scale runs. The update introduces a gather stage before the communication to reduce the number of MPI ranks that actually participates to the global all-to-all communication. The simplified communication pattern improved the reliability of runs via reducing the number of MPI messages posted to the communication network.

Finally, we made extensive use of the Data Warp Burst Buffer device on Cori, which provides very high IO bandwidth and IO operation throughput. We find a huge speed up (more than a factor of 10) switching from the scratch space (lustre file system) to burst buffer. The speed up increased the fraction of communication overhead of the IO module, and motivated our change to {\tt bigfile} that added the new gather stage. The updated version of {\tt bigfile} further reduced the IO time by a factor of 10. With the new software, IO took a small 7\% of the total wall-clock time of HV10240 runs. Since the IO / Computation ratio of FastPM is already higher than typical full N-body simulations and hydro-dynamical simulations, this finding suggests that on-the-fly compressed IO will not meaningfully reduce the wall clock time for N-body simulations on systems equipped with similar fast on-line IO devices. We also note that the bandwidth utilization is only a small fraction of the peak of the burst buffer, and further improvements are still possible.

The HV10240 runs were performed on the Cori computer at NERSC with 8192 KNL computing nodes and 524,288 MPI ranks. Utilizing 90\% of the full Cori KNL partition, and operating with half a million MPI ranks, requires special tweaks to the CrayMPI run-time environment.  We explored several and summarize our findings here:
\begin{enumerate}
\item
enabling huge pages of size 2 MB reduces the run time by 10\% comparing to not using huge pages.
\item 
increasing the all-reduce block size to 8 MB significantly improved the speed of the global reduction operation. ({\tt MPICH\_ALLREDUCE\_BLK\_SIZE})
\item increasing the number of Cray GNI message boxes to the number of MPI ranks
 negatively affected the reliability of the runs due to an exhaustion of 2 MB huge pages used for the message boxes. This is in contrast to our experience with BlueTides-II (81,000 ranks on NCSA BlueWaters), where the simulation could only proceed with the parameter set to the number of MPI ranks. ( \texttt{MPICH\_GNI\_MBOXES\_PER\_BLOCK})
\item allowing GNI to fall back to regular page sizes improved the reliability of the runs.\\
({\tt MPICH\_GNI\_MALLOC\_FALLBACK})
\end{enumerate}

\begin{figure}
    \centering
    \resizebox{\columnwidth}{!}{\includegraphics{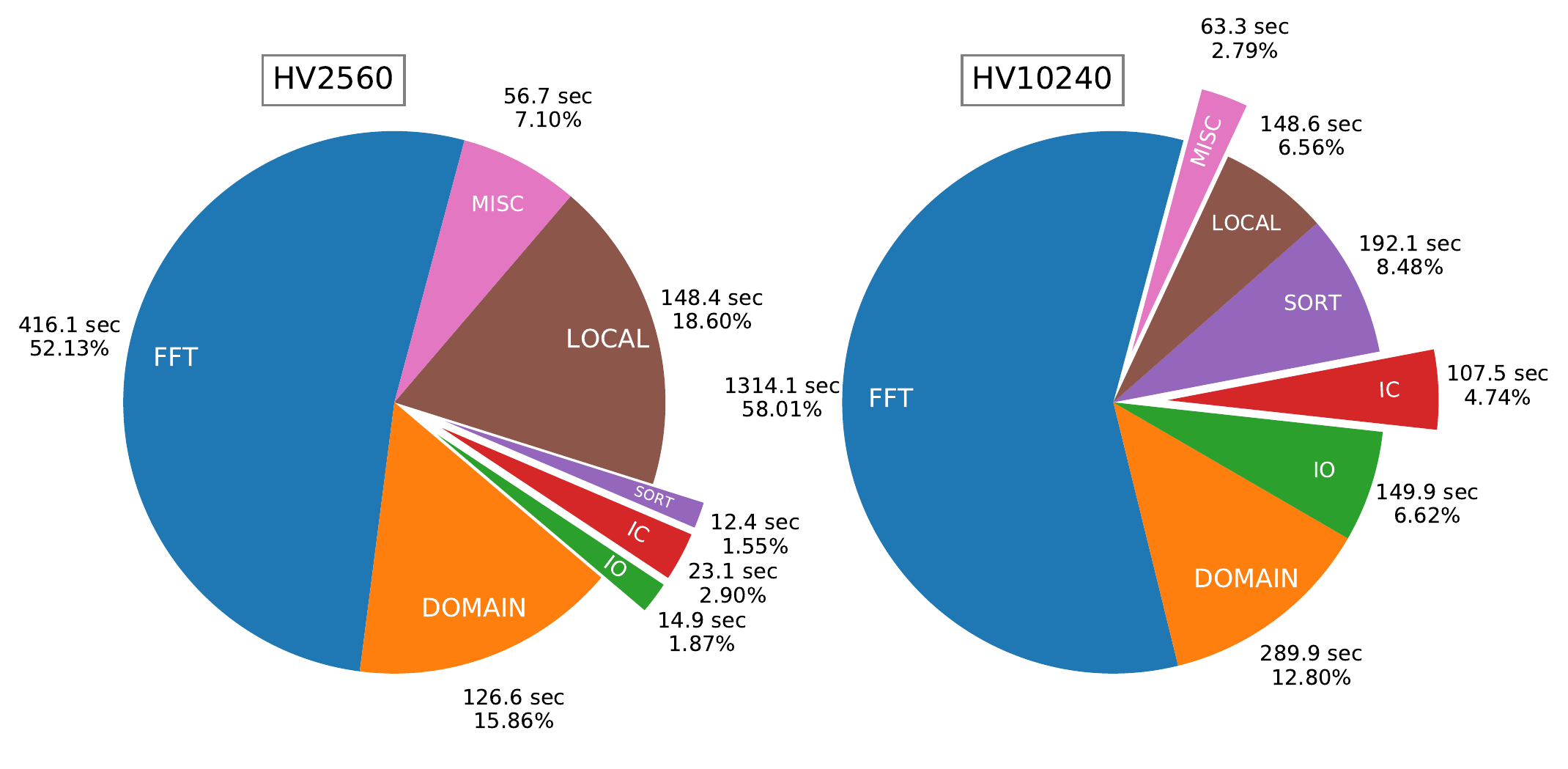}}
    \caption{The fraction of time in each component of the application for the HV2560 and HV10240 simulations. The HV2560 simulation was run on 128 nodes with 8,192 MPI ranks. The HV10240 simulation was run on 8192 nodes with 524,288 MPI ranks.}
    \label{fig:wallclock-pies}
\end{figure}

In Figure \ref{fig:wallclock-pies}, we show the wall clock time in different components of the HV2560 (128 nodes) and HV10240 runs (8192 nodes). Both simulations have the same load per MPI rank, and we see that the LOCAL operations took the same amount of computing time. We also see that operations that use network communication are penalized in the larger (HV10240) run compared to the smaller (HV2560) run:
\begin{itemize}
\item Fourier transform in the Poisson solver (FFT), $3\times$; 
\item decomposition and ghost particles (DOMAIN), $2\times$; 
\item sorting snapshots (SORT) and writing snapshots (IO), $10\times$; 
\item initial condition generator (IC), $5\times$. 
\end{itemize}
The penalty in the FFT is expected as network communication becomes limited by the bandwidth of nodes further away in the topology. The IC penalty is more severe than the FFT penalty; we believe this is because the IC module samples the full $x-y$ plane for random seeds, which is 16 times larger in HV10240 than in HV2560. There is a penalty in DOMAIN, because the 2-D pencil beams become narrower in HV10240, increasing the surface to volume ratio. The penalty in SORT and IO are more severe than for the FFT, suggesting there is still space for further optimization in future work.  However, we note that any gains in SORT and IO will only bring marginal improvement to the run time, as each currently uses less than 10\% of the total wall clock time. 

%%%%%%%%%%%%%%%%%%%%%%%%%%%%%%%%%%%%%%%%%
\section{Satellite model}
\label{app:satellites}

We allow some of the HI in our halos to be hosted by satellites, in addition to centrals.  As there is no observational data set with which to constrain the manner in which HI traces mass in satellite galaxies as a function of host halo mass and redshift, we shall opt for a simple model.  Our simulation does not resolve halo substructure directly, so we add satellites in post-processing.  We assume that the number of satellites is self-similar with \cite{Reed05,Angulo09,Brook14}
\begin{equation}
     N_{\rm sat}(>M_{\rm sat}) = \left(\frac{M_1}{M_{\rm sat}}\right)^{0.8}
     \qquad \mathrm{with}\qquad 0.1\,M_{\rm cut}<M_{\rm sat}<0.1\,M_h
\label{eqn:msat_dist}
\end{equation}
where $M_1=0.03\,M_h$ corresponds roughly to the mass of the heaviest subhalo \cite{Angulo09}. The slope of the HI-hosting subhalo mass function is not well known, but we have assumed it to be $0.8$, slightly shallower than that of all satellites which is $\sim 0.9$  \cite{Angulo09}.  The lower mass limit of $0.1\,M_{\rm cut}$ is purely for convenience as lower mass subhalos are essentially free of HI. Since the rate of HI mass loss within subhalos is not well constrained, we opt to use an instantaneous assigned subhalo mass rather than attempting to correct for differential matter-HI mass loss. 

Eq.~\ref{eqn:msat_dist} implies that the total number of subhalos more massive than $0.1\,M_{\rm cut}$ is $\left(M_1/0.1\,M_{\rm cut}\right)^{0.8}$.  We take this number of to be fixed and introduce stochasticity by randomly assigning masses to these satellites following Eq.~\ref{eqn:msat_dist}.  Each satellite is distributed within the halo following an NFW \cite{NFW} profile with a concentration of 7 and given an additional line-of-sight velocity drawn from a Gaussian whose width equals the halo velocity disperion, $\sigma_h$. We use two velocity dispersions, one corresponding to matter (from \cite{Evrard08}) and one with $\sigma_h$ reduced by two-thirds. The latter is motivated by Table 4 of \cite{VN18} who find that the dispersion of HI is lower than that of matter.  The details of the spatial distribution of subhalos significantly affects the clustering only on scales smaller than are of interest for proposed interferometric 21-cm experiments.  The line-of-sight velocity affects the redshift-space power spectrum, leading to an additional damping of power at high $k_\parallel$.

\section{UV Background and ionization model}
\label{app:uvbg}

In this appendix, we discuss the way we generate the ionizing UV background, as well as details of MHR model for photo-ionization. 

\subsection{Generating ionizing background}

To generate the UV ionization background field, we follow ref.~\cite{ConWhite13} to populate our galaxies with black holes and convert the black hole mass to QSO luminosity. Then we propagate these photons assuming a fixed mean free path $\lambda_{\rm mfp}^{912}$ (i.e.~neglecting any density-dependence of $\lambda_{\rm mfp}^{912}$). We use the relation in ref.~\cite{Worseck14} for the (comoving) mean free path at redshifts $2.3<z<5.5$, who provide a power law fit:
\begin{equation}
        \lambda_{\rm mfp}^{912} = 150\,h^{-1}{\rm Mpc}\ \left[\frac{1+z}{5}\right]^{-4.4}
        \qquad ({\rm comoving}) .
\end{equation}
For simplicity, we assume that the same relation holds at $z=6$. Since the mean free path is larger than our box size at $z\le 3$ we focus our attention on higher redshifts. 

To model the QSO sources of UV photons we assign our mock galaxies a black hole mass \cite{ConWhite13} 
\begin{equation}
    \frac{M_{\rm BH}}{10^{10}M_\odot} = 10^{\alpha}\Big(\frac{M_{\rm gal}}{10^{10}M_\odot}\Big)^{\beta}
\end{equation}
with $(\alpha, \beta) = (-3.5, 1)$, $(-4, 0.75)$, $(-3.7, 0.8)$, $(-3.6, 0.8)$ and $(-3.65, 0.7)$ at $z=6$, 5, 4.5, 4 and $3.5$ respectively. 
We adopt an $M_{\rm gal}$-independent scatter in this relation of $0.3\,$dex.
We assume a fraction, $f_{\rm on} = 0.01$, of the black holes are radiating as QSOs with
\begin{equation}
    L_Q = 3.3\, \times \, 10^4L_\odot\ \eta\ \frac{M_{\rm BH}}{10^{10}M_\odot}
\end{equation}
and sample the Eddington ratio, $\eta\equiv L/L_{\rm Edd}$, from a log-normal distribution with redshift independent mean $\eta_0=0.1$ and dispersion $0.1\,$dex.  We convert $L_Q$ to magnitudes via $M_{1450} = 72.79 - 2.5\, {\rm log}_{10}\, L_Q$.

In figure \ref{fig:qlf}, we show that the QSO luminosity function matches recent measurements where there is overlap, though our simulation volume is small and observations do not probe faint QSOs.  With the QSOs in hand, the UV background is generated from the QSO luminosity field $\rho_L$ in our simulation as
\begin{equation}
    \rho_\Gamma({\bf x}) = \int d^3r \frac{\rho_L({\bf x+r})e^{-r/\lambda}}{r^2}
    \qquad\Rightarrow\qquad
    \rho_\Gamma({\bf k}) = \rho_L({\bf k})\frac{{\rm arctan}(k\lambda)}{k\lambda}
\end{equation}
Since we do not attempt to model small-scale radiative transfer within halos or other small-scale effects, we additionally smooth the UV background by $1\,h^{-1}$Mpc to tame the $r^{-2}$ divergence above.

The contribution from the stellar background follows a similar approach, except we assume a constant stellar-mass-to-light ratio so $L_Q$ above is replaced by $M_{\rm \star}$ as estimated from fit of ref. \cite{Moster13}.  This is added to $\rho_\Gamma$ with amplitude $\langle \rho_\Gamma \rangle$ times the ratio of the contribution from star forming galaxies to that of QSOs. From figure 1 of ref.~\cite{Faucher19} we take this ratio to be $1$, 2, 4, 7, 10 and $100$ at redshift $z = 3$, 3.5, 4, 4.5, 5 and $6$ respectively.

\subsection{MHR model}

Given a UV background we modulate the HI content of every galaxy as a power-law in $\Gamma$ at its location.  While this model can be taken as phenomenological, it can also be motivated by the MHR model \cite{Miralda00, McQuinn11,Sanderbeck18}.  To see this assume that there is a critical density, $\Delta_S$, above which HI is self-shielded.  Further assume that this critical density is set by the by the photo-ionization rate with the relation
\begin{equation}
    \Gamma \propto \Delta_S^{(7-\gamma)/3} \propto \Delta_S^{-\kappa} 
    \label{eqn:gamma_2}
\end{equation}
where $\gamma$ is the power-law index of the volume weighted gas density distribution $P(\Delta)$ for $\Delta\gg 1$ and we define $\kappa = (7-\gamma)/3$ for convenience. If the density profile of HI inside the galaxies is assumed to be a power-law with index $\alpha$ ($\rho \propto r^{-\alpha}$) then $\alpha=3/(\gamma-1)$.
The mass of HI inside a galaxy is $M_{\rm HI} \propto R_{\Delta_S}^{3-\alpha}$ where $R_{\Delta_S} \propto \Delta_S^{-1/\alpha}$ is the radius corresponding to the critical self-shielding density. Combining these equations, we find that there is a power-law relation between the mass of HI inside galaxies and the ionizing background:
\begin{equation}
    M_{\rm HI} \propto \Gamma^p \qquad{\rm with}\quad p=\frac{3-\alpha}{\alpha \kappa}
\end{equation}
To estimate a plausible range for $p$, consider $\alpha$ varying from $2$, corresponding to the isothermal case, through $2.9$, which is roughly the slope of HI profile in the halos as found by ref.~\cite{VN18}.  This gives $0.02<p<1/3$.
We assume that in the case of a uniform mean background, ${\Gamma_0}$, there is a fiducial mass of HI, $M_{\rm HI, 0}$, that will be hosted inside each galaxy (which we take to be the HI mass assigned to the galaxies in our Model A).  We then modulate this mass as a power of the local $\Gamma$ as in Eq.~\ref{eqn:mh1modulate}.

\bibliographystyle{JHEP}
\bibliography{main}

\providecommand{\href}[2]{#2}\begingroup\raggedright\begin{thebibliography}{100}

\bibitem{Kovetz17}
E.~D. {Kovetz}, M.~P. {Viero}, A.~{Lidz}, L.~{Newburgh}, M.~{Rahman},
  E.~{Switzer} et~al., \emph{{Line-Intensity Mapping: 2017 Status Report}},
  {\emph{ArXiv e-prints} (2017) }
  [\href{https://arxiv.org/abs/1709.09066}{{\ttfamily 1709.09066}}].

\bibitem{CVDE-21cm}
{Cosmic Visions 21 cm Collaboration}, R.~{Ansari}, E.~J. {Arena}, K.~{Bandura},
  P.~{Bull}, E.~{Castorina} et~al., \emph{{Inflation and Early Dark Energy with
  a Stage II Hydrogen Intensity Mapping experiment}}, {\emph{ArXiv e-prints}
  (2018) } [\href{https://arxiv.org/abs/1810.09572}{{\ttfamily 1810.09572}}].

\bibitem{CHIME}
K.~{Bandura}, G.~E. {Addison}, M.~{Amiri}, J.~R. {Bond}, D.~{Campbell-Wilson},
  L.~{Connor} et~al., \emph{{Canadian Hydrogen Intensity Mapping Experiment
  (CHIME) pathfinder}},  in \emph{Ground-based and Airborne Telescopes V},
  vol.~9145 of \emph{\procspie}, p.~914522, July, 2014,
  \href{https://arxiv.org/abs/1406.2288}{{\ttfamily 1406.2288}},
  \href{https://doi.org/10.1117/12.2054950}{DOI}.

\bibitem{HIRAX}
L.~B. {Newburgh}, K.~{Bandura}, M.~A. {Bucher}, T.-C. {Chang}, H.~C. {Chiang},
  J.~F. {Cliche} et~al., \emph{{HIRAX: a probe of dark energy and radio
  transients}},  in \emph{Ground-based and Airborne Telescopes VI}, vol.~9906
  of \emph{\procspie}, p.~99065X, Aug., 2016,
  \href{https://arxiv.org/abs/1607.02059}{{\ttfamily 1607.02059}},
  \href{https://doi.org/10.1117/12.2234286}{DOI}.

\bibitem{Bingo}
R.~{Battye}, I.~{Browne}, T.~{Chen}, C.~{Dickinson}, S.~{Harper}, L.~{Olivari}
  et~al., \emph{{Update on the BINGO 21cm intensity mapping experiment}},
  {\emph{ArXiv e-prints} (2016) }
  [\href{https://arxiv.org/abs/1610.06826}{{\ttfamily 1610.06826}}].

\bibitem{Tianlai}
X.~{Chen}, \emph{{The Tianlai Project: a 21CM Cosmology Experiment}},  in
  \emph{International Journal of Modern Physics Conference Series}, vol.~12 of
  \emph{International Journal of Modern Physics Conference Series},
  pp.~256--263, Mar., 2012, \href{https://arxiv.org/abs/1212.6278}{{\ttfamily
  1212.6278}}, \href{https://doi.org/10.1142/S2010194512006459}{DOI}.

\bibitem{SKACosmo}
F.~B. {Abdalla}, P.~{Bull}, S.~{Camera}, A.~{Benoit-L{\'e}vy}, B.~{Joachimi},
  D.~{Kirk} et~al., \emph{{Cosmology from HI galaxy surveys with the SKA}},
  {\emph{Advancing Astrophysics with the Square Kilometre Array (AASKA14)}
  (2015) 17} [\href{https://arxiv.org/abs/1501.04035}{{\ttfamily 1501.04035}}].

\bibitem{CasWhi19}
E.~{Castorina} and M.~{White}, \emph{{Measuring the growth of structure with
  intensity mapping surveys}}, {\emph{arXiv e-prints} (2019) }
  [\href{https://arxiv.org/abs/1902.07147}{{\ttfamily 1902.07147}}].

\bibitem{Dave13}
R.~{Dav{\'e}}, N.~{Katz}, B.~D. {Oppenheimer}, J.~A. {Kollmeier} and D.~H.
  {Weinberg}, \emph{{The neutral hydrogen content of galaxies in cosmological
  hydrodynamic simulations}},
  \href{https://doi.org/10.1093/mnras/stt1274}{\emph{\mnras} {\bfseries 434}
  (2013) 2645} [\href{https://arxiv.org/abs/1302.3631}{{\ttfamily 1302.3631}}].

\bibitem{Eagle}
A.~{Rahmati}, J.~{Schaye}, R.~G. {Bower}, R.~A. {Crain}, M.~{Furlong},
  M.~{Schaller} et~al., \emph{{The distribution of neutral hydrogen around
  high-redshift galaxies and quasars in the EAGLE simulation}},
  \href{https://doi.org/10.1093/mnras/stv1414}{\emph{\mnras} {\bfseries 452}
  (2015) 2034} [\href{https://arxiv.org/abs/1503.05553}{{\ttfamily
  1503.05553}}].

\bibitem{VN18}
F.~{Villaescusa-Navarro}, S.~{Genel}, E.~{Castorina}, A.~{Obuljen}, D.~N.
  {Spergel}, L.~{Hernquist} et~al., \emph{{Ingredients for 21cm intensity
  mapping}}, {\emph{ArXiv e-prints} (2018) }
  [\href{https://arxiv.org/abs/1804.09180}{{\ttfamily 1804.09180}}].

\bibitem{Padmanabhan17}
H.~{Padmanabhan}, A.~{Refregier} and A.~{Amara}, \emph{{A halo model for
  cosmological neutral hydrogen : abundances and clustering}},
  \href{https://doi.org/10.1093/mnras/stx979}{\emph{\mnras} {\bfseries 469}
  (2017) 2323} [\href{https://arxiv.org/abs/1611.06235}{{\ttfamily
  1611.06235}}].

\bibitem{Castorina17}
E.~{Castorina} and F.~{Villaescusa-Navarro}, \emph{{On the spatial distribution
  of neutral hydrogen in the Universe: bias and shot-noise of the H I power
  spectrum}}, \href{https://doi.org/10.1093/mnras/stx1599}{\emph{\mnras}
  {\bfseries 471} (2017) 1788}
  [\href{https://arxiv.org/abs/1609.05157}{{\ttfamily 1609.05157}}].

\bibitem{Padmanabhan15}
H.~{Padmanabhan}, T.~R. {Choudhury} and A.~{Refregier}, \emph{{Theoretical and
  observational constraints on the H I intensity power spectrum}},
  \href{https://doi.org/10.1093/mnras/stu2702}{\emph{\mnras} {\bfseries 447}
  (2015) 3745} [\href{https://arxiv.org/abs/1407.6366}{{\ttfamily 1407.6366}}].

\bibitem{Obuljen2019}
A.~{Obuljen}, D.~{Alonso}, F.~{Villaescusa-Navarro}, I.~{Yoon} and M.~{Jones},
  \emph{{The H I content of dark matter haloes at z {\ensuremath{\approx}} 0
  from ALFALFA}}, \href{https://doi.org/10.1093/mnras/stz1118}{\emph{\mnras}
  {\bfseries 486} (2019) 5124}
  [\href{https://arxiv.org/abs/1805.00934}{{\ttfamily 1805.00934}}].

\bibitem{Masui13}
K.~W. {Masui}, E.~R. {Switzer}, N.~{Banavar}, K.~{Bandura}, C.~{Blake}, L.-M.
  {Calin} et~al., \emph{{Measurement of 21 cm Brightness Fluctuations at $z\sim
  0.8$ in Cross-correlation}},
  \href{https://doi.org/10.1088/2041-8205/763/1/L20}{\emph{\apjl} {\bfseries
  763} (2013) L20} [\href{https://arxiv.org/abs/1208.0331}{{\ttfamily
  1208.0331}}].

\bibitem{Switzer13}
E.~R. {Switzer}, K.~W. {Masui}, K.~{Bandura}, L.-M. {Calin}, T.-C. {Chang},
  X.-L. {Chen} et~al., \emph{{Determination of $z\sim 0.8$ neutral hydrogen
  fluctuations using the 21 cm intensity mapping autocorrelation}},
  \href{https://doi.org/10.1093/mnrasl/slt074}{\emph{\mnras} {\bfseries 434}
  (2013) L46} [\href{https://arxiv.org/abs/1304.3712}{{\ttfamily 1304.3712}}].

\bibitem{Anderson18}
C.~J. {Anderson}, N.~J. {Luciw}, Y.-C. {Li}, C.~Y. {Kuo}, J.~{Yadav}, K.~W.
  {Masui} et~al., \emph{{Low-amplitude clustering in low-redshift 21-cm
  intensity maps cross-correlated with 2dF galaxy densities}},
  \href{https://doi.org/10.1093/mnras/sty346}{\emph{\mnras} {\bfseries 476}
  (2018) 3382} [\href{https://arxiv.org/abs/1710.00424}{{\ttfamily
  1710.00424}}].

\bibitem{Rafols}
I.~{P{\'e}rez-R{\`a}fols}, A.~{Font-Ribera}, J.~{Miralda-Escud{\'e}},
  M.~{Blomqvist}, S.~{Bird}, N.~{Busca} et~al., \emph{{The SDSS-DR12
  large-scale cross-correlation of damped Lyman alpha systems with the Lyman
  alpha forest}}, \href{https://doi.org/10.1093/mnras/stx2525}{\emph{\mnras}
  {\bfseries 473} (2018) 3019}
  [\href{https://arxiv.org/abs/1709.00889}{{\ttfamily 1709.00889}}].

\bibitem{TMS17}
A.~R. {Thompson}, J.~M. {Moran} and G.~W. {Swenson}, Jr., \emph{{Interferometry
  and Synthesis in Radio Astronomy, 3rd Edition}}. 2017.

\bibitem{ZalFurHer04}
M.~{Zaldarriaga}, S.~R. {Furlanetto} and L.~{Hernquist}, \emph{{21 Centimeter
  Fluctuations from Cosmic Gas at High Redshifts}},
  \href{https://doi.org/10.1086/386327}{\emph{\apj} {\bfseries 608} (2004) 622}
  [\href{https://arxiv.org/abs/astro-ph/0311514}{{\ttfamily
  astro-ph/0311514}}].

\bibitem{McQ06}
M.~{McQuinn}, O.~{Zahn}, M.~{Zaldarriaga}, L.~{Hernquist} and S.~R.
  {Furlanetto}, \emph{{Cosmological Parameter Estimation Using 21 cm Radiation
  from the Epoch of Reionization}},
  \href{https://doi.org/10.1086/505167}{\emph{\apj} {\bfseries 653} (2006) 815}
  [\href{https://arxiv.org/abs/astro-ph/0512263}{{\ttfamily
  astro-ph/0512263}}].

\bibitem{Seo2010}
H.-J. {Seo}, S.~{Dodelson}, J.~{Marriner}, D.~{Mcginnis}, A.~{Stebbins},
  C.~{Stoughton} et~al., \emph{{A Ground-based 21 cm Baryon Acoustic
  Oscillation Survey}},
  \href{https://doi.org/10.1088/0004-637X/721/1/164}{\emph{\apj} {\bfseries
  721} (2010) 164} [\href{https://arxiv.org/abs/0910.5007}{{\ttfamily
  0910.5007}}].

\bibitem{Bull2015}
P.~{Bull}, P.~G. {Ferreira}, P.~{Patel} and M.~G. {Santos}, \emph{{Late-time
  Cosmology with 21 cm Intensity Mapping Experiments}},
  \href{https://doi.org/10.1088/0004-637X/803/1/21}{\emph{\apj} {\bfseries 803}
  (2015) 21} [\href{https://arxiv.org/abs/1405.1452}{{\ttfamily 1405.1452}}].

\bibitem{SeoHir16}
H.-J. {Seo} and C.~M. {Hirata}, \emph{{The foreground wedge and 21-cm BAO
  surveys}}, \href{https://doi.org/10.1093/mnras/stv2806}{\emph{\mnras}
  {\bfseries 456} (2016) 3142}
  [\href{https://arxiv.org/abs/1508.06503}{{\ttfamily 1508.06503}}].

\bibitem{Wol17}
L.~{Wolz}, C.~{Blake} and J.~S.~B. {Wyithe}, \emph{{Determining the HI content
  of galaxies via intensity mapping cross-correlations}}, {\emph{ArXiv
  e-prints} (2017) } [\href{https://arxiv.org/abs/1703.08268}{{\ttfamily
  1703.08268}}].

\bibitem{Alonso17}
D.~{Alonso}, P.~G. {Ferreira}, M.~J. {Jarvis} and K.~{Moodley},
  \emph{{Calibrating photometric redshifts with intensity mapping
  observations}}, {\emph{ArXiv e-prints} (2017) }
  [\href{https://arxiv.org/abs/1704.01941}{{\ttfamily 1704.01941}}].

\bibitem{White17}
M.~{White} and N.~{Padmanabhan}, \emph{{Matched filtering with interferometric
  21 cm experiments}},
  \href{https://doi.org/10.1093/mnras/stx1682}{\emph{\mnras} {\bfseries 471}
  (2017) 1167} [\href{https://arxiv.org/abs/1705.09669}{{\ttfamily
  1705.09669}}].

\bibitem{Obuljen18}
A.~{Obuljen}, E.~{Castorina}, F.~{Villaescusa-Navarro} and M.~{Viel},
  \emph{{High-redshift post-reionization cosmology with 21cm intensity
  mapping}}, \href{https://doi.org/10.1088/1475-7516/2018/05/004}{\emph{\jcap}
  {\bfseries 5} (2018) 004} [\href{https://arxiv.org/abs/1709.07893}{{\ttfamily
  1709.07893}}].

\bibitem{Chen19}
S.-F. {Chen}, E.~{Castorina}, M.~{White} and A.~{Slosar}, \emph{{Synergies
  between radio, optical and microwave observations at high redshift}},
  {\emph{ArXiv e-prints} (2018) }
  [\href{https://arxiv.org/abs/1810.00911}{{\ttfamily 1810.00911}}].

\bibitem{Cohn16}
J.~D. {Cohn}, M.~{White}, T.-C. {Chang}, G.~{Holder}, N.~{Padmanabhan} and
  O.~{Dor{\'e}}, \emph{{Combining galaxy and 21-cm surveys}},
  \href{https://doi.org/10.1093/mnras/stw108}{\emph{\mnras} {\bfseries 457}
  (2016) 2068} [\href{https://arxiv.org/abs/1511.07377}{{\ttfamily
  1511.07377}}].

\bibitem{Furlanetto06}
S.~R. {Furlanetto}, S.~P. {Oh} and F.~H. {Briggs}, \emph{{Cosmology at low
  frequencies: The 21 cm transition and the high-redshift Universe}},
  \href{https://doi.org/10.1016/j.physrep.2006.08.002}{\emph{\physrep}
  {\bfseries 433} (2006) 181}
  [\href{https://arxiv.org/abs/astro-ph/0608032}{{\ttfamily
  astro-ph/0608032}}].

\bibitem{Shaw14}
J.~R. {Shaw}, K.~{Sigurdson}, U.-L. {Pen}, A.~{Stebbins} and M.~{Sitwell},
  \emph{{All-sky Interferometry with Spherical Harmonic Transit Telescopes}},
  \href{https://doi.org/10.1088/0004-637X/781/2/57}{\emph{\apj} {\bfseries 781}
  (2014) 57} [\href{https://arxiv.org/abs/1302.0327}{{\ttfamily 1302.0327}}].

\bibitem{Pober15}
J.~C. {Pober}, \emph{{The impact of foregrounds on redshift space distortion
  measurements with the highly redshifted 21-cm line}},
  \href{https://doi.org/10.1093/mnras/stu2575}{\emph{\mnras} {\bfseries 447}
  (2015) 1705} [\href{https://arxiv.org/abs/1411.2050}{{\ttfamily 1411.2050}}].

\bibitem{Seo16}
H.-J. {Seo} and C.~M. {Hirata}, \emph{{The foreground wedge and 21-cm BAO
  surveys}}, \href{https://doi.org/10.1093/mnras/stv2806}{\emph{\mnras}
  {\bfseries 456} (2016) 3142}
  [\href{https://arxiv.org/abs/1508.06503}{{\ttfamily 1508.06503}}].

\bibitem{Shaw15}
J.~R. {Shaw}, K.~{Sigurdson}, M.~{Sitwell}, A.~{Stebbins} and U.-L. {Pen},
  \emph{{Coaxing cosmic 21 cm fluctuations from the polarized sky using m -mode
  analysis}}, \href{https://doi.org/10.1103/PhysRevD.91.083514}{\emph{\prd}
  {\bfseries 91} (2015) 083514}
  [\href{https://arxiv.org/abs/1401.2095}{{\ttfamily 1401.2095}}].

\bibitem{Byrne18}
R.~{Byrne}, M.~F. {Morales}, B.~{Hazelton}, W.~{Li}, N.~{Barry}, A.~P.
  {Beardsley} et~al., \emph{{Fundamental Limitations on the Calibration of
  Redundant 21 cm Cosmology Instruments and Implications for HERA and the
  SKA}}, {\emph{ArXiv e-prints} (2018) }
  [\href{https://arxiv.org/abs/1811.01378}{{\ttfamily 1811.01378}}].

\bibitem{Datta10}
A.~{Datta}, J.~D. {Bowman} and C.~L. {Carilli}, \emph{{Bright Source
  Subtraction Requirements for Redshifted 21 cm Measurements}},
  \href{https://doi.org/10.1088/0004-637X/724/1/526}{\emph{\apj} {\bfseries
  724} (2010) 526} [\href{https://arxiv.org/abs/1005.4071}{{\ttfamily
  1005.4071}}].

\bibitem{Morales12}
M.~F. {Morales}, B.~{Hazelton}, I.~{Sullivan} and A.~{Beardsley}, \emph{{Four
  Fundamental Foreground Power Spectrum Shapes for 21 cm Cosmology
  Observations}},
  \href{https://doi.org/10.1088/0004-637X/752/2/137}{\emph{\apj} {\bfseries
  752} (2012) 137} [\href{https://arxiv.org/abs/1202.3830}{{\ttfamily
  1202.3830}}].

\bibitem{Parsons12}
A.~R. {Parsons}, J.~C. {Pober}, J.~E. {Aguirre}, C.~L. {Carilli}, D.~C.
  {Jacobs} and D.~F. {Moore}, \emph{{A Per-baseline, Delay-spectrum Technique
  for Accessing the 21 cm Cosmic Reionization Signature}},
  \href{https://doi.org/10.1088/0004-637X/756/2/165}{\emph{\apj} {\bfseries
  756} (2012) 165} [\href{https://arxiv.org/abs/1204.4749}{{\ttfamily
  1204.4749}}].

\bibitem{Liu14}
A.~{Liu}, A.~R. {Parsons} and C.~M. {Trott}, \emph{{Epoch of reionization
  window. II. Statistical methods for foreground wedge reduction}},
  \href{https://doi.org/10.1103/PhysRevD.90.023019}{\emph{\prd} {\bfseries 90}
  (2014) 023019} [\href{https://arxiv.org/abs/1404.4372}{{\ttfamily
  1404.4372}}].

\bibitem{FengChuEtAl16}
Y.~{Feng}, M.-Y. {Chu}, U.~{Seljak} and P.~{McDonald}, \emph{{FASTPM: a new
  scheme for fast simulations of dark matter and haloes}},
  \href{https://doi.org/10.1093/mnras/stw2123}{\emph{\mnras} {\bfseries 463}
  (2016) 2273} [\href{https://arxiv.org/abs/1603.00476}{{\ttfamily
  1603.00476}}].

\bibitem{CAMB}
A.~Lewis, A.~Challinor and A.~Lasenby, \emph{{Efficient computation of CMB
  anisotropies in closed FRW models}},
  \href{https://doi.org/10.1086/309179}{\emph{\apj} {\bfseries 538} (2000) 473}
  [\href{https://arxiv.org/abs/astro-ph/9911177}{{\ttfamily
  astro-ph/9911177}}].

\bibitem{Angulo16}
R.~E. {Angulo} and A.~{Pontzen}, \emph{{Cosmological N-body simulations with
  suppressed variance}},
  \href{https://doi.org/10.1093/mnrasl/slw098}{\emph{\mnras} {\bfseries 462}
  (2016) L1} [\href{https://arxiv.org/abs/1603.05253}{{\ttfamily 1603.05253}}].

\bibitem{Chuang18}
C.-H. {Chuang}, G.~{Yepes}, F.-S. {Kitaura}, M.~{Pellejero-Ibanez},
  S.~{Rodriguez-Torres}, Y.~{Feng} et~al., \emph{{UNIT project: Universe
  \$N\$-body simulations for the Investigation of Theoretical models from
  galaxy surveys}}, {\emph{arXiv e-prints} (2018) arXiv:1811.02111}
  [\href{https://arxiv.org/abs/1811.02111}{{\ttfamily 1811.02111}}].

\bibitem{Ding}
Z.~{Ding}, H.-J. {Seo}, Z.~{Vlah}, Y.~{Feng}, M.~{Schmittfull} and
  F.~{Beutler}, \emph{{Theoretical systematics of Future Baryon Acoustic
  Oscillation Surveys}},
  \href{https://doi.org/10.1093/mnras/sty1413}{\emph{\mnras} {\bfseries 479}
  (2018) 1021} [\href{https://arxiv.org/abs/1708.01297}{{\ttfamily
  1708.01297}}].

\bibitem{TreePM}
M.~{White}, \emph{{The Mass Function}},
  \href{https://doi.org/10.1086/342752}{\emph{\apjs} {\bfseries 143} (2002)
  241} [\href{https://arxiv.org/abs/astro-ph/0207185}{{\ttfamily
  astro-ph/0207185}}].

\bibitem{Stark15a}
C.~W. {Stark}, M.~{White}, K.-G. {Lee} and J.~F. {Hennawi}, \emph{{Protocluster
  discovery in tomographic Ly {$\alpha$} forest flux maps}},
  \href{https://doi.org/10.1093/mnras/stv1620}{\emph{\mnras} {\bfseries 453}
  (2015) 311} [\href{https://arxiv.org/abs/1412.1507}{{\ttfamily 1412.1507}}].

\bibitem{Stark15b}
C.~W. {Stark}, A.~{Font-Ribera}, M.~{White} and K.-G. {Lee}, \emph{{Finding
  high-redshift voids using Lyman {$\alpha$} forest tomography}},
  \href{https://doi.org/10.1093/mnras/stv1868}{\emph{\mnras} {\bfseries 453}
  (2015) 4311} [\href{https://arxiv.org/abs/1504.03290}{{\ttfamily
  1504.03290}}].

\bibitem{ST99}
R.~K. {Sheth} and G.~{Tormen}, \emph{{Large-scale bias and the peak background
  split}},
  \href{https://doi.org/10.1046/j.1365-8711.1999.02692.x}{\emph{\mnras}
  {\bfseries 308} (1999) 119}
  [\href{https://arxiv.org/abs/astro-ph/9901122}{{\ttfamily
  astro-ph/9901122}}].

\bibitem{Crighton15}
N.~H.~M. {Crighton}, M.~T. {Murphy}, J.~X. {Prochaska}, G.~{Worseck},
  M.~{Rafelski}, G.~D. {Becker} et~al., \emph{{The neutral hydrogen
  cosmological mass density at z = 5}},
  \href{https://doi.org/10.1093/mnras/stv1182}{\emph{\mnras} {\bfseries 452}
  (2015) 217} [\href{https://arxiv.org/abs/1506.02037}{{\ttfamily
  1506.02037}}].

\bibitem{DESI}
{DESI Collaboration}, A.~{Aghamousa}, J.~{Aguilar}, S.~{Ahlen}, S.~{Alam},
  L.~E. {Allen} et~al., \emph{{The DESI Experiment Part I: Science,Targeting,
  and Survey Design}}, {\emph{ArXiv e-prints} (2016) }
  [\href{https://arxiv.org/abs/1611.00036}{{\ttfamily 1611.00036}}].

\bibitem{Behroozi18}
P.~{Behroozi}, R.~{Wechsler}, A.~{Hearin} and C.~{Conroy},
  \emph{{UniverseMachine: The Correlation between Galaxy Growth and Dark Matter
  Halo Assembly from z=0-10}}, {\emph{arXiv e-prints} (2018) arXiv:1806.07893}
  [\href{https://arxiv.org/abs/1806.07893}{{\ttfamily 1806.07893}}].

\bibitem{Moster13}
B.~P. {Moster}, T.~{Naab} and S.~D.~M. {White}, \emph{{Galactic star formation
  and accretion histories from matching galaxies to dark matter haloes}},
  \href{https://doi.org/10.1093/mnras/sts261}{\emph{\mnras} {\bfseries 428}
  (2013) 3121} [\href{https://arxiv.org/abs/1205.5807}{{\ttfamily 1205.5807}}].

\bibitem{Matsubara08}
T.~{Matsubara}, \emph{{Resumming cosmological perturbations via the Lagrangian
  picture: One-loop results in real space and in redshift space}},
  \href{https://doi.org/10.1103/PhysRevD.77.063530}{\emph{\prd} {\bfseries 77}
  (2008) 063530} [\href{https://arxiv.org/abs/0711.2521}{{\ttfamily
  0711.2521}}].

\bibitem{Carlson13}
J.~{Carlson}, B.~{Reid} and M.~{White}, \emph{{Convolution Lagrangian
  perturbation theory for biased tracers}},
  \href{https://doi.org/10.1093/mnras/sts457}{\emph{\mnras} {\bfseries 429}
  (2013) 1674} [\href{https://arxiv.org/abs/1209.0780}{{\ttfamily 1209.0780}}].

\bibitem{Vlah16}
Z.~{Vlah}, E.~{Castorina} and M.~{White}, \emph{{The Gaussian streaming model
  and convolution Lagrangian effective field theory}},
  \href{https://doi.org/10.1088/1475-7516/2016/12/007}{\emph{\jcap} {\bfseries
  12} (2016) 007} [\href{https://arxiv.org/abs/1609.02908}{{\ttfamily
  1609.02908}}].

\bibitem{Foreman16}
S.~{Foreman}, H.~{Perrier} and L.~{Senatore}, \emph{{Precision comparison of
  the power spectrum in the EFTofLSS with simulations}},
  \href{https://doi.org/10.1088/1475-7516/2016/05/027}{\emph{\jcap} {\bfseries
  5} (2016) 027} [\href{https://arxiv.org/abs/1507.05326}{{\ttfamily
  1507.05326}}].

\bibitem{Modi17}
C.~{Modi}, M.~{White} and Z.~{Vlah}, \emph{{Modeling CMB lensing cross
  correlations with CLEFT}},
  \href{https://doi.org/10.1088/1475-7516/2017/08/009}{\emph{\jcap} {\bfseries
  8} (2017) 009} [\href{https://arxiv.org/abs/1706.03173}{{\ttfamily
  1706.03173}}].

\bibitem{Vlah15}
Z.~{Vlah}, M.~{White} and A.~{Aviles}, \emph{{A Lagrangian effective field
  theory}}, \href{https://doi.org/10.1088/1475-7516/2015/09/014}{\emph{\jcap}
  {\bfseries 9} (2015) 014} [\href{https://arxiv.org/abs/1506.05264}{{\ttfamily
  1506.05264}}].

\bibitem{Zel70}
Y.~B. {Zel'dovich}, \emph{{Gravitational instability: An approximate theory for
  large density perturbations.}}, {\emph{\aap} {\bfseries 5} (1970) 84}.

\bibitem{Wilson19}
M.~{Wilson} and M.~{White}, \emph{{Cosmology with dropout selection I:
  Straw-man surveys and CMB lensing}}, {\emph{preprints} (2018) }.

\bibitem{Kai87}
N.~{Kaiser}, \emph{{Clustering in real space and in redshift space}},
  \href{https://doi.org/10.1093/mnras/227.1.1}{\emph{\mnras} {\bfseries 227}
  (1987) 1}.

\bibitem{H98}
A.~J.~S. {Hamilton}, \emph{{Linear Redshift Distortions: a Review}},  in
  \emph{The Evolving Universe} (D.~{Hamilton}, ed.), vol.~231 of
  \emph{Astrophysics and Space Science Library}, p.~185, 1998,
  \href{https://arxiv.org/abs/astro-ph/9708102}{{\ttfamily astro-ph/9708102}},
  \href{https://doi.org/10.1007/978-94-011-4960-0\_17}{DOI}.

\bibitem{CasWhi18a}
E.~{Castorina} and M.~{White}, \emph{{Beyond the plane-parallel approximation
  for redshift surveys}},
  \href{https://doi.org/10.1093/mnras/sty410}{\emph{\mnras} {\bfseries 476}
  (2018) 4403} [\href{https://arxiv.org/abs/1709.09730}{{\ttfamily
  1709.09730}}].

\bibitem{CasWhi18b}
E.~{Castorina} and M.~{White}, \emph{{The Zeldovich approximation and
  wide-angle redshift-space distortions}},
  \href{https://doi.org/10.1093/mnras/sty1437}{\emph{\mnras} {\bfseries 479}
  (2018) 741} [\href{https://arxiv.org/abs/1803.08185}{{\ttfamily
  1803.08185}}].

\bibitem{Vlah18}
Z.~{Vlah} and M.~{White}, \emph{{Exploring redshift-space distortions in
  large-scale structure}}, {\emph{arXiv e-prints} (2018) }
  [\href{https://arxiv.org/abs/1812.02775}{{\ttfamily 1812.02775}}].

\bibitem{Weinberg13}
D.~H. {Weinberg}, M.~J. {Mortonson}, D.~J. {Eisenstein}, C.~{Hirata}, A.~G.
  {Riess} and E.~{Rozo}, \emph{{Observational probes of cosmic acceleration}},
  \href{https://doi.org/10.1016/j.physrep.2013.05.001}{\emph{\physrep}
  {\bfseries 530} (2013) 87} [\href{https://arxiv.org/abs/1201.2434}{{\ttfamily
  1201.2434}}].

\bibitem{Bharadwaj98}
S.~{Bharadwaj}, \emph{{Perturbative Growth of Cosmological Clustering. II. The
  Two-Point Correlation}}, \href{https://doi.org/10.1086/176950}{\emph{\apj}
  {\bfseries 460} (1996) 28}
  [\href{https://arxiv.org/abs/astro-ph/9511085}{{\ttfamily
  astro-ph/9511085}}].

\bibitem{Meiksin99}
A.~{Meiksin}, M.~{White} and J.~A. {Peacock}, \emph{{Baryonic signatures in
  large-scale structure}},
  \href{https://doi.org/10.1046/j.1365-8711.1999.02369.x}{\emph{\mnras}
  {\bfseries 304} (1999) 851}
  [\href{https://arxiv.org/abs/astro-ph/9812214}{{\ttfamily
  astro-ph/9812214}}].

\bibitem{ESW07}
D.~J. {Eisenstein}, H.-J. {Seo} and M.~{White}, \emph{{On the Robustness of the
  Acoustic Scale in the Low-Redshift Clustering of Matter}},
  \href{https://doi.org/10.1086/518755}{\emph{\apj} {\bfseries 664} (2007) 660}
  [\href{https://arxiv.org/abs/astro-ph/0604361}{{\ttfamily
  astro-ph/0604361}}].

\bibitem{Smith08}
R.~E. {Smith}, R.~{Scoccimarro} and R.~K. {Sheth}, \emph{{Motion of the
  acoustic peak in the correlation function}},
  \href{https://doi.org/10.1103/PhysRevD.77.043525}{\emph{\prd} {\bfseries 77}
  (2008) 043525} [\href{https://arxiv.org/abs/astro-ph/0703620}{{\ttfamily
  astro-ph/0703620}}].

\bibitem{Crocce08}
M.~{Crocce} and R.~{Scoccimarro}, \emph{{Nonlinear evolution of baryon acoustic
  oscillations}}, \href{https://doi.org/10.1103/PhysRevD.77.023533}{\emph{\prd}
  {\bfseries 77} (2008) 023533}
  [\href{https://arxiv.org/abs/0704.2783}{{\ttfamily 0704.2783}}].

\bibitem{Seo08}
H.-J. {Seo}, E.~R. {Siegel}, D.~J. {Eisenstein} and M.~{White},
  \emph{{Nonlinear Structure Formation and the Acoustic Scale}},
  \href{https://doi.org/10.1086/589921}{\emph{\apj} {\bfseries 686} (2008) 13}
  [\href{https://arxiv.org/abs/0805.0117}{{\ttfamily 0805.0117}}].

\bibitem{White14}
M.~{White}, \emph{{The Zel'dovich approximation}},
  \href{https://doi.org/10.1093/mnras/stu209}{\emph{\mnras} {\bfseries 439}
  (2014) 3630} [\href{https://arxiv.org/abs/1401.5466}{{\ttfamily 1401.5466}}].

\bibitem{White15}
M.~{White}, \emph{{Reconstruction within the Zeldovich approximation}},
  \href{https://doi.org/10.1093/mnras/stv842}{\emph{\mnras} {\bfseries 450}
  (2015) 3822} [\href{https://arxiv.org/abs/1504.03677}{{\ttfamily
  1504.03677}}].

\bibitem{ES3}
D.~J. {Eisenstein}, H.-J. {Seo}, E.~{Sirko} and D.~N. {Spergel},
  \emph{{Improving Cosmological Distance Measurements by Reconstruction of the
  Baryon Acoustic Peak}}, \href{https://doi.org/10.1086/518712}{\emph{\apj}
  {\bfseries 664} (2007) 675}
  [\href{https://arxiv.org/abs/astro-ph/0604362}{{\ttfamily
  astro-ph/0604362}}].

\bibitem{FVN2017}
F.~{Villaescusa-Navarro}, D.~{Alonso} and M.~{Viel}, \emph{{Baryonic acoustic
  oscillations from 21 cm intensity mapping: the Square Kilometre Array case}},
  \href{https://doi.org/10.1093/mnras/stw3224}{\emph{\mnras} {\bfseries 466}
  (2017) 2736} [\href{https://arxiv.org/abs/1609.00019}{{\ttfamily
  1609.00019}}].

\bibitem{McQuinn11a}
M.~McQuinn, L.~Hernquist, A.~Lidz and M.~Zaldarriaga, \emph{{The signatures of
  large-scale temperature and intensity fluctuations in the Lyman $\alpha$
  forest}},
  \href{https://doi.org/10.1111/j.1365-2966.2011.18788.x}{\emph{Monthly Notices
  of the Royal Astronomical Society} {\bfseries 415} (2011) 977}
  [\href{https://arxiv.org/abs/http://oup.prod.sis.lan/mnras/article-pdf/415/1/977/3134997/mnras0415-0977.pdf}{{\ttfamily
  http://oup.prod.sis.lan/mnras/article-pdf/415/1/977/3134997/mnras0415-0977.pdf}}].

\bibitem{Pontzen14}
A.~{Pontzen}, \emph{{Scale-dependent bias in the
  baryonic-acoustic-oscillation-scale intergalactic neutral hydrogen}},
  \href{https://doi.org/10.1103/PhysRevD.89.083010}{\emph{\prd} {\bfseries 89}
  (2014) 083010} [\href{https://arxiv.org/abs/1402.0506}{{\ttfamily
  1402.0506}}].

\bibitem{Cabass}
G.~{Cabass} and F.~{Schmidt}, \emph{{A new scale in the bias expansion}},
  {\emph{arXiv e-prints} (2018) arXiv:1812.02731}
  [\href{https://arxiv.org/abs/1812.02731}{{\ttfamily 1812.02731}}].

\bibitem{Gontcho14}
S.~Gontcho A~Gontcho, J.~Miralda-Escudé and N.~G. Busca, \emph{{On the effect
  of the ionizing background on the Ly$\alpha$ forest autocorrelation
  function}}, \href{https://doi.org/10.1093/mnras/stu860}{\emph{Mon. Not. Roy.
  Astron. Soc.} {\bfseries 442} (2014) 187}
  [\href{https://arxiv.org/abs/1404.7425}{{\ttfamily 1404.7425}}].

\bibitem{Meiksin18}
A.~Meiksin and M.~McQuinn, \emph{{Time-dependent fluctuations in the
  metagalactic photoionization background}},
  \href{https://doi.org/10.1093/mnras/sty2907}{\emph{Mon. Not. Roy. Astron.
  Soc.} {\bfseries 482} (2019) 4777}
  [\href{https://arxiv.org/abs/1809.08645}{{\ttfamily 1809.08645}}].

\bibitem{Sanderbeck18}
P.~{Upton Sanderbeck}, V.~{Irsic}, M.~{McQuinn} and A.~{Meiksin},
  \emph{{Estimates for the impact of Ultraviolet Background fluctuations on
  galaxy clustering measurements}}, {\emph{arXiv e-prints} (2018) }
  [\href{https://arxiv.org/abs/1810.12321}{{\ttfamily 1810.12321}}].

\bibitem{Miralda00}
J.~{Miralda-Escud{\'e}}, M.~{Haehnelt} and M.~J. {Rees}, \emph{{Reionization of
  the Inhomogeneous Universe}},
  \href{https://doi.org/10.1086/308330}{\emph{\apj} {\bfseries 530} (2000) 1}
  [\href{https://arxiv.org/abs/astro-ph/9812306}{{\ttfamily
  astro-ph/9812306}}].

\bibitem{McGreer18}
I.~D. {McGreer}, X.~{Fan}, L.~{Jiang} and Z.~{Cai}, \emph{{The Faint End of the
  z = 5 Quasar Luminosity Function from the CFHTLS}},
  \href{https://doi.org/10.3847/1538-3881/aaaab4}{\emph{\aj} {\bfseries 155}
  (2018) 131} [\href{https://arxiv.org/abs/1710.09390}{{\ttfamily
  1710.09390}}].

\bibitem{Kulkarni18}
G.~{Kulkarni}, G.~{Worseck} and J.~F. {Hennawi}, \emph{{Evolution of the AGN UV
  luminosity function from redshift 7.5}}, {\emph{ArXiv e-prints} (2018) }
  [\href{https://arxiv.org/abs/1807.09774}{{\ttfamily 1807.09774}}].

\bibitem{ConWhite13}
C.~{Conroy} and M.~{White}, \emph{{A Simple Model for Quasar Demographics}},
  \href{https://doi.org/10.1088/0004-637X/762/2/70}{\emph{\apj} {\bfseries 762}
  (2013) 70} [\href{https://arxiv.org/abs/1208.3198}{{\ttfamily 1208.3198}}].

\bibitem{Faucher19}
{Faucher-Giguere} and C.~{-A.}, \emph{{A 2019 Cosmic UV/X-ray Background Model
  Update}}, {\emph{arXiv e-prints} (2019) arXiv:1903.08657}
  [\href{https://arxiv.org/abs/1903.08657}{{\ttfamily 1903.08657}}].

\bibitem{Modi16}
C.~{Modi}, E.~{Castorina} and U.~{Seljak}, \emph{{Halo bias in Lagrangian
  space: estimators and theoretical predictions}},
  \href{https://doi.org/10.1093/mnras/stx2148}{\emph{\mnras} {\bfseries 472}
  (2017) 3959} [\href{https://arxiv.org/abs/1612.01621}{{\ttfamily
  1612.01621}}].

\bibitem{Meiksin09}
A.~A. {Meiksin}, \emph{{The physics of the intergalactic medium}},
  \href{https://doi.org/10.1103/RevModPhys.81.1405}{\emph{Reviews of Modern
  Physics} {\bfseries 81} (2009) 1405}
  [\href{https://arxiv.org/abs/0711.3358}{{\ttfamily 0711.3358}}].

\bibitem{McQuinn16}
M.~{McQuinn}, \emph{{The Evolution of the Intergalactic Medium}},
  \href{https://doi.org/10.1146/annurev-astro-082214-122355}{\emph{\araa}
  {\bfseries 54} (2016) 313}
  [\href{https://arxiv.org/abs/1512.00086}{{\ttfamily 1512.00086}}].

\bibitem{DEFW}
M.~{Davis}, G.~{Efstathiou}, C.~S. {Frenk} and S.~D.~M. {White}, \emph{{The
  evolution of large-scale structure in a universe dominated by cold dark
  matter}}, \href{https://doi.org/10.1086/163168}{\emph{\apj} {\bfseries 292}
  (1985) 371}.

\bibitem{FengModi18}
Y.~{Feng} and C.~{Modi}, \emph{{A fast algorithm for identifying
  friends-of-friends halos}},
  \href{https://doi.org/10.1016/j.ascom.2017.05.004}{\emph{Astronomy and
  Computing} {\bfseries 20} (2017) 44}
  [\href{https://arxiv.org/abs/1607.03224}{{\ttfamily 1607.03224}}].

\bibitem{Gadget2}
V.~{Springel}, \emph{{The cosmological simulation code GADGET-2}},
  \href{https://doi.org/10.1111/j.1365-2966.2005.09655.x}{\emph{\mnras}
  {\bfseries 364} (2005) 1105}
  [\href{https://arxiv.org/abs/astro-ph/0505010}{{\ttfamily
  astro-ph/0505010}}].

\bibitem{mpsort}
Y.~{Feng}, M.~{Straka}, T.~{Di Matteo} and R.~{Croft}, \emph{Parallel sorting
  at scale on bluewaters in cosmological simulations},  in \emph{Cray User
  Group 2015}, April, 2015.

\bibitem{bigfile}
Y.~Feng, S.~Bird and F.~Lanusse, \emph{rainwoodman/bigfile 0.1.39},  Nov.,
  2017.
\newblock 10.5281/zenodo.1051252.

\bibitem{Reed05}
D.~{Reed}, F.~{Governato}, T.~{Quinn}, J.~{Gardner}, J.~{Stadel} and G.~{Lake},
  \emph{{Dark matter subhaloes in numerical simulations}},
  \href{https://doi.org/10.1111/j.1365-2966.2005.09020.x}{\emph{\mnras}
  {\bfseries 359} (2005) 1537}
  [\href{https://arxiv.org/abs/astro-ph/0406034}{{\ttfamily
  astro-ph/0406034}}].

\bibitem{Angulo09}
R.~E. {Angulo}, C.~G. {Lacey}, C.~M. {Baugh} and C.~S. {Frenk}, \emph{{The fate
  of substructures in cold dark matter haloes}},
  \href{https://doi.org/10.1111/j.1365-2966.2009.15333.x}{\emph{\mnras}
  {\bfseries 399} (2009) 983}
  [\href{https://arxiv.org/abs/0810.2177}{{\ttfamily 0810.2177}}].

\bibitem{Brook14}
C.~B. {Brook}, A.~{Di Cintio}, A.~{Knebe}, S.~{Gottl{\"o}ber}, Y.~{Hoffman},
  G.~{Yepes} et~al., \emph{{The Stellar-to-halo Mass Relation for Local Group
  Galaxies}}, \href{https://doi.org/10.1088/2041-8205/784/1/L14}{\emph{\apjl}
  {\bfseries 784} (2014) L14}
  [\href{https://arxiv.org/abs/1311.5492}{{\ttfamily 1311.5492}}].

\bibitem{NFW}
J.~F. {Navarro}, C.~S. {Frenk} and S.~D.~M. {White}, \emph{{A Universal Density
  Profile from Hierarchical Clustering}},
  \href{https://doi.org/10.1086/304888}{\emph{\apj} {\bfseries 490} (1997) 493}
  [\href{https://arxiv.org/abs/astro-ph/9611107}{{\ttfamily
  astro-ph/9611107}}].

\bibitem{Evrard08}
A.~E. {Evrard}, J.~{Bialek}, M.~{Busha}, M.~{White}, S.~{Habib}, K.~{Heitmann}
  et~al., \emph{{Virial Scaling of Massive Dark Matter Halos: Why Clusters
  Prefer a High Normalization Cosmology}},
  \href{https://doi.org/10.1086/521616}{\emph{\apj} {\bfseries 672} (2008) 122}
  [\href{https://arxiv.org/abs/astro-ph/0702241}{{\ttfamily
  astro-ph/0702241}}].

\bibitem{Worseck14}
G.~{Worseck}, J.~X. {Prochaska}, J.~M. {O'Meara}, G.~D. {Becker}, S.~L.
  {Ellison}, S.~{Lopez} et~al., \emph{{The Giant Gemini GMOS survey of z$_{em}$
  \&gt; 4.4 quasars - I. Measuring the mean free path across cosmic time}},
  \href{https://doi.org/10.1093/mnras/stu1827}{\emph{\mnras} {\bfseries 445}
  (2014) 1745} [\href{https://arxiv.org/abs/1402.4154}{{\ttfamily 1402.4154}}].

\bibitem{McQuinn11}
M.~{McQuinn}, S.~P. {Oh} and C.-A. {Faucher-Gigu{\`e}re}, \emph{{On Lyman-limit
  Systems and the Evolution of the Intergalactic Ionizing Background}},
  \href{https://doi.org/10.1088/0004-637X/743/1/82}{\emph{\apj} {\bfseries 743}
  (2011) 82} [\href{https://arxiv.org/abs/1101.1964}{{\ttfamily 1101.1964}}].

\end{thebibliography}\endgroup
\end{document}